\documentclass[12pt,cite,epsfig]{article}
\usepackage{times,epsfig}
\newcommand{\nc}{\newcommand}
\nc{\bb}{\bibitem}
\nc{\be}{\begin{equation}}
\nc{\ee}{\end{equation}}
\nc{\pa}{\partial}
\nc{\parsym} {\stackrel{\leftrightarrow}{\pa}}
\nc{\ra}{\rightarrow}
\nc{\la}{\leftarrow}
\nc{\etp}{{\eta^\prime}}
\nc{\omg}{\omega}
\nc{\ggam}{\gamma \gamma}
\nc{\gam}{\gamma }
\nc{\bea}{\begin{eqnarray}}
\nc{\eea}{\end{eqnarray}}
\nc{\beas}{\begin{eqnarray*}}
\nc{\eeas}{\end{eqnarray*}}
\nc{\non}{\nonumber}

\def\hhhb{\rule[-3.mm]{0.mm}{9.mm}}
\def\hhhc{\rule[-3.mm]{0.mm}{3.mm}}
\def\hhhd{\rule[-3.mm]{0.mm}{2.mm}}

\def\hhhu{\rule[-3.mm]{0.mm}{12.mm}}
\def\hhhv{\rule[-3.mm]{0.mm}{9.mm}}

\begin{document}
\begin{titlepage}
\vbox{~~~ \\
                                   \null \hfill LPNHE 2007--05\\
                                  \null \hfill  FERMILAB-PUB-07-597-BSS\\

\title{The Dipion Mass Spectrum 
In $e^+e^-$ Annihilation and $\tau$ Decay~:  
A Dynamical ($\rho$, $\omg$, $\phi$) Mixing 
 Approach
   }
\author{
M.~Benayoun$^{(a)}$, P.~David$^{(a)}$, L.~ DelBuono$^{(a)}$, \\
O.~Leitner$^{(a,b)}$  and H.~B.~O'Connell$^{(c)}$ \\
\small{$^{(a)}$ LPNHE Paris VI/VII, IN2P3/CNRS, F-75252 Paris, France }\\
\small{$^{(b)}$  Laboratori Nazionali di Frascati, INFN, I-00044 Frascati (Roma) Italy}\\
\small{$^{(c)}$ Fermilab,  Batavia IL 60510, USA.}\\
}
\date{\today}
\maketitle
\begin{abstract}
We readdress the problem of finding a simultaneous description of the pion form factor data
in $e^+e^-$ annihilations and in $\tau$ decays. For this purpose, we work in the framework of
 the Hidden Local Symmetry (HLS) Lagrangian and modify the vector meson mass term by including
the pion and kaon loop contributions. This leads us to define the physical $\rho$, $\omg$ and
$\phi$  fields as linear combinations of their ideal partners, with coefficients 
being meromorphic functions of $s$, the square of the 4--momentum flowing into the vector
meson lines. This allows us to define a dynamical, {\it i.e.} $s$-dependent, vector meson mixing scheme.
The model is overconstrained by extending the framework in order to include the description of 
all meson radiative ($VP\gamma$  and $P\gamma \gamma$ couplings) and leptonic 
($Ve^+e^-$ couplings) decays and also the isospin breaking ($\omg/\phi \ra \pi^+ \pi^-)$ 
decay modes. The model provides a simultaneous, consistent and good description of the 
$e^+e^-$ and $\tau$ dipion spectra. The expression for pion form factor in the latter case is 
derived from those in the former case by switching off the isospin breaking 
effects specific to $e^+e^-$ and switching on those for  $\tau$ decays.
Besides, the model also provides a good account of all decay
modes of the form $VP\gamma$, $P\gamma \gamma$  as well as the isospin breaking decay modes.
It leads us to propose new reference values for the $\rho^0 \ra e^+ e^-$ and 
$\omg \ra \pi^+ \pi^-$ partial widths which are part of our description of the pion form factor.
Other topics ($\phi \ra K \overline{K}$, the $\rho$ meson mass and width parameters) are briefly
discussed. 
As the $e^+e^-$ data are found perfectly consistent with $\tau$ data up to identified
isospin breaking effects, one finds no reason to cast any doubt on them and, therefore, 
on the theoretical estimate of the muon
anomalous moment $a_\mu$ derived from them. Therefore, our work turns out to confirm
the relevance of the reported  3.3 $\sigma$ discrepancy between this theoretical 
estimate of $a_\mu$ and its  direct BNL measurement.

\end{abstract}
}
\end{titlepage}

\section{Introduction}
\label{introduction}
\indent \indent In order to study the phenomenology of light flavor mesons below 1 GeV, 
like partial decay widths or meson form factors, one needs a framework which includes in a well 
defined manner the lowest mass nonets of pseudocalar (P) and vector (V) mesons. Such a framework 
is well represented by the Hidden Local symmetry (HLS) model \cite{HLSOrigin,HLSRef}. In this approach,
vector mesons are gauge bosons of a spontaneously  broken  hidden local symmetry which generates their
(Higgs--Kibble, HK) 
masses. Besides the non--anomalous sector, this model has an anomalous sector, hereafter called
FKTUY Lagrangian \cite{FKTUY}, which aims at describing couplings of the form $VVP$, $VP \gamma$
$P \gamma \gamma $, $VPPP$ for light flavor mesons.

In its original form, the full (non--anomalous and anomalous) 
bare HLS Lagrangian fulfills the $U(3)$ symmetry,
as it possesses both Nonet Symmetry and $SU(3)$ flavor symmetry. As such,  
the HLS model covers a limited phenomenological scope. In order to broaden this scope,
especially in order to account for $VP \gamma$ and $P\gamma \gamma$ couplings as derived
from measured partial widths, the original HLS model should be supplemented with  
symmetry breaking mechanisms.  

Several  $SU(3)$ symmetry breaking schemes have been proposed \cite{BGP,BGPbis,Heath1} following
the original idea of Bando, Kugo and Yamawaki (BKY) \cite{BKY}. It has been shown
\cite{rad,mixing}  that the most successful  variant is the so--called ``new scheme" 
of Ref. \cite{Heath1} briefly recalled in Appendix \ref{CC}. However,  breaking only the $SU(3)$ symmetry is insufficient
\cite{rad,mixing} in order to reach a satisfactory description of the data on  $VP \gamma$ and 
$P\gamma \gamma$ couplings which requires one to also break  Nonet Symmetry. 
 This was first performed in an {\it ad hoc} manner in \cite{rad}
with the aim of recovering the radiative decay couplings of  O'Donnell \cite{ODonnell}
which is the most general set  fulfilling only $SU(3)$ symmetry~; therefore, the model developed
in \cite{rad} was indeed in agreement with general group theoretical considerations with
additionally a mechanism smoothly breaking $SU(3)$ flavor symmetry.

Slightly later, it was shown \cite{mixing,WZWChPT} that an appropriate mechanism for Nonet Symmetry breaking 
can be produced by adding determinant terms \cite{tHooft} to the HLS Lagrangian~; the result
was shown to meet all properties of Extended Chiral  Perturbation 
Theory (EChPT) \cite{leutw,leutwb,feldmann} at leading order in the breaking parameters. 
Additionally, 
it was also proved \cite{WZWChPT} that the Nonet Symmetry breaking mechanism proposed in \cite{rad} 
was indeed an appropriate approximation of the (rigorous) mechanism derived from 
adding the determinant 
terms to the bare HLS Lagrangian.

However, breaking  $SU(3)$ and Nonet symmetries is still not enough to be in position of describing
fully the whole set of radiative decays  $VP \gamma$~; indeed, a process like $\phi \ra \pi^0 \gamma$  
requires including a $\omg-\phi$ mixing scheme~; additionally, any global fit
of all available $VP \gamma$ transitions cannot be successful without introducing such a mixing. 
Traditionally, the   $\omg-\phi$ mixing is described \cite{rad,mixing} by rotating the fields $\omg_I$
and $\phi_I$ which are the entries of the bare vector field matrix, generally
called  ideal fields. An angle involving the mixing of the $\eta$ and $\eta^\prime$  mesons is also 
required which has been shown \cite{WZWChPT} to vanish in the limit of exact $SU(3)$ symmetry
of the HLS Lagrangian. 

A brief account of the HLS model in its anomalous (FKTUY) and non--anomalous sectors is given in 
Appendices \ref{AA} and \ref{BB}, mostly focused on the subject of this work. Appendix \ref{CC} 
describes shortly but fully the various symmetry breaking procedures and the field renormalization scheme,
except for the $SU(2)$ breaking mechanism which is the main subject of this paper.
Appendix \ref{YM} deals with some specific features of the Yang--Mills piece of the HLS Lagrangian
which provides the kinetic energy part of the vector meson fields and should also undergo
SU(3) symmetry breaking.

These Appendices illustrate  that the model depends on a very few number of free parameters
as clear from Appendix \ref{DD} where they all appear.
Among these free parameters, some are specific to our HLS based model, but some others are constrained.
For instance, one of the two SU(3) breaking parameters is the ratio of decay constants $f_K/f_\pi$ 
and can be constrained by the corresponding measured value.

On another hand, decay processes like $\omg/\phi \ra \pi^+ \pi^-$ can hardly be understood
without some scheme for $\rho^0-\omg$ and  $\rho^0-\phi$ mixings. Moreover, the decay behaviour
of $\rho^0$ and $\rho^\pm$ cannot differ at the coupling constant level.
It happens that the HLS model at one loop order provides a mechanism which allows us
to perform a full $\rho^0-\omg-\phi$ mixing starting from the corresponding ideal (bare) fields.
Within the non--anomalous HLS Lagrangian, the precise mechanism, kaon loop contributions, has been already 
described in \cite{su2}. However, we shall see that the (FKTUY) anomalous sector provides
$K^* K$ loops as additional mechanism and that another one ($K^*\overline{K}^*$ loops) 
is provided by the Yang--Mills piece.

Within the HLS model, as recalled below, $ V_i \leftrightarrow V_j$ transitions
among the ideal $\rho^0_I$, $\omg_I$ and $\phi_I$  fields are generated by loop effects~; 
these transitions generally 
have not a constant amplitude but rather depend on the squared 4--momentum flowing through
the $V_i$ (and $V_j$) line(s). We show in Section \ref{OneLoop} that charged and neutral
kaon, $K^* K$ and $K^*\overline{K}^*$ loops come through their sum in  the $ \omg_I\leftrightarrow \phi_I$ amplitude, 
while they come 
through their difference in $\rho^0_I \leftrightarrow \phi_I$ and $\rho^0_I \leftrightarrow \omg_I$ 
amplitudes. This also means that the $\omg - \phi$ mixing proceeds from quantum effects rather 
than symmetry breaking effects, in contrast with the $\rho^0-\omg$ or $\rho^0-\phi$ mixings. 
These transition amplitudes are given by Dispersion Relations which should
be subtracted  in order to make the integral convergent (see, for instance, Appendix A in
\cite{mixing}). This gives rise to polynomials with real coefficients to be fixed 
using external renormalization conditions. 

In the exact SU(3) symmetry limit, charged and neutral kaons carry the same mass
and one can expect (or require) the polynomials associated with the charged and neutral
kaon loops to coincide. Likewise, the charged and neutral $K^* K$ loop functions can be made equal.
In this case, the $ \omg_I\leftrightarrow \phi_I$ amplitude
survives with its renormalization polynomial, while the $\rho^0_I \leftrightarrow \phi_I$ 
and $\rho^0_I \leftrightarrow \omg_I$ transition amplitudes exactly vanish. If
one breaks the SU(3) flavor symmetry leaving conserved the $(u,d)$ quark sector, the same 
conclusion holds. However, if one introduces a breaking of $SU(2)$ flavor symmetry,
the mass difference beween $u$ and $d$ quarks generates a  mass difference between
the charged and neutral kaons (and $K^*$'s). Then,
the three possible transition amplitudes do not identically vanish any longer and, moreover,
they depend on the invariant mass  associated by the 4-momentum flowing through the 
vector  meson lines. Stated otherwise~: the $\rho^0_I$, $\omg_I$ and $\phi_I$
mixing into the physical $\rho^0$, $\omg$ and $\phi$ fields should also be invariant mass
dependent. As already noted in \cite{mixing,su2} this implies that the vector squared
mass matrix, which has to be diagonalized in order to define the physical fields, is also invariant mass
dependent and that the notion of mass for the physical vector fields becomes unclear
as soon as one goes beyond tree level.

The property that $\rho^0-\omg$ mixing should be invariant mass dependent has been the subject of several
studies  in the framework of general local effective field theories \cite{RhoOmg0}, then in
Vector Dominance Models \cite{RhoOmg1,RhoOmg2} where it was pointed out that the mixing amplitude
should vanish  at $s=0$, as the $\rho^0$ and $\omg$ self--masses, in order to preserve gauge invariance.
 One may also quote other studies
going in the same direction \cite{RhoOmg3,RhoOmg4} attributing the mixing to  finite width effects,
or quark loops (and pion loops). Using only the pion form factor data available at that time, 
Ref. \cite{ RhoOmg5}
derived an approximate expression for the $\rho^0-\omg$ mixing amplitude~; however, 
limiting that much the  kind  of data used, one cannot really observe a clear mass dependence effect. 
On the other hand, one  should also note that Ref. \cite{RhoOmg6} proved  that isospin violation effects
describing the $\rho^0-\omg$ mixing
are not accounted for by the low energy constants (LEC) of Chiral Perturbation Theory (ChPT) but
are generated near threshold by the difference between charged and neutral kaon loops~;
it was also shown that these effects are tiny at the two--pion threshold ($10^{-4}$) while they
are known to be at the percent level in resonance peak region. 
This illustrates the 
invariant mass dependence of the $\rho^0-\omg$ mixing.
However, one should note that the absence of the LECs in this calculation
suggests a one-loop result may be unreliable.

Therefore, the question we address is to consider the effects of loops on 
$\rho^0,~\omg,~\phi$ mixing. However, in order to have some chance to single out the
mass dependent behaviour, one clearly has to treat simultaneously the pion form factor data
in the spacelike and timelike regions together with the largest possible set of light
meson decays (radiative, leptonic and  isospin violating strong decays) within a single
framework. As argued above, the HLS model, equiped with suitable symmetry breaking mechanisms,
seems able to provide such a framework. 
We shall not try to include Isospin Symmetry breaking effects
into the coupling constants  which would help by providing more parameter freedom in the fit
procedures. If really needed, it can certainly be done along the lines of the BKY breaking scheme 
\cite{BKY,Heath1} as illustrated by \cite{Hashimoto}.

\vspace{0.5cm}
\indent \indent 
The question of having a unified description of the largest possible set of low energy
data is by itself interesting. However, this also addresses the puzzling 
and long--standing problem of the difference between the (estimated) 
isospin 1 part of the pion form factor in $e^+e^-$ annihilations and in $\tau$ decays
which are related through the Conserved Vector Current assuption (CVC). The importance of the problem is 
enhanced by its
implication for the predicted value  of the muon anomalous magnetic moment $a_\mu$ to be
compared with the direct BNL  measurement \cite{BNL}~; referring to the latest
account by M. Davier \cite{Davier2007} the estimate of the hadronic vacuum polarization 
(which enters the theoretical estimate)
derived from $e^+e^-$ data provides a 3.3 $\sigma$ disagreement between the theoretical estimation
of $a_\mu$ and the BNL direct measurement \cite{BNL}~;  moreover, the $\tau$ data estimate of the hadronic vacuum 
polarization provides a value of $a_\mu$ very close to its direct measurement \cite{Davier2007}. 

Except for an experimental problem with $e^+e^-$ annihilation data (which seems by now
unlikely)  in the  data recently collected at Novosibirsk
\cite{CMD2-1995corr,CMD2-1998-1,CMD2-1998-2,SND-1998}, or some new (or unidentified) physics effect,
the disagreement between $e^+e^-$ and $\tau$ data \cite{Aleph,Opal,Cleo} is hard to explain. Indeed, {\it a priori}
the single difference between these two channels, should be due to Isospin Symmetry breaking (IB) 
effects. However, the comparison has been already performed with  IB effects accounted for
in both $e^+e^-$ and $\tau$ data. This includes \cite{DavierPrevious1,DavierPrevious2,DavierPrevious3}
pion mass values in kinematical factors, (a parametrization of the) $\rho-\omg$ mixing, charged
and neutral $\rho$  mass and width differences, short range \cite{Marciano}
and long range \cite{Cirigliano1,Cirigliano2,Cirigliano3,Mexico1,Mexico2,Mexico3} IB effects in the
$\tau$ partial decay width to two pions. 

This persistent disagreement may point towards new physics effects \cite{Fred1}~; however,
one should also note that the way some IB effects are accounted for  
has been questioned several times. For instance, effects due to the charged and neutral $\rho$ pole positions \cite{Fred2,Fred3}
were considered, but have not been found sufficient in order to solve the observed discrepancy
\cite{DavierPrevious2,DavierPrevious3}~;  $\rho-\omega$ mixing effects may also have been poorly estimated
\cite{Maltman1}. However, based on sum rules derived  using an OPE input, 
K. Maltman \cite{Maltman2,Maltman3} concluded there is inconsistency of the (presently) estimated
isospin 1 part of $e^+e^-$ data with expectation, while $\tau$ data provide a nice agreement.

We plan to address this question once again by building an effective model relying on the (symmetry broken)
HLS model. In this approach, we plan to have a framework giving simultaneously an account of the
partial decay widths of light mesons decays (radiative, leptonic, isospin breaking decay modes), of
the pion form factor in $e^+e^-$ data (both timelike and close spacelike regions) 
and in $\tau$ decay. By construction, the
corresponding expressions of the pion form factor will be such that they will solely differ
from each other by Isospin Symmetry breaking effects, mostly located in the $(\rho,~\omega,~\phi)$
mixing scheme which, of course, has no counterpart in  $\tau$ decay. 
Stated otherwise, our model is built in such a way that going from
the pion form factor expression in $e^+e^-$ annihilations to the expression valid for $\tau$ decays
is performed by switching off IB effects specific to $e^+e^-$ and switching on those specific
to $\tau$ decays.

At start, the model we built is rendered complicated by the large number of possible loops
involved. Fortunately, it can be somewhat simplified without loosing too much physics insight.
Of course, this model depends
on some parameters to be fixed in a fitting procedure~; we should define a fitting procedure
flexible enough that $\tau$ decay data can be removed or kept. Stated otherwise, the  
light mesons decays and the pion form factor in $e^+e^-$ data are expected to fix practically
the U(3)/SU(3)/SU(2) breaking model. In this approach, we may get a prediction of the $\tau$
decay 2--pion spectrum which can be compared  with the existing measurements. 
Including $\tau$ decay data should only refine
the values of the fitted parameters.

\vspace{0.5cm}
\indent \indent The paper is organized as follows~: In Section \ref{HLS1}, we derive the Lagrangian pieces
of relevance in order to deal with the pion form factor in $e^+e^-$ annihilations and $\tau$ decays, while
in Section \ref{FFunbrk} we give the pion form factor expressions without loop corrections and
symmetry breaking effects, mostly for illustration.  In Section \ref{OneLoop} we discuss the loop corrections 
which modify the vector meson mass matrix and perform already some simplifications. The modified vector meson 
squared mass matrix we propose is
given in Section \ref{diagonal} with the diagonalization procedure and the relation between physical
and ideal vector meson fields. The method used in order to renormalize the loop functions defining
the self--energies and the transition amplitudes is sketched in Section \ref{Renormalization}. The 
form factor functions used for $\tau$ decays and $e^+e^-$ annihilations are given in resp.
Sections \ref{ModelFFtau} and \ref{ModelFFee}.  A necessary
ingredient affecting the  pion form factor in $e^+e^-$ annihilations is the photon vacuum polarization (VP),
which is discussed in Section \ref{ModelFFgVP}. Fitting with the partial width expressions 
is briefly discussed in Section \ref{DecayWidths}~; more
details can be found in \cite{su2}, where a practically identical method is used with constant
mixing functions, however. The way to deal with the various kinds of data used in the fit procedures is 
described in Section \ref{FullDataSet}, with a special emphasis on 
our dealing with some correlation phenomena present in the existing data.
In Section \ref{DataFitting}, we fully describe the fit procedures we worked out under various
conditions and the results and conclusions we reach~; we also comment on the numerical and physical properties of our model.
Finally, Section \ref{summary} is devoted to a summary of our conclusions.

As already commented upon in course of the Introduction, several Appendices gather results
of ours or others already published. They are given in an attempt to be as self contained
as possible~; they are placed outside the main text in the interest of clarity and ease of reading.
However, Appendix \ref{YM} contains new results concerning the Yang--Mills 
piece of the HLS Lagrangian.

\section{The HLS Lagrangian Model}
\label{HLS1}
\indent \indent We outline  in the Appendices the main features of the HLS Model
in both the non--anomalous (Appendix A) and anomalous (Appendix B) sectors. This allows us
to derive the leading
terms of the non--anomalous Lagrangian of interest for the present paper by expanding the
exponentials defining the $\xi$ fields. The breaking of flavour symmetries, SU(3) and the Nonet
Symmetry is sketched in Appendix \ref{CC}. 

Several pieces of the HLS Lagrangian of relevance for our problem will be given explicitly in the main text~; 
first, the part describing the photon sector (traditional VMD) is~:

\be
\begin{array}{ll}
{\cal L}_{VMD} &=  \displaystyle ie (1-\frac{a}{2})  A \cdot \pi^- \parsym \pi^+
+ \displaystyle i\frac{e}{z_A} (z_A-\frac{a}{2} -b) A \cdot K^- \parsym K^+
+ \displaystyle i\frac{e}{z_A} b A \cdot K^0 \parsym  \overline{K}^0
 \\[0.5cm]
~&\displaystyle 
+\frac{ia g}{2} \rho^0_I \cdot \pi^- \parsym \pi^+ + \displaystyle \frac{ia g}{4 z_A}
(\rho^0_I + \omg_I -\sqrt{2}  z_V \phi_I ) K^- \parsym K^+ 
+\displaystyle \frac{ia g}{4z_A}
(\rho^0_I - \omg_I +\sqrt{2}  z_V \phi_I ) K^0 \parsym \overline{K}^0 \\[0.5cm]
~& \displaystyle -e a g f^2_\pi \left[
\displaystyle \rho^0_I + \frac{1}{3} \omg_I - \frac{\sqrt{2}}{3} z_V \phi_I
\right] \cdot A + 
\displaystyle \frac{1}{9} a f^2_\pi e^2 (5+z_V) A^2 
+  \frac{a f^2_\pi g^2 }{2}
\left[\displaystyle (\rho^0_I)^2 +\omg^2_I + z_V \phi^2_I \right]
\end{array}
\label{Lag_ee}
\ee
\noindent  limiting oneself to vector mesons, pion and kaon fields. Flavor symmetries have been broken
and, as noted in \cite{BKY,Heath1}, this implies a pseudoscalar field renormalization.
The pseudoscalar field renormalization
has been performed (following the prescription given by Eqs. (\ref{CC3}) or, rather, by Eqs. (\ref{CC7}) which include
Nonet Symmetry breaking). The free Lagrangian  of the vector meson fields is 
standard \cite{HLSOrigin,HLSRef}, 
as well as the (canonical) pseudoscalar kinetic energy piece, the leptonic (see Eq.(\ref{AA11})) and photonic 
free Lagrangian pieces. 

The parameter   $g$ is the traditional universal vector meson coupling constant. On the other hand,
the parameter  $a$ is specific of the  HLS model and fulfills $a=2$ in standard VMD approaches~; 
however such a stringent condition is not mandatory and
several phenomenological studies involving pion form factor data on the one hand \cite{ffVeryOld,ffOld}
and light meson decays on the other hand \cite{rad,mixing} concluded that  a much better favored value 
is $a \simeq 2.4 \div 2.5$. This opens a way to a direct coupling of photons to pseudoscalar pairs within VMD--like approches. 
One should remark the presence of a photon mass term of order $e^2$
which is traditionally removed by field redefinition \cite{HLSRef,Klingl} (see also \cite{Tony} and the discussion
concerning the photon pole position). It can also be removed by renormalization conditions at one loop order.

The parameter $b$ in Eq. (\ref{Lag_ee}) is $b=a(z_V-1)/6$ where $z_V$ is  the SU(3) 
breaking parameter of the ${\cal L}_V$ part of the HLS Lagrangian, while 
  $z_A=[f_K/f_\pi]^2=1.495 \pm 0.031$ \cite{RPP2006} is the SU(3) breaking parameter of 
its ${\cal L}_A$ part \cite{HLSOrigin,HLSRef}. $z_A$ is almost fixed numerically, while $z_V$ is the major origin
of the HK mass difference between the $\phi$ meson and the $(\omg,\rho^0)$ system~; 
it has to be fitted
as the relation between vector meson masses determined experimentally and the (Higgs--Kibble) masses occuring
in Lagrangians is unclear \cite{Klingl,mixing}. On the other hand, the value for  $f_K/f_\pi$
provided by the Review of Particle Properties (RPP) \cite{RPP2006} can be added to the set of 
measurements to be fit.

A subscript $I$   on the fields, standing for ``ideal'',
affects the neutral vector meson fields. It indicates that the corresponding
 fields occuring in the Lagrangian are not the physical fields.

One should note \cite{Heath1} that the SU(3) breaking of the HLS Lagrangian generates a non--resonant
coupling of the photon to neutral kaon pairs~; this is a property common to all breaking procedures 
of the HLS Lagrangian proposed so far \cite{BKY,Heath1,BGP,BGPbis}.

On the other hand, still limiting oneself to pions and kaon terms, the Lagrangian piece of relevance for 
$\tau$ decay after symmetry breaking and field renormalization is given by~:

\be
\begin{array}{ll}
{\cal L}_{\tau}& =- \displaystyle \frac{i g_2}{2}  V_{ud} W^+\cdot
\left[ (1-\frac{a}{2}) \pi^- \parsym \pi^0 + 
\displaystyle (z_A-\frac{a}{2}) \frac{1}{z_A \sqrt{2}} K^0 \parsym K^-
\right]\\[0.5cm]
~&  -  \displaystyle  \frac{af_\pi^2 g g_2}{2} V_{ud} W^+ \cdot \rho^- 
-\displaystyle \frac{iag}{2} \rho^- 
\left[ \pi^0 \parsym \pi^+ -\displaystyle \frac{1}{z_A \sqrt{2}} \overline{K}^0 \parsym K^+
\right]\\[0.5cm]
~& \displaystyle + f^2_\pi g_2^2 \left \{ \frac{1+a}{4} 
\left[\displaystyle z_A|V_{us}|^2 + |V_{ud}|^2\right] +
\frac{a}{4}[\sqrt{z_V}-z_A] |V_{us}|^2  \right \} W^+ \cdot W^-
\displaystyle  + a f^2_\pi g^2 \rho^+\rho^- 
\end{array}
\label{Lag_tau}
\ee
\noindent plus the conjugate of the interaction term (the $W^-$ term, not displayed). 
 This Lagrangian piece depends on $g_2$ (which is fixed by
its relation with the Fermi constant (see Eq. (\ref{AA12}))), on the CKM matrix element
$V_{ud}= 0.97377 \pm 0.00027$ \cite{RPP2006}, on the universal coupling $g$ and on the breaking
parameters $z_A$  and $z_V$ already defined. One should note, balancing the photon mass
term in ${\cal L}_{VMD}$, a small mass term complementing the $W$ mass of the Standard Model
which could be removed by appropriate field redefinitions.

Finally, the effective Lagrangian of the Model  we use in order to describe low energy physics
is~:
\be
{\cal L}={\cal L}_{VMD} +{\cal L}_{\tau} + {\cal L}_{anomalous}+ {\cal L}_{YM}
\label{Lag_tot}
\ee
\noindent where ${\cal L}_{anomalous}$ is given by Eq. ({\ref{CC10}).
Its $VVP$ part is not given in the Appendices, but can be found fully expanded in 
the Appendices of \cite{rad}. The last term is the Yang--Mills piece given by Eq. (\ref{un})
which undergoes SU(3) symmetry breaking as depicted in Appendix \ref{YM}~; the main effect
of this on the low energy phenomenology we deal with concerns the $K^*$ radiative decay widths
(see the discussion around Eq. (\ref{CC10})). 

The first two terms in Eq. (\ref{Lag_tot})
allow us to build up the pion form factor in $e^+e^-$ interactions and $\tau$ decay and
the leptonic widths of neutral vector mesons, while the anomalous decays will be dealt with
starting from the third piece. All breaking 
parameters are common to all pieces of our ${\cal L}$~; more precisely, all parameters of 
our model, except those of the vector meson  mixing, could be fixed from only ${\cal L}_{anomalous}$
and the leptonic decays of vector mesons. This was proved in \cite{rad,mixing,su2}
by adding various $(\rho, \omg ,\phi)$ mixing schemes including the most traditional $(\omg ,\phi)$ mixing
in isolation \cite{rad}. The case of the parameter $z_T$, which is important 
only for $K^*$ radiative decays, is special~; this comes out as the SU(3) breaking parameter
of ${\cal L}_{YM}$ which shows up in the anomalous Lagrangian of Eq. (\ref{CC10}) when replacing there
the bare vector field matrix by its renormalized partner (see Appendix \ref{YM}).

\section{The Pion Form Factor Without Symmetry Breaking}
\label{FFunbrk}
\indent \indent The Lagrangian given in Eq. (\ref{Lag_tot}) allows us to construct the pion form 
factor in $e^+e^-$ annihilation  and in $\tau$ decay. One has ($m_\pi \equiv m_{\pi^\pm}$)~:
 \be
\begin{array}{ll}
F_\pi(s) & = \displaystyle (1-\frac{a}{2}) - \frac{a^2 g^2 f_\pi^2}{2}
\frac{1}{D_V(s)}
\end{array}
\label{FF1}
\ee
\noindent  for both processes involving intermediate photon and W boson. We also have~:
 \be
 \left \{
\begin{array}{ll}
\sigma(e^+ e^- \ra \pi^+ \pi^-) & = \displaystyle \frac{8 \pi \alpha^2}{3
s^{5/2}} |F_\pi(s)|^2 q_\pi^3 \\[0.5cm]
\displaystyle \frac{d\Gamma}{ds}(s) & = \displaystyle
\frac{|V_{ud}|^2 G_F^2}{64 \pi^3 m_\tau^3} |F_\pi(s)|^2 
[G_0(s) + \epsilon^2 G_2(s)]\\[0.5cm]
\end{array}
\right .
\label{FF2}
\ee
\noindent with\footnote{Of course, in the SU(2) symmetry limit, we have $\epsilon=0$ and $q_\pi=Q_\pi$.}~:
 \be
 \left \{
\begin{array}{lll}
G_0(s) &= \displaystyle \frac{4}{3} \frac{(m_\tau^2-s)^2(m_\tau^2+2 s)}{s^{3/2}} Q_\pi^3\\[0.5cm]
G_2(s) &= \displaystyle - \frac{(m_\tau^2-s)^2(4 s-m_\tau^2)}{s^{5/2}}m_\pi^4 Q_\pi
\end{array}
\right .
\label{FF3}
\ee
\noindent and~:
\be
 \left \{
\begin{array}{lll}
\epsilon &= \displaystyle \frac{m_{\pi^0}^2 -m_{\pi^+}^2}{m_{\pi^+}^2} \simeq -0.06455 
& ,~~~~~(m_\pi \equiv m_{\pi^+})\\[0.5cm]
q_\pi &= \displaystyle \frac{1}{2} \sqrt{s-4 m_\pi^2}\\[0.5cm]
 Q_\pi &= \displaystyle \frac{\sqrt{[s-(m_{\pi^0}+m_{\pi^+})^2][s-(m_{\pi^0}-m_{\pi^+})^2]}}{2\sqrt{s}}
\end{array}
\right .
\label{FF4}
\ee
\noindent where one accounts for the pion mass difference. The $G_2(s)$ term gives a completely negligible 
contribution to the form factor and will be cancelled out from now on. The bare inverse propagator
$D_V= s - m_\rho^2$ has to be
modified for self--mass effects which fortunately shift the $\rho$ meson pole off the physical region
by giving it an imaginary part. At this stage, there is also
no inclusion of loop effects in $\gamma \rho$ or $W \rho$ transition amplitudes in the expression for
the pion form factor itself.

Additionally, there is clearly no interplay of the $\omega$ or $\phi$ mesons as can be seen from inspecting
the various pieces of the full Lagrangian in Eq. (\ref{Lag_tot})~; this should come
from Isospin Symmetry breaking.

Including self--mass effects for the $\rho$ (and adding the $\omega/\phi$ meson contributions
for $e^+e^-$ annihilation),  these expressions provide the usual HLS based framework for pion form factor
fitting of  $e^+e^-$  data \cite{ffOld,ffVeryOld}. Even if never done, in principle, Eq. (\ref{FF1})  
applies to $\tau$ data, again after shifting the $\rho^\pm$ singularity off the physical region by means
of a varying width Breit-Wigner amplitude, for instance.

\section{Including One--Loop Effects In The HLS Lagrangian}
\label{OneLoop}
\indent \indent From the expressions given in the previous Section, the $\rho$ meson occurs as
a pole on the physical region~; this is moved off the real axis by self--mass (loop) effects which 
essentially turn out to provide a width to the $\rho$ through the imaginary part of the pion loop. 
However, besides this effect, a closer look at our
${\cal L}$ allows us to see that loop effects contribute to generate self--masses to
all vector mesons, {\it and} transition amplitudes among all neutral vector mesons. Assuming from now on
SU(3) and SU(2) breaking effects, 
the charged and neutral pion and kaon masses become different. One can see that kaon loops
and the anomalous $VVP$ piece of the Lagrangian give the following transition amplitudes for
neutral vector mesons (ideal fields are understood, {\it i.e.} $\Pi_{\rho \rho}(s)$
should be understood as $\Pi_{\rho_I \rho_I}(s)$, for instance)~:
\be
\left \{
\begin{array}{ll}
\Pi_{\rho \rho}(s) = g_{\rho \pi \pi}^2 \Pi^\prime(s) + g_{\rho K K}^2 (\Pi_+(s)+\Pi_0(s))
 &+ \left [ g^2_{\rho \omega  \pi} \Pi_{\omega \pi}(s) +\cdots
\right ] \\[0.5cm]
\Pi_{\omega \omega}(s)= g_{\omega K K}^2 (\Pi_+(s)+\Pi_0(s)) 
 &+ \left [ 
g^2_{\rho  \omega  \pi} \Pi_{\rho  \pi}(s)
+\cdots \right ] 
\\[0.5cm]
\Pi_{\phi \phi}(s)= g_{\phi K K}^2 (\Pi_+(s)+\Pi_0(s)) 
 &+ \left [  
g^2_{\phi  K^{*} K}  \Pi_{ K^{*+}  K^-}(s)
+\cdots \right ] 
\\[0.5cm]
\Pi_{\omega \phi}(s)= -g_{\omega K K} g_{\phi K K} (\Pi_+(s)+\Pi_0(s)) 
 &+ \left [  
2~g_{\phi  K^{*} K}g_{\omg  K^{*} K} (\Pi_{ K^{*\pm}  K^\mp}(s) + \Pi_{ K^{*0}  K^0}(s) ) ~
\right ] 
\\[0.5cm]
\Pi_{\rho \omega}(s)= g_{\rho K K} g_{\omega K K} (\Pi_+(s)-\Pi_0(s))
 &+ \left [  
2 ~g_{\omega K^* K}g_{\rho K^* K} (\Pi_{ K^{*\pm}  K^\mp}(s) -\Pi_{ K^{*0}  K^0}(s) ) ~
\right ] 
\\[0.5cm]
\Pi_{\rho \phi}(s)= -g_{\rho K K} g_{\phi K K} (\Pi_+(s)-\Pi_0(s))
 &+ \left [  
2 ~ g_{\phi K^* K}g_{\rho K^* K} (\Pi_{ K^{*\pm}  K^\mp}(s) -\Pi_{ K^{*0}  K^0}(s) ) ~
\right ] 
\\[0.5cm]
\end{array}
\label{SelfMasses}
\right .
\ee
\noindent where we have defined $g_{\rho \pi \pi}=ag/2$, $g_{\rho K K}=g_{\omega K K} =ag/(4z_A)$
and $ g_{\phi K K}=\sqrt{2} ag z_V/(4z_A)$.
 $\Pi^\prime(s)$ is the charged
pion loop, while $\Pi_+(s)$ and $\Pi_0(s)$ are the charged and neutral kaon loops,
each amputated from their couplings to vector mesons ({\it i.e.} loops carrying unit coupling constants). 

The contributions of the anomalous loops have been displayed between square brackets.
The anomalous FKTUY Lagrangian gives several terms contributing to the self--masses
$\Pi_{\rho \rho}(s)$, $\Pi_{\omega \omega}(s)$ and $\Pi_{\phi \phi}(s)$. They
can easily be constructed from the $VVP$ Lagrangian given in Appendix 4 of \cite{rad}~; for these,
we have displayed in Eqs. (\ref{SelfMasses}) only one representative of the full list which includes always
$ K^{*\pm}  K^\mp$, $ K^{*0}  \overline{K}^0$,$ \overline{K}^{*0}  K^0$ and, depending on the
particular self--mass considered, contributions from  $\omg \pi^0$, $\rho \pi^0$, $\omg \eta$, $\rho \eta$, 
$\phi \eta$, $\omg \eta^\prime$, $\rho \eta^\prime$ or $\phi \eta^\prime$ loops. 

 The anomalous parts of all  transition amplitudes have been entirely displayed, as 
they exhibit an interesting correspondence with the non-anomalous contributions.
We have identified to each other both $K^{*\pm}  K^\mp$ loops on the one hand, 
and $ K^{*0}  \overline{K}^0$ with $ \overline{K}^{*0}  K^0$ on the other hand. Using the present set 
of notations, we have defined $g_{\rho K^* K}=g_{\omega K^* K}=\sqrt{z_T/z_A}G_{anom}/2 $ and
$g_{\phi K^* K}=G_{anom}/\sqrt{2 z_A z_T}$ with \cite{rad}  $G_{anom}=-3 g^2/(8 \pi^2 f_\pi)$.
We have also denoted by $\Pi_{ K^{*\pm}  K^\mp}(s)$ and $\Pi_{ K^{*0}  K^0}(s)$ the amputated
$K^{*\pm}  K^\mp$ and $ K^{*0}  K^0$ loop functions.

In the exact Isospin Symmetry limit, one has $\Pi_+(s)=\Pi_0(s)$ on the one hand,
and $\Pi_{ K^{*\pm}  K^\mp}(s)=\Pi_{ K^{*0}  K^0}(s)$  on the other hand.
Then, all transition amplitudes vanish except for $\omg \phi$. 

 Therefore, within the HLS model, the $\omg \phi$ mixing is a natural feature
generated by loop effects and not from some breaking mechanism. In contrast, the $\rho \phi$ mixing 
and the prominent $\rho \omg$ mixing are fully due to Isospin Symmetry breaking.
Including the anomalous sector does not change the picture.

Actually, as emphasized in Appendix \ref{YM}, the Yang--Mills sector of the HLS Lagrangian
\cite{HLSRef} provides a further mechanism which comes in supplementing the kaon and $K^*K$ loop
mechanism just described. This additional mechanism is produced by charged and neutral 
$K^* \overline{K}^*$ loops~; these still come in by their sum in the $\omg \phi$ 
transition amplitude, while it is their difference which takes place in the $\rho \omg$ 
and $\rho \phi$ transition amplitudes. 

One may wonder that, within VMD--like approaches, all  mechanisms 
contributing to the vector meson  mixing  at order $g^2$
always involve loops with a pair of mesons carrying 
open strangeness. This was true for the non--anomalous HLS Lagrangian ($K   \overline{K}$) and
for the anomalous HLS Lagrangian ($K^* K$)~; we also find it true 
for the Yang--Mills piece ($K^* \overline{K}^*$).

\vspace{0.5cm}

If one denotes by $\Pi_{+/0}(s)$ the amputated $K^+K^0$ loop and by $\Pi^{\prime \prime}(s)$
the amputated $\pi^\pm \pi^0$ loop, the charged $\rho$  self--mass reads~:
\be
\Pi_{\rho \rho}^\prime(s) = g_{\rho \pi \pi}^2 \Pi^{\prime \prime}(s) + 2 g_{\rho K K}^2 \Pi_{+/0}(s)
 + \left [g^2_{\rho^\pm \omega \pi^\pm} \Pi_{\omega \pi}(s) + \cdots \right ]
\label{SelfMasses2}
\ee
\noindent with a partial display of the anomalous loop contributions between the square brackets. The Yang--Mills term
introduces also $\rho \rho$ loops (not displayed as they are of little importance at our energies).
This expression actually differs little from the neutral $\rho$ self--mass~; indeed, the effect of having
different masses for neutral and charged particles in these loop computations is tiny. Of course,
in the Isospin Symmetry limit, we have $\Pi_{\rho \rho}^\prime(s)=\Pi_{\rho \rho}(s)$ and then
the $\rho^0$ and $\rho^\pm$ propagators (and their poles) coincide.

\vspace{0.5cm}

As is clear from the Lagrangian pieces given by Eqs. (\ref{Lag_ee}) and (\ref{Lag_tau}), the fields
$\rho^I$, $\omg^I$, $\phi^I$ as well as $\rho^\pm$ are certainly mass eigenstates at tree level. This statement
remains  true for $\rho^\pm$ at one--loop order as there is no transition loop from this meson to any other one. 
This is, however, clearly not true for  $\rho^I$, $\omg^I$, $\phi^I$ fields which undergo mixing with each other,
as can be seen from Eqs. (\ref{SelfMasses}). Moreover, as for the self--masses, these transition amplitudes
are invariant mass dependent as already noted \cite{mixing,su2}~! 

This implies that physical fields associated with the $\rho^0$, $\omg$, $\phi$ mesons do not coincide with 
their ideal combinations as soon as one--loop corrections are considered. Moreover, the precise content
of the physical fields in terms of ideal fields varies with $s$, or more precisely with the invariant mass
flowing  through the physical field under consideration. This does not prevent in standard
approaches to use $\rho^I$, $\omg^I$, $\phi^I$ in physical amplitudes \cite{KSfit}. However, 
as one loop effects have certainly to be considered even only in order to
shift the vector meson poles off the physical region, they should legitimately be considered 
also for field mixing.
 
 We raise the question of  taking these loop effects properly into account and proceeding to the appropriate
field redefinition to physical fields. 
In order to deal with this problem, let us define as effective Lagrangian
the Lagrangian in Eq. (\ref{Lag_tot}) supplemented with the self--masses and transition terms
occuring at one--loop order~; this turns out to replace the simple vector meson mass term in the HLS Lagrangian
 by ($m^2=a g^2 f_\pi^2$):
\be
{\cal L}_{mass} = \left \{
\begin{array}{lll}
\displaystyle \frac{1}{2} \left \{
[m^2 +\Pi_{\rho \rho}(s)] \rho_I^2 +  [m^2 +\Pi_{\omega \omega}(s)] \omega_I^2+
[z_V m^2 +\Pi_{\phi \phi }(s)] \phi_I^2 \right . \\[0.5cm]
\left . +2 \Pi_{\rho \omega}(s) \rho_I \omega_I + 2 \Pi_{\rho \phi}(s) \rho_I \phi_I
+2 \Pi_{\omega \phi}(s) \omega_I \phi_I \right \} +[m^2 +\Pi_{\rho \rho}^\prime(s)] \rho^+ \rho^-
\end{array}
\right .
\label{SelfMasses3}
\ee

The  $K^*$ mass term, which should also be modified correspondingly, is not shown as it plays
no role in the present problem.  
\vspace{0.5cm}

Even  if anomalous $VP$ contributions seem  to play 
some role visible \cite{ffOld}  (and, nevertheless, marginal) in pion form factor data, 
qualitatively  their explicit form is really active only above the $\omg \pi$ threshold, which is the lowest
mass $VP$ threshold~; all others are far above the GeV region\footnote{Their threshold masses are spread out  between
$\simeq 1.3$  GeV and $\simeq 2 $ GeV. The situation is similar for the loops generated by the Yang--Mills term.}. 
Below the threshold, the main effect is due to their subtraction
polynomials which can be well absorbed in the subtraction polynomials of the accompanying pion and kaon loops 
in order to put the poles of the $\rho$ propagator at the place requested by the data.

Beside the (non--anomalous) pion and kaon loops, all transition amplitudes involve 
$K^*K$ and $K^* \overline{K}^*$ loops, the thresholds of which being resp. at $\simeq 1.4$ GeV and $\simeq 1.8$ GeV. 
This means that, besides their subtraction 
polynomials (minimally of degree 2), in the region below the GeV,  their behavior \cite{mixing} is 
 a real logarithmic function (below $\sqrt{s} \simeq 0.4$ GeV) or an arctangent  function
 ($0.4 \leq \sqrt{s} \leq 1.4$ GeV). 
This also can be numerically absorbed in a fitted subtraction polynomial.

Therefore, there is some sense in neglecting the (explicit) contributions of the anomalous and Yang--Mills loops,
being understood that their effect is mostly concentrated in their subtraction polynomials. Moreover,
as these come always together with pion and kaon loops, they can be accounted for by simply letting
the (free) pion and kaon loop subtraction polynomials to be second degree. Therefore, 
we shall neglect  their (explicit) contributions, pointing at 
the appropriate places to their possible influence. Thus, from now on, the self--energies and transition
amplitudes should be understood as Eqs. (\ref{SelfMasses}--\ref{SelfMasses2}) amputated from 
the square bracket terms and without the Yang--Mills contributions depicted in Appendix \ref{YM}.

The use of the modified HLS Lagrangian has already been discussed in \cite{mixing} where it was shown, 
for instance, that this method
allows one to recover vector meson propagators usually derived through the Schwinger--Dyson
resummation procedure, which turns out to sum up an infinite series, which is not necessarily convergent.
However, we show shortly  that introducing this modified mass term allows us to also account for the 
other transition effects which would be more difficult
to derive from the   Schwinger--Dyson resummation procedure 
(of course, this should be possible, merely tedious).

\section{Mass Matrix Diagonalization And Physical Fields}
\label{diagonal}
\indent \indent As clear  from Eq. (\ref{SelfMasses3}), at one loop order, the mass term
is diagonal in the charged vector meson sector and will not be discussed any longer.
 In the neutral vector meson sector, however, the mass matrix is not diagonal and
 the effective Lagrangian mass term is~:

\be
{\cal L}_{mass} = \displaystyle \frac{1}{2}\widetilde{U} M^2(s) U~~~~{\rm with~~~}
\widetilde{U} =(\rho^I,\omg^I,\phi^I)
\label{Mass1}
\ee
\noindent (the ideal fields being supposed real) and\footnote{For ease of reading, $\epsilon_1$ $\epsilon_2$ 
are not written 
with their explicit $s$ dependence which is (or may be)  understood throughout this paper.}~:
\be
M^2(s) =\left ( 
\begin{array}{lll}
m^2 + \Pi_{\pi \pi}(s) + \epsilon_2  & ~~~~~~~~\epsilon_1        & ~~~~~~-\mu \epsilon_1\\[0.5cm]
 ~~~~~~~\epsilon_1               &  m^2 + \epsilon_2 &  ~~~~~~-\mu \epsilon_2\\[0.5cm]
 ~~~~~-\mu \epsilon_1	 & ~~-\mu \epsilon_2 	 & z_V m^2 + \mu^2 \epsilon_2
\end{array}
\right )
~~~~~~~ ({\rm with} ~~~ \mu \equiv z_V \sqrt{2}~~)
\label{Mass2}
\ee
\noindent where we have defined~:
\be
\left \{
\begin{array}{lll}
\epsilon_1=g_{\rho K K}^2 (\Pi_+(s)-\Pi_0(s))\\[0.5cm]
\epsilon_2=g_{\rho K K}^2 (\Pi_+(s)+\Pi_0(s))\\[0.5cm]
\Pi_{\pi \pi}(s) = g_{\rho \pi \pi}^2 \Pi^\prime(s)
\end{array}
\right .
\label{Mass3}
\ee

In the region where we work -- invariant masses bounded essentially by the two--pion threshold and 
the $\phi$ mass-- , the functions $\epsilon_1$ and $\epsilon_2$ are small and can be treated as 
perturbation parameters{\footnote{Actually, from their very expressions in terms of kaon 
 and $K^*K$ loops, one may
expect $\epsilon_1$  to be sensitively smaller than $\epsilon_2$ in absolute magnitude. }}~;
moreover, they are real for real $s$ up to the two--kaon threshold region. In  contrast, 
$\Pi_{\pi \pi}(s)$ is complex starting from the two--pion production threshold and is 
not expected to be small enough to be consistently treated as a perturbation parameter.

The physical vector meson mass eigenstates are the ($s$--dependent) eigenvectors of $M^2(s)$ and their 
masses are the corresponding eigenvalues, which are also $s$ dependent~! Expressed this way, the notion
of vector meson mass looks a little bit paradoxical, however, it is not really new~: writing, as usual, the
inverse $\rho$ dressed propagator $D_V(s)=s - m^2 - \Pi_{\rho \rho}(s)$ can be interpreted as stating that
the $\rho$ mass squared is $m^2 + \Pi_{\rho \rho}(s)$ and includes an imaginary part.
From a physics point of view, what is important is that the pole position associated with the
$\rho$ is always a zero of $s - m^2 - \Pi_{\rho \rho}(s)$ located on the unphysical sheet, 
close to the physical region\footnote{The upper lip of the physical region-- the $s  \geq 4 m_\pi^2$ semi--axis --
located on the physical sheet is topologically close to the lower lip in the unphysical sheet of the Riemann surface~;
in contrast, the lower lip in the physical sheet is topologically far from the upper lip in the same sheet.}. 
The $\rho$ pole position has been fitted long ago by \cite{Bernicha} in $e^+e^-$ data, and
more recent fit values can be found in \cite{Melikhov2}~; this piece of
information is actually highly model independent, in contrast with any other definition \cite{Tony}.
We shall revisit this issue with our fit results.

One may wonder about the hermiticity properties of the Lagrangian modified as proposed. As below
the two--pion threshold, the loops defined above are all real for real $s$, we still indeed have
${\cal L} (s) = {\cal L}^\dagger (s)$, however, above this point, the hermiticity should be redefined as
${\cal L} (s) = {\cal L}^\dagger (s^*)$.
This property known as hermitian analyticity \cite{ELOP}  is fulfilled by our modified Lagrangian as it is
already fulfilled by the loop functions.

Now, in order to define the physical $\rho$, $\omg$, $\phi$ in terms of their ideal partners, 
one has  to find the eigenstates of the squared mass matrix given by Eq. (\ref{Mass2}). Let us
take advantage of the smallness of $\epsilon_1$ and $\epsilon_2$ to solve the problem perturbatively
in order to avoid dealing with untractable expressions.
Let us split up the squared mass matrix into two pieces and write it $M^2= M^2_0 +\epsilon B $ with~:
\be
M^2_0 =
\left ( 
\begin{array}{lll}
m^2 +\Pi_{\pi \pi}(s) & ~~~~~~~~0 & ~~~~~~~~0 \\[0.5cm]
~~~~~~~~0 & m^2 + \epsilon_2  &  ~~~~~~~~ 0  \\[0.5cm]
~~~~~~~~0 & ~~~~~~~~0 & z_V m^2 +\mu^2 \epsilon_2
\end{array}
\right )
~~~,~~~
\epsilon B=\left ( 
\begin{array}{lll}
\epsilon_2 &  \epsilon_1  & -\mu \epsilon_1 \\[0.5cm]
\epsilon_1 & 0 & -\mu \epsilon_2 \\[0.5cm]
-\mu \epsilon_1& -\mu \epsilon_2  &  0 
\end{array}
\right )
\label{Mass4}
\ee

In this splitting up, we have found appropriate to leave a part of the actual perturbation inside 
 $M^2_0$. In this way, we avoid to some extent the problem of having the unperturbed eigenvalues degenerated twice 
or even  three times (when $z_V=1$) for some values of $s$ and some $z_V$. However, while assuming
that $\Pi_{\pi \pi}(s)$, $\epsilon_1(s)$ and $\epsilon_2(s)$ vanish  at origin, one cannot avoid
to have a twofold degeneracy at $s \equiv 0$~; this degeneracy is resolved as soon as $s$ departs
from zero by an arbitrary small quantity. This issue, which affects strictly the point $s=0$ (where
the exact solution is trivial~!), 
does not  raise any problem with our data which are all at $s \ne 0$, even if close to zero, as
the NA7 spacelike form factor data \cite{NA7}.
Another solution to this problem would be that the HK masses for $\rho_I$ and $\omg_I$
would be slightly different~; such a mechanism remains to be found\footnote{ A way to get it would have been to
use as breaking matrix $X_V={\rm Diag}(1+\varepsilon_u,1+\varepsilon_d, \sqrt{z_V})$
instead of $X_V={\rm Diag}(1,1,\sqrt{z_V})$ while computing $\cal{L}_{\rm V}$ (see Appendix \ref{CC}).
This, actually, generates a mass difference between $\rho^0$ and $\rho^\pm$, but
the HK mass for the $\omg$ meson remains equal to that of the  $\rho^0$ meson.
Additionally, the coupling constants of the charged and neutral $\rho$ mesons
to a pion pair differ only by terms of order $\varepsilon_{u/d}^2$
and thus can be kept equal.  
\label{rpr0mass}
}.


The unperturbed solution is then trivial, as the eigenvectors are the canonical ideal combinations
of the neutral vector meson fields,
with eigenvalues as can be read off the diagonal of  $M^2_0$. Then, one has to solve the following system
for the perturbations $\delta v_i$ and $\delta \lambda_i$~:

\be
\left \{
\begin{array}{lll}
M^2_0~  v_i =\lambda_i v_i ~~~, \tilde{v_i} \cdot v_i =1,~~~~(i=1,2,3) \\[0.5cm]
M^2 (v_i +\delta v_i)=(\lambda_i +\delta \lambda_i)(v_i +\delta v_i)~~~,~~~{\rm with \cite{BF}~~~:~~~ }
\tilde{v_i} \cdot \delta v_i =0
\end{array}
\right . 
\label{Mass5}
\ee
\noindent for each $i=(\rho,\omg,\phi)=(1,2,3)$. The solution can be written~:
\be
\left (
\begin{array}{lllll}
\rho^0\\[0.5cm]
\omega\\[0.5cm]
\phi
\end{array}
\right ) 
=
R(s)
\left (
\begin{array}{lll}
\rho^0_I\\[0.5cm]
\omega_I\\[0.5cm]
\phi_I
\end{array}
\right ) ~~~~~~~,~~~~~
\left (
\begin{array}{lll}
\rho^0_I\\[0.5cm]
\omega_I\\[0.5cm]
\phi_I
\end{array}
\right ) 
=
\widetilde{R}(s)
\left (
\begin{array}{lll}
\rho^0\\[0.5cm]
\omega\\[0.5cm]
\phi
\end{array}
\right )
\label{Mass6}
\ee
\noindent where (recall $\epsilon_i \equiv \epsilon_i(s)$ are analytic functions of $s$)~:
\be
R =
\left (
\begin{array}{lll}
 ~~~~~~~~~1 & \displaystyle \frac{\epsilon_1}{\Pi_{\pi \pi}(s)-\epsilon_2} &
\displaystyle - \frac{\mu \epsilon_1 }{
(1-z_V) m^2 + \Pi_{\pi \pi}(s) -\mu^2 \epsilon_2}\\[0.5cm]
\displaystyle - \frac{\epsilon_1}{\Pi_{\pi \pi}(s)-\epsilon_2} &  ~~~~~~~~~1 &
\displaystyle - \frac{\mu \epsilon_2 }{
(1-z_V) m^2 + (1-\mu^2) \epsilon_2}\\[0.5cm]
\displaystyle   \frac{\mu \epsilon_1 }{(1-z_V) m^2 + \Pi_{\pi \pi}(s) -\mu^2 \epsilon_2}
& \displaystyle\frac{\mu \epsilon_2 }{ (1-z_V) m^2 + (1-\mu^2) \epsilon_2} & \hspace{3.cm} 1
\end{array}
\right ) 
\label{Mass7}
\ee

The matrix $R$ is orthogonal up to (neglected) second order terms (see Section 6 in \cite{su2}) and
its elements are, actually,  meromorphic functions of $s$~; this, for instance, means that one
has to check that they do not develop singularities in the region of physical interest for our model.

On the other hand, one may wonder
getting $\widetilde{R}(s)$ with no complex conjugation in the field transformation ($\Pi_{\pi \pi}(s)$ is complex and
fulfills the real analyticity condition $\Pi_{\pi \pi}(s)=\Pi_{\pi \pi}^*(s^*)$). 
This is due to the unitarity condition  which writes \cite{mixing}~:
\be
R(s+i\varepsilon) \cdot R^\dagger((s+i\varepsilon)^*)=R(s+i\varepsilon) \cdot  R^\dagger(s-i\varepsilon) 
=1
\label{Mass7a}
\ee
for real $s$ above threshold and $\varepsilon >0$. The real analyticity property fulfilled
by the matrix function $R$ then gives $R^*(s-i\varepsilon)=R(s+i\varepsilon) $ and then
Eq. (\ref{Mass7a}) becomes~:
\be
R(s+i\varepsilon) \widetilde{R}(s+i\varepsilon)=1
\label{Mass7b}
\ee
\noindent as can be checked directly with the $R$ matrix above.

At first order, the corrections for eigenvalues
are not changed with respect to their unperturbed values for $\omg$ and $\phi$ , while for
$\rho^0$, the first order correction is such that the eigenvalue is restored to 
$m^2 + \Pi_{\pi \pi}(s) + \epsilon_2 $  and is formally identical to the $\rho^\pm$ 
mass squared{\footnote{For $\rho^\pm$ , the mass squared value contains what was
named $\Pi_{\pi \pi}^\prime(s)$ and $\epsilon_2 \ra 2 g_{\rho K K}^2 \Pi_{+/0}(s) $.
}}.
Therefore, in order to deal with the physical eigenstates $\rho^0$, $\omg$ and $\phi$,
one has to introduce in the Lagrangian  (\ref{Lag_tot}) above, the physical fields as defined by
 Eqs. (\ref{Mass6}) using Eqs. (\ref{Mass7}). For coupling constants, one has to perform exactly as explained in 
Section 6.3 of \cite{su2}, using the $R$ matrix above and, where appropriate, the ideal coupling 
constants given in Appendix \ref{DD}.

In order that this solution should be considered valid, one has to check that the non--diagonal
elements of the matrix $R$ are small compared to 1 in the whole range of application of our model.
As they depend on fit parameters, this check can only be performed with the fit solution. 

\vspace{0.5cm}
\indent \indent
We will not go into more details with expressing the full Lagrangian (\ref{Lag_tot})
in the basis of physical neutral vector meson fields, as formulae become really complicated 
(even if they can be readily and tediously written down).  Let us only give the most interesting piece
in terms of physical fields  for illustrative purposes~:
\be
\displaystyle \frac{ia g}{2} \rho^I \cdot \pi^- \parsym \pi^+ = 
\frac{ia g}{2} \left [
\rho^0    - \frac{\epsilon_1}{\Pi_{\pi \pi}(s)-\epsilon_2} \omega
+  \frac{\mu \epsilon_1 }{ (1-z_V) m^2 +  \Pi_{\pi \pi}(s)  -\mu^2 \epsilon_2}\phi 
\right] \cdot \pi^- \parsym \pi^+
\label{Mass8}
\ee

This clearly shows how kaon loops generate couplings of the {\it physical}
$\omega$ and $\phi$ fields to $\pi^- \pi^+$ which vanish (with $\epsilon_1$) in the Isospin Symmetry limit.
As $\Pi_{\pi \pi}(s)$ has a large imaginary part, it is clear that the phase of the $\omg$ coupling
compared with $\rho^0$ will be very large at the $\rho$ peak. It should also be mentioned that
the couplings shown here (and the matrix elements of $R$) have all a finite limit 
at $s=0$ even if the loops individually vanish at $s=0$ as the pseudoscalar pairs couple
to conserved currents \cite{Klingl,RhoOmg0}.

The effects of the neglected loops could be briefly mentioned here. The most important effect
in the expression for $R(s)$ (Eq. (\ref{Mass7})) is on the denominators of $R_{12}(s)$ and  $R_{21}(s)$
where the difference of the anomalous contributions to self--energies for the $\rho$ and $\omg$ mesons
will add to the present $\Pi_{\pi \pi}(s) - \epsilon_2(s)$~; this could change a little bit
the behaviour near $s=0$ where all loops tend to zero.

As stated above, at first order in perturbations, the squared mass eigenvalues are the entries in the diagonal of
$M^2(s)$ given in Eq. (\ref{Mass2}). For further use, let us also give the second order corrections to the
eigenvalues (and thus to the running squared meson masses)~:
\be
\left \{
\begin{array}{lll}
\displaystyle  \delta_2 \lambda_\rho= 
\frac{\epsilon_1^2}{\Pi_{\pi \pi}(s) - \epsilon_2}
+\frac{\mu^2 \epsilon_1^2}{(1-z_V) m^2+ \Pi_{\pi \pi}(s) -\mu^2 \epsilon_2 }\\[0.5cm]
\displaystyle  \delta_2 \lambda_\omega=  
-\frac{\epsilon_1^2}{\Pi_{\pi \pi}(s) -\epsilon_2}
+ \frac{\mu^2 \epsilon_2^2}{(1-z_V) m^2+ (1-\mu^2) \epsilon_2 }\\[0.5cm]
\displaystyle  \delta_2 \lambda_\phi=  
-\frac{\mu^2 \epsilon_1^2}{(1-z_V) m^2+ \Pi_{\pi \pi}(s) -\mu^2 \epsilon_2 }
-\frac{\mu^2 \epsilon_2^2}{(1-z_V) m^2+ (1-\mu^2) \epsilon_2 }
\end{array}
\right .
\label{Mass9}
\ee

In the mass range where we are working (from the two--pion threshold to the $\phi$ mass), the mass eigenvalues
at first order are real{\footnote{Actually, near the accepted $\phi$ mass our perturbation parameters start
to carry a tiny imaginary part.}} for the $\omg$ and $\phi$, excluding a width. At second order, one observes that
the pion loop generates an imaginary part to these mass eigenvalues. Let us remind the reader that, as the pole positions
are the solutions of $s-\lambda_i(s)=0$, one expects to find the $\rho$ pole position close to the value found by
\cite{Bernicha}. However there is little chance that the $\omg$ width happens to carry the correct width
value as this should be generated by the anomalous $\rho \pi$ loop with additional insertion of a pion loop
on the $\rho$ leg (or simply considering the dressed $\rho$ propagator) or directly through
a possible $\omega\to 3\pi\to \omega$ (double) loop. Finally, in the model we use, it is only at second order
that a difference between neutral and charged $\rho$ propagators (and thus masses) occurs and this is a net
(and small) effect of the neutral vector meson mixing. This comes in addition to  other sources of
$\rho^0-\rho^\pm$ mass difference (see footnote \ref{rpr0mass}).

\section{Renormalization Conditions On Loops}
\label{Renormalization}
\indent \indent
With the approximations we choosed (neglecting the anomalous loop contributions),
the loop expressions needed in order to construct the pion form factor $F_\pi(s)$ are only the
 $\pi \pi$ and $K \overline{K}$ loops. They can be computed by means of
Dispersion Relations \cite{Klingl,mixing} and can be derived without explicit integration,
relying only on properties of analytic functions, especially the uniqueness property of the analytic
continuation (see{\footnote{In this reference, the loop expressions for equal mass pseudoscalar
meson pairs and vector--peudoscalar pairs are already given and are correct~; the 
function given for unequal mass pseudoscalar meson pairs is not correct as the contribution
of the gauge term has been omitted~; we apologize for this  error 
and correct for in the present paper (see Appendix  \ref{EE}).} }  Appendix A in \cite{mixing}). 

From general principles, any loop $\Pi(s)$ is a so--called real analytic function (see the Section
just above), the imaginary part of which is calculable using the Cutkosky rules, or in the
simple case of single loops, using the partial width of the process $V \ra P P^\prime$. Indeed~:
\be
{\rm Im} \Pi(s) = -\sqrt{s} \Gamma(s)~~~~~~~,~~ s>s_0
\label{loop1}
\ee
\noindent where $s$ is the pair invariant mass squared, $s_0$ the threshold mass squared of
the pair and $\Gamma$ the partial width of the decay. With this at hand, the full loop is the solution
of the integral equation~:
\be
\displaystyle \Pi(s) = P_{n-1}(s) + \frac{s^n}{\pi} \int_{s_0}^\infty 
\frac{{\rm Im} \Pi(z)}{z^n(z-s+i\epsilon)} dz
\label{loop2}
\ee
\noindent where $P_{n-1}(s)$ is a polynomial of degree $n-1$ with real coefficients and the integral 
runs over the right-hand cut (the physical region). These coefficients should be fixed by means
of condition to be fulfilled by $\Pi(s)$, the so--called subtraction polynomial, 
which are nothing but renormalization conditions chosen
externally and depending on  the problem at hand. 

A priori, the number of subtractions, {\it i.e.} the number of conditions can be arbitrary, however,
in order that the integral in Eq. (\ref{loop2}) converges, there is a minimal number of subtractions
to perform~: For $PP^\prime$ and $VV$ loops $n \geq 2$, for $VP$ loops  $n \geq 3$.

In the most general form of the pion form factor following from the Lagrangian (\ref{Lag_tot}),
and using the modified one--loop mass term (\ref{SelfMasses3}) the relevant basic loops are only the pion
loops and the kaon loops. These imply that  at least  $n = 2$~; however, the very existence of $VP$ neglected 
loops implies that we are still minimally subtracting with using  $n = 3$ for all loop
functions in this paper. As discussed in Section \ref{OneLoop}, in this way, the subtraction 
polynomials carry some (unknown) information on the anomalous loop contribution.

Additionnally,
we request all polynomials  $P_{n-1}(s)$ to fulfill $P_{n-1}(0)=0$ reflecting this way
current conservation \cite{Klingl,RhoOmg0,Melikhov1,ffOld} when needed and an appropriate constraint
otherwise{\footnote{In this case, the constant term  in the $\rho$ propagators is the squared (HK) mass
occuring in the Lagrangian with no modification.}}.

In usual approaches\cite{Melikhov1,Melikhov2,Klingl}, the renormalization conditions are defined  
from start and, then,  one leaves free other parameters like meson mass and width in order to
accomodate the experimental data. As already done in \cite{ffOld}, we proceed
in the opposite way~: as masses and couplings are fixed consistently in our Lagrangian, we leave
free the subtraction polynomials  in the loops $\Pi_{\pi \pi}(s)$ , $\epsilon_1(s)$ and $ \epsilon_2(s)$.
This way  allows the full data set to contribute to fixing the subtraction constants.
The basic (pion and kaon) loop expressions are given in Appendix E and are used only subtracted
once (in order that they vanish at $s=0$)~; they are supplemented with  second degree polynomials 
vanishing at $s=0$ and  having coefficients to be fixed by fitting the data. 

\section{ The Model Pion Form Factor In $\tau$ Decay}
\label{ModelFFtau}
\indent \indent Introducing pion and kaon loop effects gives the $\rho^\pm$ a self--mass, 
but, nevertheless, the $\rho^\pm$ fields  remain mass eigenstates. To stay consistent with using
$\rho$ self--mass, one has also to account for  loops in the $W-\rho$
transition amplitude. In $\tau$  decay the relevant loop effects, while neglecting anomalous $VP$ loops,
are the $\pi^\pm \pi^0$ and  $K^0 K^\pm$ contributions. Accounting for this modifies 
Eq. (\ref{FF1}) to~:
\be
\displaystyle
F_\pi^\tau(s) = \left [
(1-\frac{a}{2}) - F_\rho^\tau g_{\rho \pi \pi} \frac{1}{D_\rho(s)}
\right]
\label{Model1}
\ee
with~:
\be
\left \{
\begin{array}{lll}
F_\rho^\tau =f_\rho^\tau - \Pi_{W}(s)\\[0.5cm]
D_\rho(s)=s-m^2 -\Pi_{\rho \rho}^\prime(s)\\[0.5cm]
f_\rho^\tau = ag f_\pi^2
\end{array}
\right .
\label{Model2}
\ee
\noindent where  $m^2=a g^2 f_\pi^2$ and the charged $\rho$ self--mass $\Pi_{\rho \rho}^\prime(s)$ 
has been defined in Section
\ref{OneLoop}  by Eq. (\ref{SelfMasses2}) and used in Eq. (\ref{SelfMasses3}) (recall we neglect VP loops). 
One should note that Eqs. (\ref{Model1}) and (\ref{Model2}) 
are not affected by any breaking mechanism. The diagrams contributing to the pion form factor in $\tau$ 
decays are sketched  in Figure \ref{PlotFF}. 

Let us denote for a moment the pion and kaon amputated ({\it i.e.} computed with unit coupling constants)
loops by $\ell_\pi(s)$ and $\ell_{K}(s)$, assuming they are
already subtracted once in order that they identically vanish at $s=0$ (see Appendix \ref{EE}).
The $W-\rho$ transition amplitude and the $\rho^\pm$ self--energy occuring in the pion form factor
have the following expressions in terms of pion and kaon amputated loops~:
\be
\left \{
\begin{array}{lll}
\displaystyle \Pi_{W}(s)=g_{\rho \pi \pi}
\left [ (1-\frac{a}{2}) \ell_\pi(s) + \frac{1}{2 z_A^2}(z_A-\frac{a}{2}) \ell_{K}(s) 
\right ]+ P_W(s) \\[0.5cm]
\displaystyle \Pi_{\rho \rho}^\prime(s) = g_{\rho \pi \pi}^2  \left [ 
\ell_\pi(s) + \frac{1}{2 z_A^2} \ell_{K}(s)~
\right ] + P_\rho(s)
\end{array}
\right .
\label{Model3}
\ee
where $g_{\rho \pi \pi}=a g/2$ and $P_W(s)$ and $P_\rho(s)$ being subtraction 
polynomials with real coefficients to be fixed by external renormalization. 
As emphasized in \cite{ffOld} and \cite{Melikhov1}, the polynomials 
$P_W(s)$ and $P_\rho(s)$ can be chosen independent. Indeed, ${\rm Im}~\Pi_{W}(s) $ 
and ${\rm Im}~\Pi_{\rho \rho}^\prime(s)$
are even not proportional as soon as SU(3) is broken ($z_A \ne 1$)~; moreover, the transition
amplitude $\Pi_{W}(s)$ is non zero even if $a=2$ as soon as  SU(3) symmetry is 
broken. We choose to constrain $P_W(s)$ and $P_\rho(s)$ to be second degree and vanishing at $s=0$,
as discussed in Section \ref{Renormalization}.

For the sake of simplicity, we have also chosen 
to use the pseudoscalar meson loops assuming $m_{\pi^\pm} =m_{\pi^0}$ and
$m_{K^\pm} =m_{K^0}$ after having checked that this is numerically armless
while dealing with all form factor data.  
Under this approximation{\footnote{
This implies that the $\rho^0$ and $\rho^\pm$ widths
do not significantly differ. This statement is supported by 
the various experimental data collected  in \cite{RPP2006}.}}, we have 
$\Pi_{\rho \rho}^\prime(s) = \Pi_{\rho \rho}(s)$ (see Eq. (\ref{SelfMasses}))
and all symmetry breaking effects due to the pion mass difference are concentrated
in the phase space factors (see Eqs. (\ref{FF2}--\ref{FF4})) for cross sections and
partial widths where these  have a sizable effect.

Therefore, one can  rewrite Eqs. (\ref{Model3}) under a form more appropriate
for our fitting procedure~:
\be
\left \{
\begin{array}{lll}
\displaystyle \Pi_{W}(s)=\frac{1}{g_{\rho \pi \pi}} \left [
(1-\frac{a}{2}) \Pi_{\pi \pi}^W(s)+ (z_A -\frac{a}{2})\epsilon_2(s) \right ]\\[0.5cm]
\displaystyle \Pi_{\rho \rho}^\prime(s)=\Pi_{\rho \rho}(s)=\Pi_{\pi \pi}^\rho(s)+\epsilon_2(s)
\end{array}
\right .
\label{Model4}
\ee
\noindent
where $\epsilon_2(s)$, already defined in Eqs. (\ref{Mass3}),
carries its own subtraction polynomial, and having defined~:
\be
\left \{
\begin{array}{lll}
\Pi_{\pi \pi}^W(s) \equiv g_{\rho \pi \pi}^2 \ell_\pi(s) +Q_W(s)\\[0.5cm]
\Pi_{\pi \pi}^\rho(s) \equiv g_{\rho \pi \pi}^2 \ell_\pi(s) +Q_\rho(s)
\end{array}
\right .
\label{Model4a}
\ee
\noindent
where $Q_W(s)$ and $Q_\rho(s)$ are second degree polynomials\footnote{We recall here, that these 
polynomials may account for the neglected anomalous loop effects not introduced explicitly.}  (with real 
coefficients to be fitted) and vanishing at origin. Possible correlations among them, if any, 
would be an outcome of the fit procedure and can be detected from inspecting the fit error  covariance matrix.
Finally, one can check that the condition $F_\pi^\tau(0) = 1$ is  automatically fulfilled

Before turning to the pion form factor in $e^+ e^-$ annihilations, let us also remind the reader that 
the $\tau$ partial width expression in Eqs. (\ref{FF2}) has to be further corrected for
isospin breaking effects by multiplying it by\footnote{Actually, this numerical value has been derived for
the pion final state~; in practical applications, it is usually assumed that this value holds also for the $\rho$ final state
-- see for instance \cite{DavierPrevious1,DavierPrevious2,DavierPrevious3}. }
 $S_{EW}=1.0232$ which accounts for short range radiative corrections \cite{Marciano}.
Long range radiative corrections have been derived in \cite{Cirigliano1,Cirigliano2,Cirigliano3}
and come as a further  factor $G_{EM}(s)$~; another estimate 
taking into account additional Feynman diagrams can be found \cite{Mexico1,Mexico2} and a
corresponding numerical parametrization  of $G_{EM}(s)$ has been provided in \cite{Mexico3}. 
This means that in all our fits we perform the substitution~:
\be
F_\pi^\tau(s) \Longrightarrow S_{EW} G_{EM}(s) F_\pi^\tau(s)
\label{Model4b}
\ee
which, therefore, accounts for all reported Isospin Symmetry breaking effects specific to the $\tau$ sector.
Another isospin breaking effect might have to be considered, namely a $\rho^0-\rho^\pm$ mass difference.
 This can be generated, for instance, by means of the mechanism
sketched in footnote \ref{rpr0mass}. It can be shown that this turns out to modify Eqs. (\ref{Model2}) to~:
 \be
\left \{
\begin{array}{lll}
F_\rho^\tau =f_\rho^\tau - \Pi_{W}(s)\\[0.5cm]
D_\rho(s)=s-m^2 -\delta m^2-\Pi_{\rho \rho}^\prime(s)\\[0.5cm]
\displaystyle f_\rho^\tau = ag f_\pi^2 + \frac{\delta m^2}{g}
\end{array}
\right .
\label{Model2Mod}
\ee
where $\delta m^2$ is left free. The modified Eqs.  xd(\ref{Model2Mod}) allows Eq. (\ref{Model1})
to still fulfill $F_\pi^\tau(0)=1$ automatically\footnote{Anticipating somewhat the fit results, a possible
$\delta m^2$ can be detected on ALEPH data \cite{Aleph} (not on CLEO data \cite{Cleo}) and amounts 
to  $\simeq -0.25$ GeV$^2$.  $f_\rho^\tau (\simeq 0.7 ~{\rm GeV}^2)$ is then increased by $0.5~ 10^{-3} ~{\rm GeV}^2$,
a quite negligible quantity.}.

\section{ The Model Pion Form Factor In $e^+e^-$ Annihilations}
\label{ModelFFee}
\indent \indent
 In $\tau$ decay, the pion form factor, as just seen, is free from any vector meson mixing effect.
Instead, the pion form factor $F_\pi^e(s)$
expression is sharply influenced by the vector meson mixing mechanism constructed explicitly in Section
\ref{diagonal} which leads us to make the transformation from ideal to physical vector meson fields.
After this tranformation, we get from our effective Lagrangian the diagrams shown in Figure \ref{PlotFF}
and the corresponding expression~:
\be
\displaystyle
F_\pi^e(s) = \left [ (1-\frac{a}{2}) - F_\rho^e(s) g_{\rho \pi \pi} \frac{1}{D_\rho(s)}
- F_\omega^e(s) g_{\omega \pi \pi} \frac{1}{D_\omega(s)}
- F_\phi^e(s) g_{\phi \pi \pi} \frac{1}{D_\phi(s)}
\right]
\label{Model5}
\ee
\noindent where $D_\rho(s)$ (see  Eq. (\ref{Model2}) for its charged partner), 
$D_\omg(s)$ and $D_\phi(s)$ are the inverse propagators of the corresponding (physical) vector 
mesons. We have now~: 
\be 
D_\rho(s)=s-m^2 -\Pi_{\rho \rho}(s) 
\label{Model5b}
\ee
\noindent (recall that our assumptions on pseudoscalar meson masses implies
 $\Pi_{\rho \rho}(s)=\Pi_{\rho \rho}^\prime(s)$, which reduces the number of free parameters in our model).
The vector meson couplings to a pion pair  after symmetry breaking are~:
\be
\left \{
\begin{array}{lll}
\displaystyle g_{\rho \pi \pi} = \frac{a g}{2}\\[0.5cm]
\displaystyle g_{\omega \pi \pi} = -\frac{a g}{2} \frac{\epsilon_1}{\Pi_{\pi \pi}^\rho(s)-\epsilon_2} \\[0.5cm]
\displaystyle g_{\phi \pi \pi} = \frac{a g}{2}
\displaystyle   \frac{\mu \epsilon_1 }{(1-z_V) m^2 + \Pi_{\pi \pi}^\rho(s) -\mu^2 \epsilon_2}
\end{array}
\right .
\label{Model6}
\ee
where $\Pi_{\pi \pi}^\rho(s)$ has been defined in the previous Section. One should note that the quantity
named $\Pi_{\pi \pi}(s)$ in the definition of the matrix $R(s)$ (see Section \ref{diagonal})
coincides with the presently defined $\Pi_{\pi \pi}^\rho(s)$. 

The quantities $F_V^e$ can be written~:
\be
F_V^e(s) = f_V^e -\Pi_{V\gamma}(s)
\label{LagMix18}
\ee

Collecting the various couplings of the ideal fields  suitably weighted by elements of the matrix
transformation $R(s)$ (see Eq. (\ref{Mass7})), we get~:
\be
\left \{
\begin{array}{llll}
\displaystyle f_\rho^e =a g f_\pi^2 \left[
1 + \frac{1}{3} \frac{\epsilon_1}{\Pi_{\pi\pi}^\rho(s) -\epsilon_2} 
+\frac{1}{3} \frac{\mu^2 \epsilon_1}{(1-z_V) m^2 +\Pi_{\pi\pi}^\rho(s) -\mu^2 \epsilon_2 }
\right]\\[0.5cm]
\displaystyle f_\omega^e =a g f_\pi^2 \left[
\frac{1}{3} - \frac{\epsilon_1}{\Pi_{\pi\pi}^\rho(s) -\epsilon_2} +
\frac{1}{3} \frac{\mu^2 \epsilon_2}{(1-z_V) m^2+(1-\mu^2 )\epsilon_2}
\right]\\[0.5cm]
\displaystyle f_\phi^e =a g f_\pi^2 \left[
-\frac{\mu}{3} + \frac{\mu \epsilon_1}{(1-z_V) m^2 +\Pi_{\pi\pi}^\rho(s) -\mu^2 \epsilon_2}
+\frac{\mu}{3} \frac{\epsilon_2}{(1-z_V) m^2+(1-\mu^2 )\epsilon_2}
\right]
\end{array}
\right .
\label{Model7}
\ee
and, keeping only the leading (first order) terms, the loop corrections $\Pi_{V\gamma}(s)$
(see the definitions in Eqs. (\ref{Mass3}), and the expression for $\mu$ in Eqs. (\ref{Mass2})) are~:
\be
\left \{
\begin{array}{llll}
\displaystyle \Pi_{\rho^0 \gamma}(s) = (1-\frac{a}{2}) 
\frac{\Pi_{\pi \pi}^\gamma(s)}{g_{\rho \pi \pi}}
+(z_A-\frac{a}{2}-b) \frac{\epsilon_1+\epsilon_2}{g_{\rho \pi \pi}} 
+b \frac{\epsilon_2-\epsilon_1}{g_{\rho \pi \pi}} \\[0.5cm]
\displaystyle \Pi_{\omega \gamma}(s) = -(1-\frac{a}{2}) 
\frac{\epsilon_1}{\Pi_{\pi\pi}^\rho(s) -\epsilon_2} \frac{\Pi_{\pi \pi}^\gamma(s)}{g_{\rho \pi \pi}}
+(z_A-\frac{a}{2}-b) \frac{\epsilon_1+\epsilon_2}{g_{\rho \pi \pi}} 
-b \frac{\epsilon_2-\epsilon_1}{g_{\rho \pi \pi}}\\[0.5cm]
\displaystyle \Pi_{\phi \gamma}(s) =(1-\frac{a}{2}) \frac{\mu \epsilon_1 }{(1-z_V) m^2 + 
\Pi_{\pi \pi}^\rho(s) -\mu^2 \epsilon_2}
\frac{\Pi_{\pi \pi}^\gamma(s)}{g_{\rho \pi \pi}}
-(z_A-\frac{a}{2}-b) \mu \frac{\epsilon_1+\epsilon_2}{g_{\rho \pi \pi}} 
+b \mu \frac{\epsilon_2-\epsilon_1}{g_{\rho \pi \pi}}
\end{array}
\right .
\label{Model8}
\ee

The first term for each transition loop is the pion loop contribution while the others are resp.
the charged and neutral kaon loops. Of course, the functions occuring there are the same
as for $F_\pi^\tau$. We have denoted by $\Pi_{\pi \pi}^\gamma(s)$ the transition amplitude
for $\gamma-\rho^I$, which is in correspondence with the $W-\rho^\pm$ transition amplitude introduced
in the previous Section (see Eq. (\ref{Model4a})). A priori, the subtraction polynomials
of $\Pi_{\pi \pi}^\gamma(s)$ and $\Pi_{\pi \pi}^W(s)$ might be slightly different. However,
in an attempt to reduce further the number of free parameters of the model, we assume
that they coincide, which turns out to identify the amputated $W-\rho^\pm$ and
$\gamma-\rho^I$ transition amplitudes{\footnote {This is a nothing but a strong CVC assumption.}}.
We shall see that this assumption is well accepted by the data and, moreover, make clearer
the switching to the $\tau$ form factor expression.
 
 In addition to the explicit dependence of our model on the HLS basic parameters $a$, $g$, and
 on the breaking parameters $x$, $z_A$, $z_V$,$z_T$ and $\delta m^2$,
 there is a further dependence on subtraction parameters hidden inside  $\Pi_{\pi \pi}^\rho(s)$,
 $\Pi_{\pi \pi}^{W/\gamma}(s)$,
 $\epsilon_1(s)$ and $\epsilon_2(s)$. Isospin symmetry breaking is reflected in having
 a non--zero $\epsilon_1(s)$ function. We have~:
 \be
\left \{
\begin{array}{llll}
\displaystyle \Pi_{\pi \pi}^{W/\gamma}(s) &= Q_0(s) + \ell_\pi(s)\\[0.5cm]
\displaystyle \Pi_{\pi \pi}^\rho(s) &= P_0(s) + \ell_\pi(s)\\[0.5cm]
\displaystyle \epsilon_1(s)&= P_-(s) + \ell_{K^\pm}(s) - \ell_{K^0}(s)\\[0.5cm]
\displaystyle \epsilon_2(s)&= P_+(s) + \ell_{K^\pm}(s) + \ell_{K^0}(s)
\end{array}
\right .
\label{Model9}
\ee
\noindent where $\ell_\pi(s)$, $\ell_{K^\pm}(s)$ and $\ell_{K^0}(s)$ are now the {\it non-amputated}
$\pi^+\pi^-$, $K^+K^-$ and $K^0 \overline{K}^0$ loops, subtracted 
in order that these loops vanish at the origin. The parameter
polynomials $Q_0(s)$, $P_0(s)$, $P_-(s)$ and $P_+(s)$ are chosen to be second degree with zero 
constant terms in order to stay consistent with the  Node theorem \cite{RhoOmg0,Klingl}.

One can check that $F_\pi^e(0) = 1 +{\cal O}(\epsilon_1^2)$, which could have been expected from
having neglected terms of order greater than 1 in our diagonalization procedure{\footnote{
Actually, it depends on the (1,2) rotation matrix element~: 
 $1+{\cal O}([R_{12}(s=0)]^2)$ and, numerically, the neglected term is $\simeq 1.5 ~10^{-3}$.}}.

As the form factor  data  collected at the $\phi$ are not currently available,
the last term in Eq. (\ref{Model5}) could have been removed.
 However, in order to account for tails effects, we preferred keeping it
 and use a fixed width Breit--Wigner
 expression incorporating the  Particle Data Group  mass and width recommended values \cite{RPP2006}.
Due to the narrowness of the $\omg$ mass distribution, we also
have replaced in our fits the $\omg$ propagator by a fixed width Breit--Wigner
constructed using the recommended mass and width  from \cite{RPP2006}.

Let us also recall that Eq. (\ref{Model5}) is our formula for the pion form factor $F^e_\pi(s)$
in both the spacelike and timelike regions. Indeed, for consistency, we will not remove the $\omg$
and $\phi$ meson contributions while going to  negative $s$.

\vspace{0.5cm}
\indent \indent 
Using the first order correction to the $\rho$ mass 
eigenvalue, the inverse propagator could have been written $D_{\rho^0}(s) = s - \lambda_\rho(s)$,
as the leading order squared mass eigenvalue is~:
 \be
\displaystyle \lambda_\rho(s)= m^2 + \Pi_{\pi \pi}^\rho(s)+ \epsilon_2(s)
\label{Model21}
\ee
 
 As far as $e^+e^-$ data are concerned, we shall  modify  the eigenvalue expression  to~:
\be
\displaystyle \lambda_\rho(s)= m^2 + \Pi_{\pi \pi}^\rho(s)+ \epsilon_2(s) + \delta_2 \lambda_\rho(s)
\label{Model22}
\ee
by adding the second order correction given in Eqs. (\ref{Mass9}). This does not add any more
freedom in the model, but rather allows some check of the  diagonalization method.

Therefore, the difference between $F_\pi^e(s)$ and $F_\pi^\tau(s)$ is solely concentrated in the
coupling changes from ideal to physical fields given by the varying matrix $R(s)$ (see
Eq.(\ref{Mass7})) which only affects $F_\pi^e(s)$. Stated otherwise, modifying the
function $F_\pi^\tau(s)$ in order to incorporate isospin breaking effects is strictly equivalent to
using our expression for $F_\pi^e(s)$ directly, the factor $S_{EW} G_{EM}(s)$ being removed
and $\delta m^2$ being made identically zero.

\section{The Photon Vacuum Polarization}
\label{ModelFFgVP}
\indent \indent
The raw data  on the pion form factor $F^e_\pi(s)$ should be ``undressed" by unfolding
the contributions due to radiative corrections  and to the photon vacuum polarization (VP)
before any comparison with $\tau$ data (we refer the reader to \cite{IFSVP,IFSVP2} for a comprehensive
analysis of these factors and for previous references). Quite generally, available
experimental data on $F^e_\pi(s)$ have already been unfolded from radiative corrections 
\cite{CMD2-1995corr,CMD2-1998-1,CMD2-1998-2,SND-1998,NA7,KLOE}. All the data sets just referred
to are not unfolded from photon vacuum polarization (VP) effects, except for KLOE data \cite{KLOE}.
Therefore, one has to account for VP effects by including the corresponding factor when comparing 
a pion form factor model
with experimental data. Traditionally (see for instance \cite{Bolek1} and references quoted therein),
 this results in the change{\footnote{
 If one denotes by $\Sigma(s)$ the photon self--mass, the inverse photon propagator 
 is given by $D^{-1}_\gamma(s) = s-\Sigma(s)= s(1 - \Sigma(s)/s)$. Therefore,
 compared with traditional notation, we have $\Pi_{VP}(s)=- \Sigma(s)/s$. This is not a problem but should
 be kept in mind.}}~:
\be
\displaystyle F_\pi^e(s) \longrightarrow (1-\Pi_{VP}(s)) F_\pi^e(s)
\label{Model50}
\ee
when comparing with most data sets. 

The VP function $\Pi_{VP}(s)$ contains two parts. 
The first one is the sum of the leptonic loops which can be computed in closed form  at leading order 
(In Appendix \ref{EE}, 
we recall the explicit form at order $\alpha$ and give its expression along the real $s$ axis). 
The second part is the one particle irreducible 
hadronic contribution to the photon self--energy which is derived by means of a dispersion relation~; 
at low energy,
where non--perturbative effects are dominant, this is estimated using the experimentally mesured $e^+ e^-$ 
cross section (see, for instance, \cite{Bolek1,Bolek2,Bolek3,IFSVP}), while the high energy 
tail is calculated using perturbative QCD. 

For our purpose, we use the sum of the leptonic VP as given in  Appendix \ref{EE} 
for $e^+e^-$, $\mu^+\mu^-$ and $\tau^+\tau^-$ together with a numerical parametrization
of the hadronic contribution. From the 2--pion threshold to the $\phi$ mass, we benefited from a
 parametrization\footnote{This parametrization neglects the (very small) imaginary part of the hadronic 
VP contribution.}  provided by M. Davier \cite{DavierPriv}. 
Below the 2--pion threshold and down to $s=-0.25$ GeV$^2$, we use instead a 
(real valued) parametrization provided by H. Burkhardt  \cite{HelmutPriv}.

\section{Decay Widths Of Light Mesons}
 \label{DecayWidths}
\indent \indent In order to compute decay widths and fit data, one has to define
the couplings allowing the decays of the light mesons involved. For the two--photon decays
of the $\eta$ and $\eta^\prime$ mesons, as well as for the radiative decays of the $\rho^\pm$
and $K^*$'s mesons, the couplings defined after SU(3) and Nonet Symmetry breaking (see 
 Eqs. (\ref{DD2}) and (\ref{DD5})) are the couplings coming directly in the decay widths formulae
(see Subsection \ref{PartialWidths}) and do not depend  on further Isospin Symmetry breaking effects 
than mass values in phase space factors.

The isospin breaking procedure we presented plays only for the $\rho^0$, $\omg$ and $\phi$
mesons. In order to compute the  leptonic decays of these, one has to use the full couplings
$F_V^e(s)$ as given by Eqs. (\ref{Model7}) and (\ref{Model8})
in Eq. (\ref{DD8}) and computed at the appropriate vector meson masses  $F_V^e(m_V^2)$.
As the loop functions are slowly varying, one can choose  $m_V$ as the Higgs--Kibble masses
occuring in the Lagrangian (see Eq.(\ref{Lag_ee})), which moreover simplifies the fitting procedure.
 
For the other (radiative decay) coupling constants one has to combine the ideal coupling
constants (given in Appendix \ref{DD}) 
using the transformation $R(s)$ to derive the physical coupling constants, as was described
in  \cite{su2}~; the context, compared to \cite{su2}, slightly differs  due to the fact
that, now, the mixing parameters are functions to be computed at the appropriate   $m_V^2$ values
in order to go to the vector meson mass shell. 

Traditionally, the $\rho^0$ is decoupled from mixing and treated as the $\rho^\pm$  and mixing effects
are only considered in the $(\omg-\phi)$ sector. Additionally, it is usual to treat the
$(\omg-\phi)$ mixing angle as a constant to be fit (see \cite{BGP,BGPbis,rad,mixing,su2} and the references 
quoted therein). The approach in the present study is different~: 
One considers a full $(\rho^0-\omg-\phi)$ mixing scheme (as in \cite{su2}), however
-- for the first time -- the mixing parameters are functionally related and the same functions
have to be computed at each vector meson mass. For instance, the $(\omg-\phi)$ mixing
``angle'' has not the same numerical value at the $\omg$ mass and at the $\phi$ mass.
This only reflects that the mixing is actually invariant mass dependent.
When, as for $\omg \ra \pi \pi$ \cite{ffVeryOld} and $\phi \ra \pi \pi$  \cite{OLYAPhi,SNDPhi}
data exist on the phase of the coupling constant, these phases can be introduced in the fit with
the same functions,  the modulii  of which determine the branching fractions. 
 
\section{The Full Set of Data Submitted To Fit}
\label{FullDataSet}
\indent \indent In order to work out the model presented in the Sections above, we  use 
several kinds of data sets. In this Section, we list them and give some details on the way 
they are dealt with in our fit procedure.  

\subsection{Partial Width Decays}
\indent \indent As a general statement, the decay data submitted to fits have been chosen 
as the so--called ``fit" values recommended by the Particle Data Group (PDG) in the latest (2006) issue
of the Review of Particle Properties \cite{RPP2006}.

This covers, with no exception, the leptonic decay widths of the $\rho^0$, $\omg$ and $\phi$ mesons,
the two--photon decay widths of the $\eta$ and $\eta^\prime$ mesons and the $\pi^+ \pi^-$
partial width of the $\phi$ meson. There are 
two measurements of the phase of the coupling constant $g_{\phi \pi^+ \pi^-}$ reported in
the literature~; the older one \cite{OLYAPhi} is $\psi_\phi=-42^\circ \pm 13^\circ$ and more
recently \cite{SNDPhi} $\psi_\phi=-34^\circ \pm 4^\circ\pm 3^\circ$. Summing up in quadrature
the errors, we choose as reference value in our fits $\psi_\phi=-34^\circ \pm 5^\circ$. We could have chosen
to include in our fits the 2--photon decay width of the $\pi^0$~; however, we preferred replacing this 
piece of information by the pion decay constant value $f_\pi=92.42$ MeV and did not let it vary, as 
this is supposed to carry a very small error\footnote{Possible Isospin Symmetry breaking effects
might have to be considered.} \cite{RPP2006}. 

The RPP pieces of information \cite{RPP2006} concerning the $\rho$ and $\omg$ decay width to $\pi^+ \pi^-$
and the partial width $\rho^0 \ra e^+ e^-$ are not considered as data to be submitted to fits, 
as they have all been
extracted from fitting the same pion form factor timelike data which we are included in our fit
procedure (see the Subsection below)~;
this does not prevent us from comparing our results to the RPP available information.  This is
also true for the relative phase of the couplings $g_{\omg \pi^+ \pi^-}$ to $g_{\rho \pi^+ \pi^-}$
(the so--called Orsay phase) which has been measured \cite{ffVeryOld}
with the result $\psi_\omg=104.7^\circ \pm 4.1^\circ$.

Instead, it is quite legitimate to include the $\omg \ra e^+ e^-$ partial width in our fit
procedure as, even if this mode could have been marginally influenced by the pion form factor data, 
it is mostly extracted from $e^+ e^-   \ra \pi^+ \pi^- \pi^0$ data \cite{RPP2006}. As the pion
form factor spectrum around the $\phi$ mass is not currently available, the $\phi \ra e^+ e^-$ partial 
width is quite legitimately included in our fit data set.

In the present  work, as in our previous works on the same subject \cite{rad,mixing,su2},
we do not intend to use the decay widths $K^* \ra K \pi$. Actually, as for the width $\rho \ra \pi \pi$
which is inherently fitted with the pion form factor, the choice of the mass value
for a very broad object makes the extraction of coupling constants a delicate matter. It should
be more appropriately discussed with the $K\pi$ form factor in $\tau$ decays when the corresponding
data will become available.

 The data on the two kaon partial widths of the $\phi$ meson are also left outside our fit procedure.
 In a previous work of some of us \cite{su2}, as in other works \cite{BGPter} the issue of
 accomodating the $\phi \ra K^+ K^-$ partial width was raised. 
  A recent work \cite{Fischbach} claims that the ratio of these partial widths
 can be accomodated by introducing corrections to the decay widths which
  increase both  partial widths as derived from the matrix elements of the transitions.
 As then, the problem may affect both the 
 $\phi \ra K^+ K^-$ and the $\phi \ra K^0 \overline{K}^0$ partial widths, we have preferred
 leaving both modes outside the fit procedure. We will discuss this issue below in a
 devoted  Subsection.

We also use all radiative decay partial widths of light flavor mesons of the form $V \ra P \gamma$
or $P \ra V \gamma$. As a general rule,  we chose as reference data the ``fit" values recommended
by the PDG as given in the latest RPP issue \cite{RPP2006}. There are two exceptions to this statement~: the
partial widths for $\omg \ra \eta \gamma$ and $\omg \ra \pi^0 \gamma$. 

Indeed, as already noted in \cite{su2}, there is some difficulty in accomodating the present 
$\omg \ra \eta \gamma$ ``fit" branching fraction ($(4.9\pm 0.5)~10^{-4}$) while the so--called ``average" value
\cite{RPP2006} ($(6.3~ \pm 1.3)~10^{-4}$) is much better accepted by our model fitting.
 
On the other hand, the new ``fit" value for the branching fraction  $\omg \ra \pi^0 \gamma$ 
($(8.90 \pm 0.25)~10^{-2}$) is also hard to accomodate in our model. We show
that the previous PDG ``fit" value ($[8.50\pm 0.50]~10^{-2}$) seems in better consistency
with the rest of the data we handle~; we also
checked that the value ($[8.39\pm 0.25]~10^{-2}$) produced by a fit \cite{pi0gam}
performed in a completely different context was also well accepted, pointing to a
possible overestimate of the central value for this mode\footnote{Looking at \cite{RPP2006}, 
the role of the data and analyses for the $e^+ e^- \ra \pi^0 \gamma$ process itself
to get the ``fit" value for the partial width $\omg \ra \pi^0 \gamma$ is unclear.}. The 
questions raised by the values of these two decay widths will be discussed at the appropriate place below.

Finally, we also introduce in the fit procedure the ratio of the kaon to pion decay constants
as they are reported in the latest RPP  \cite{RPP2006}. This actually coincides with our SU(3) 
breaking parameter ($z_A=[f_K/f_\pi]^2$).

\vspace{0.5cm}
\subsection{Timelike Form Factor Data in $e^+ e^-$ Annihilations}
\indent \indent Beside the decay data listed just above, we have included in our
fit all data on the pion form factor collected in $e^+ e^-$ annihilations by the OLYA
and CMD Collaborations as tabulated in \cite{Barkov} and the DM1 data \cite{DM1} collected
at ACO (Orsay). These data will be referred to globally as ``old timelike data". 
When included into a $\chi^2$ expression, systematic errors have to be combined with
the published statistical errors~; they have been first added in quadrature to the statistical
errors for OLYA data ($4\%$) and CMD ($2\%$) following expert advice \cite{simonPriv}.
 However, for sake of consistency with the new data discussed just below, we preferred extracting
the correlated part of the systematic errors, estimated \cite{simonPriv} to 1\%
 and have performed  the same treatment as for the new data (see just below). 
The accuracy of the DM1 data making the influence of this data set 
marginal, we did not add any further contribution to the published errors.
We only use the data points located below the $\phi$ meson mass in order to avoid being sensitive
to higher mass vector mesons, not included in the present model.

Four additional data sets have been collected later at Novosibirsk on the VEPP2M ring.
The first one, covering the region from about 600 to  960  MeV,
collected by the CMD2 collaboration \cite{CMD2-1995} and
recently corrected \cite{CMD2-1995corr}, is claimed to have the lowest 
systematic error ($0.6\%$) ever reached in this field.

CMD2 has collected in 1998 and recently published two additional data sets, one \cite{CMD2-1998-1}
covering the energy region from 600 to 970 MeV is claimed to reach a systematic error of 
$0.8\%$, and a second set \cite{CMD2-1998-2} covering the threshold region (from 370 to 520 MeV)
with an estimated systematic error of $0.7\%$. On the other hand, the SND collaboration has recently published 
\cite{SND-1998} a new data set covering the invariant mass region from 370 to 970 MeV with
 a systematic error of $1.3\%$ over the whole data set except for the very low mass region
where the (first) 2 points carry a systematic error of $3.2\%$.

Concerning these four data sets (which will be referred to  globally as ``new timelike data"),
we could, as per usual, add in quadrature the systematic and statistical errors and then get
a diagonal error matrix which can  be used in  $\chi^2$ fits in a trivial way.

However, an important part of the  systematic uncertainties for these data sets is expected to be a common
global scale uncertainty  \cite{simonPriv} which has been estimated to $0.4\%$ and generates bin to bin
correlated errors. In principle, one should take the latter information into account 
in fits~; this implies dealing with systematic and statistical errors in a way slightly more elaborate
than simply adding  in quadrature statistical and systematic errors. 

Firstly, the (bin per bin) uncorrelated part of the systematic error is derived
by subtracting in  quadrature $0.4\%$  from the already quoted systematic errors. 
This uncorrelated part of the systematic error 
 ({\it i.e.}  $\sqrt{\sigma_{syst.}^2-(0.4\%)^2}$, depending on the data set considered)
can certainly be added in
quadrature to the statistical error bin per bin, giving a combined standard deviation named $\sigma_i$
for the measurement $m_i$ in the energy bin $i$~; the $\sigma_i$ are
uncorrelated errors and define a diagonal error matrix. The question then becomes how to redefine 
the full covariance matrix for each experiment, being understood that the quantity 
to be compared with the theoretical pion form factor $f^{th}_i$ for each energy bin $i$
is related with the measurement $m_i$ by~:
\be
f^{th}_i \longmapsto m^\prime_i=(1 + \delta \lambda) m_i
\label{data1}
\ee

\noindent where $\delta \lambda$ is considered a gaussian random variable with zero mean and
standard deviation $ \lambda =0.4\times 10^{-2}$. With this assumption it is possible to
model reasonably well the covariance matrix, which is no longer diagonal.

Secondly, one has to treat these correlations.
The quoted correlated systematic error is a conservative estimate of the accuracy 
of radiative corrections performed on the four data sets using the same Monte Carlo 
generator \cite{simonPriv}. Therefore,  the fit parameter introduced
in order to optimize the absolute scale should be the same for all data sets~; in statistical
terms, this fit parameter value can be considered  as only $one$ sampling of the gaussian random variable 
$\delta \lambda$ defined just above and should be valid for all data collected by the CMD2 and
SND Collaborations. 

If the correlated part of the systematic error was strictly zero, the error covariance matrix 
for each data set would simply be given by~:
\be
V_{ij} = \sigma_i^2 \delta_{ij}
\label{data2}
\ee
\noindent where $i$ and $j$ label energy bins in the data set. In the case of existing
correlations, having defined $\sigma_i$ as the sum in quadrature of the statistical error and
the uncorrelated systematic error in the $i^{th}$ bin , the error covariance matrix elements can be written~:
\be
V_{ij} = \sum_{k,l}M_{ik} W_{kl}M_{lj} ~~~~,
\label{data3}
\ee
\noindent where~:
\be
\begin{array}{lll}
M_{ij} =& \sigma_i \delta_{ij}~~~~~ {\rm and} ~~~~~,
W_{ij}=&\delta_{ij} + \lambda^2 e_i e_j
\end{array}
\label{data4}
\ee
\noindent $\lambda =0.4~10^{-2}$ being the standard deviation of the
correlated error function and the vector $e$ being defined by its components on the various
energy bins $i$ as the ratio of the corresponding measurement to its uncorrelated error~:
\be
e_i=m_i/\sigma_i~~~~~, ~~~~~  \forall i \in [1, \cdots n_{measur.}]
\label{data5}
\ee

However, what is relevant for $\chi^2$ fitting, is not so much the covariance matrix
Eq. (\ref{data4}) as its inverse. It happens that the matrix $W$ can be inverted in closed
form~:

\be
\begin{array}{lll}
W_{ij}^{-1}=&\delta_{ij} - \mu^2 e_i e_j~~~~~ {\rm and} ~~~~~ \displaystyle
\mu^2=\frac{\lambda^2}{1+\lambda^2 \sum_{i=1}^{n_{measur.}} e_i^2}
\end{array}
\label{data6}
\ee
\noindent and the full error covariance matrix is also inverted in closed form~:
 \be
V_{ij}^{-1} =\sum_{k,l} M_{ik}^{-1} W_{kl}^{-1} M_{lj}^{-1} ~~~~.
\label{data7}
\ee

This is, together with the measured values, the main ingredient of the $\chi^2$ calculation which will
be performed with the four new timelike data  sets. Finally, while fitting the new data, a term has to be added
to the $\chi^2$~; naming $\lambda_{fit}$ the fit parameter for the global scale common to all the new Novosibirsk
data sets and $\lambda_{exp}=0.4\times 10^{-2}$ the scale uncertainty on the measured form factor estimated by the 
experiments, this additional contribution to the $\chi^2$ is simply $[\lambda_{fit}/\lambda_{exp}]^2$. 

{\it Mutatis mutandis}, the same method  has been applied to the old Novosibirsk data sets using another global
scaling factor $\lambda_{fit}^\prime$ with $\lambda_{exp}^\prime=1.0~10^{-2}$, as recommended by informed people
\cite{simonPriv},  and the same procedure to construct the final inverse covariance matrix to be used in fits.

A new data set has been recently collected by the KLOE collaboration \cite{KLOE} using the Radiative Return
Method. Existing analyses (see, for instance, the short account in \cite{Davier2007}) however report a disagreement 
between KLOE data and the recently collected data sets at Novosibirsk due to some systematic effect presently
not understood. A recent study of a parametrization of the pion factor \cite{OmgShift} argues about a possible
systematic energy shift in the data which would be detected by fitting the $\omega$ mass. In view of this unclear
situation, we have found it appropriate to postpone including the existing KLOE set among our fitting data samples.
 
\subsection{Spacelike Pion Form  Factor Data}
\indent \indent
In order to further constrain the pion form factor in the timelike region, information on the close
spacelike region is valuable.
Reliable data on the pion form factor in the negative $s$ region are somewhat old \cite{NA7,fermilab2}.
The Fermilab data set \cite{fermilab2} consists of 14 measurements of $|F_\pi(s)|^2$ between
$s=-0.039$ GeV$^2/c$ and $s=-0.092$ GeV$^2/c$ with $2 \div 7$ \% statistical error~;
an estimated systematic error (overall normalization) of 1 \%  is provided in
\cite{fermilab2}. The NA7 data cover the region between
$s=-0.015$ GeV$^2/c$ and $s=-0.253$ GeV$^2$ with 45 measurement points and an overall statistical precision better
than those of the Fermilab data. However, NA7 data are also claimed to undergo
an overall scale error of 0.9 \% rms.

One will use these two data sets and treat these correlated systematic errors exactly as explained above
for the timelike  pion form factor data. 

Data have more recently been collected at the Jefferson Accelerator Facility \cite{Jefferson}
and reanalyzed in order to optimize the extraction of the pion form factor data in the region
for $s$ between $-0.60$ and $-1.60$  GeV$^2$ with a quoted uncertainty of about 10\%. 
No precise information about the correlated--uncorrelated sharing of the systematic error is reported.
Including these data involves some more studies and modelling which goes beyond the main task 
of the present work,  namely, to check the consistency of $e^+e^-$ and $\tau$ data.

\subsection{Phase Pion Form  Factor Data}
\indent \indent There are several data sets available which provide
measurements of the isospin 1 part of the $\pi \pi$ amplitude phase shift. The most precise set
is the CERN/Munich one \cite{Ochs}, but the older Fermilab data set \cite{Protopescu} is still
useful. However,  systematic errors here are not completely controlled.  Moreover, as we neglect
vertex corrections at the $\pi \pi$ vertex and $t$-channel resonance exchanges which may carry
some unknown imaginary part, one cannot draw firm conclusions when comparing the phase information
of our pion form factor with phase shift data. Therefore, we have left these data outside our fitting 
procedure and limited ourselves to simply compare graphically with the phase of our pion form factor.

\subsection{Pion Form  Factor Data From $\tau$ decays}
\indent \indent There are presently three available data sets concerning
the pion form factor. These have been collected at LEP by ALEPH \cite{Aleph}
and OPAL  \cite{Opal} Collaborations and at much lower energy by the CLEO Collaboration
\cite{Cleo}.

The data provided by the  ALEPH Collaboration \cite{Aleph} include the covariance matrices for
statistical and systematic errors which should be added before inversion in order to be used
in a $\chi^2$ minimization.
There is some disagreement between ALEPH \cite{Aleph} and OPAL data \cite{Opal} which has led
most works to discard this data set~; we shall do likewise.

The CLEO data \cite{Cleo} on the pion form factor are also provided with their 
full error matrix, but one that accounts for statistical errors only. 
Statistical errors dominate most of the systematic uncertainties
except for those contributing to the absolute energy scale for determining
$\sqrt{s}$ \cite{CleoPriv}.  These were quantified by CLEO as a systematic
uncertainty on the value of the $\rho^\pm$  mass obtained in their fits to
form factor models, estimated to be 0.9 MeV. This error, not accounted
for by the CLEO error covariance matrix, is a systematic error which
correlates the various bin energy values. In contrast with the Novosibirsk data,
it is not easy to rigorously account for this correlated systematic 
 error\footnote{It affects the position of the measurement, not the measured value 
itself.}. As an approximation we allow the central bin $\sqrt{s}$
value to vary by some $\varepsilon$ MeV and add $[\varepsilon/0.9]^2$ to the CLEO
data $\chi^2$.  This approach provides a simple and reasonable way to deal
with the data and errors \cite{CleoPriv}.

In order to stay consistent with our dealing with $e^+e^-$ data, we have limited
our fitting range to the $\phi$ mass and then removed all points above $s=0.9$ GeV$^2$.
This leaves us with 33 measurements from ALEPH and 25 from CLEO, largely  unaffected by higher 
mass vector meson effects, as will be checked.

\section{The Main Global Fit To The Data Sets }
\label{DataFitting}
\subsection{General Comments About The Fits} 
\indent \indent Our global model has seven parameters carrying an obvious physical meaning~:
\begin{itemize}
\item 
 The universal vector coupling $g$, 
 \item
 the SU(3) breaking parameter $z_A$ (expected to coincide
 with $[f_K/f_\pi]^2$ within errors), 
 \item
 the Nonet Symmetry breaking parameter $x$, 
 \item
 the basic HLS parameter $a$ (expected close to 2), 
 \item
 the parameter $z_V$ which mostly governs 
 the mass difference between the $\rho^0-\omg$ system and the $\phi$ meson but also plays a
 role in some coupling constants, 
  \item
 $z_T$ which affects only the $K^*$ radiative
 decay sector in the data used, 
 \item
 and, finally, the $\rho^0-\rho^\pm$ squared mass shift $\delta m^2$.
 \end{itemize} 
 
\indent \indent  These have been already fitted in isolation in related previous works
\cite{rad,mixing,su2,ffOld,box} and we expect to find fit values close to the already published ones.
Within our approach, the pseudoscalar mixing angle is not free, but is derived
from the previous parameters using Eq. (\ref{DD6b}) and is expected close to $-10.5^\circ$
degrees from previous fits \cite{rad,su2}. This has been found in perfect agreement with the 
two-angle formulation \cite{WZWChPT} expressed in the framework of Extended Chiral Perturbation
Theory \cite{leutw,leutwb}.

Beside these parameters carrying a clear physical meaning, one has the subtraction polynomial
of the pion loop  (mostly associated with the $\rho$ meson self--energy),
 assumed to be written $c_1 s + c_2 s^2$  with $c_1$ and $c_2$ to be fitted. 
Two additional subtraction polynomials carrying the same form and associated with
the difference ($\epsilon_1(s)$) and the sum ($\epsilon_2(s)$) of the $K^+K^-$ and $K^0 \overline{K}^0$
loops introduce 4 more parameters\footnote{We note that we approximate the $K^\pm K^0$ loop
by the average value of the $K^+K^-$ and $K^0 \overline{K}^0$ loops  in order to limit
the number of free parameters.} to be fit.  
Finally, two more subtraction parameters come from the specific subtraction of the $\gamma \rho$ (or
$W\rho^\pm$) transition amplitude. We thus end up
with 15 parameters\footnote{It will be emphasized later on that one among these subtraction
parameters does not influence the fit and can be safely fixed to zero.}
 for a number of data of 344 (18 decay modes, 127 data points from 
the new timelike pion form factor data, 82 from the old timelike pion form factor data, 59 data
points in the spacelike region, 33 data points coming from  ALEPH data and 25 from CLEO).

In addition to these parameters which define our model, we have to account for
correlated systematic errors in several experiments by fitting the corresponding
scale factors and using the experimental pieces of information as constraints. 
These additional degrees of freedom are therefore exactly compensated in number by the constraints.
This covers the global scale factor of the former Novosibirsk experiments  
as reported in \cite{Barkov} (estimated to 1.0\% r.m.s.), the global scale factor of the new 
Novosibirsk experiments as reported in \cite{CMD2-1995corr,CMD2-1998-1,CMD2-1998-2,SND-1998}
(estimated to 0.4\% r.m.s), the scale factor for the NA7 \cite{NA7} and Fermilab \cite{fermilab2} data 
(estimated respectively to 0.9 \% and 1.0 \% r.m.s.). Finally, the CLEO data set is expected
to carry a systematic energy shift which will be fitted and is expected \cite{CleoPriv} of 
the order 0.9 MeV.

\vspace{0.5cm}
\indent \indent We have  performed various kinds of fits. In all of them, as detailed
above, we have introduced all usual symmetry breaking effects in the value of meson masses,
the prominent effects of $\rho^0-\omg-\phi$ mixing (for $e^+e^-$ data) and the long--
and short--range \cite{Marciano} IB correction factors (for the $\tau$ spectra). We have observed
that the two proposed ways to account for long range corrections by either of 
\cite{Cirigliano1,Cirigliano2,Cirigliano3} and \cite{Mexico1,Mexico2,Mexico3} approaches provide 
quite similar effects and that,
on the basis of probabilities, the difference was never observed significant in any of our fits.
For definiteness, we choose to use the function of \cite{Cirigliano1,Cirigliano2,Cirigliano3}
for all results presented here.
\begin{table}[!htb]

\begin{tabular}{|| c  | c  | c | c  | c | c ||}
\hline
\hline
\hhhu Data Set  & Without VP &  \multicolumn{4}{|c|}{With Vacuum poliarisation (VP)}\\
\hline
\hhhu $\sharp$ (data $+$ conditions) & Full Fit & Full Fit  & No $\tau$ & No Spacelike 
& No$\rho$  mass shift\\
\hline
\hline
Decays (18+1)\hhhu       & $11.46   $   & $11.13 $ &   $11.52 $ & $11.48  $  & $11.25$ \\
\hline
New  \hhhu &  & & & & \\
Timelike (127+1) & $132.81 $   & $128.10 $ &  $122.02 $&  $125.76 $ &  $132.23 $\\
\hline
Old  \hhhu &  & & & & \\
Timelike (82+1) & $62.22 $    & $59.05 $ &  $54.68 $ &  $55.20 $  &  $60.15 $ \\
\hline
Spacelike(59+2) \hhhu    & $68.53 $ & $65.70  $ &  $55.20$ & \bf{ 89.82/(59)} &  $65.13$ \\
\hline
$\tau$ ALEPH (33)\hhhu & $27.06 $  &  $23.86 $ & \bf{ 42.27/(33)} &  $20.80 $ &  $24.48 $ \\
\hline
$\tau$ CLEO (25+1)\hhhu  & $25.53 $&  $26.06 $ & \bf{26.16/(25)}&  $29.72$ &  $28.55$   \\
\hline
\hline
$\chi^2/\rm{dof}$ \hhhu &327.40/331  & 313.83/331 &  257.73/274 & 238.81/272& 321.75/332\\
Probability  & 54.6 \%  & 74.3 \%  &  75.2\% & 92.7\% & 64.7\%\\
\hline
\hline
\end{tabular}
\caption{
\label{T1} The first column lists the subset named as defined in the text together
with its number of measurements and condition(s) if any.     
Each row displays the corresponding $\chi^2$ contribution
under the condition quoted in the title of the data column.
The last row  gives the total $\chi^2$/(number of degree of freedom), followed by the fit probability.
Information written boldface indicates the $\chi^2$ distance of the fit function
to a data set left outside from the fit procedure together with its number of data points.
In this case, the condition parameter associated with the corresponding  data set (scale or mass shift) is fixed to 
the value returned by the full global fit reported in the second data column and given in Table \ref{T2a}.
}
 
\end{table}

On the other hand, it was useful to check the effect of excluding
the photon vacuum polarisation (VP), by fixing  the corresponding factor
 to 1. We also found it of interest to perform fits by excluding either
the $\tau$ data or the spacelike data~; this gives information on the effect of
these on the global fit quality and on the stability of the fit parameter values.  
Finally, it has also been of interest to check the mass shift effect between the $\rho^0$ and the $\rho^\pm$
mesons, by fixing the corresponding parameter to zero.
The results summarizing the statistical qualities are gathered in Table \ref{T1} and the 
fit parameter values can be found in the appropriate data column of Tables \ref{T2a} and \ref{T2b}.

\begin{table}[!htb]
\begin{tabular}{|| c  | c  | c | c  | c  ||}
\hline
\hline
\hhhu Parameter  & Full Fit &  No $\tau$ &  No Spacelike & No $\rho$ mass shift\\ 
\hline
\hline
Scale New Timelike \hhhb & $1.006 \pm 0.004$ &$1.000 \pm 0.003$ & $1.004 \pm 0.003$
&  $1.007 \pm 0.003$ \\
\hline
Scale Old Timelike  \hhhb & $1.012 \pm 0.009$  &  $1.010 \pm 0.009$ & $1.011 \pm 0.009$
 & $1.013 \pm 0.009$ \\
\hline
Scale NA7  \hhhb & $1.008\pm 0.007$  & $1.008 \pm 0.007$ &  {\bf 1.008}
&  $1.011 \pm 0.006$  \\
\hline
Scale  Fermilab \hhhb &$1.006 \pm 0.007$  &$1.006 \pm 0.008$ & {\bf 1.006} 
& $1.008 \pm 0.007$\\
\hline
CLEO  Shift (MeV)\hhhb &$0.40 \pm 0.52$ &  {\bf 0.40} &  $0.36 \pm 0.52$
 &  $1.37 \pm 0.39$  \\
\hline
$\delta m^2$ ($10^2$ GeV$^2$)\hhhb & $-0.268 \pm 0.095$ & {\bf -0.268}& $-0.285 \pm 0.096$ &  {\bf 0}\\
\hline
\hline
$a$\hhhb & $2.303 \pm 0.012$ & $2.297 \pm 0.012$ & $2.292 \pm 0.012$ & $2.306 \pm 0.012$ \\
\hline
$g$\hhhb & $5.576 \pm 0.015$& $5.578 \pm 0.017$ & $5.597 \pm 0.016$ & $5.573 \pm 0.015$ \\
\hline
$x$\hhhb & $0.903 \pm 0.013$ & $0.902 \pm 0.013$  & $0.902 \pm 0.013$ & $0.903 \pm 0.013$   \\
\hline
$z_A$\hhhb & $1.503 \pm 0.010$ &  $1.505 \pm 0.010$ &  $1.507 \pm 0.010$&  $1.503 \pm 0.010$ \\
\hline
$z_V$\hhhb &$1.459 \pm 0.014$&  $1.466 \pm 0.014$ &  $1.453 \pm 0.014$&  $1.460 \pm 0.014$  \\
\hline
$z_T$\hhhb & $1.246 \pm 0.049$ & $1.245 \pm 0.049$ & $1.243 \pm 0.049$& $1.246 \pm 0.049$  \\
\hline
\hline
\end{tabular}
 
\caption{
\label{T2a}
Parameter values in fits performed including  photon VP. Three data columns are associated
with all data (first data column), removing only the $\tau$ data (second data column)
and removing only the spacelike data (third data column). The last data column
reports parameter values returned while fitting all data sets by fixing $\delta m^2 \equiv 0$. 
Information written boldface displays values not allowed to vary in the fit procedure.}
 
\end{table}

\subsection{Discussion Of The Fit Information}
  \indent \indent
  Table \ref{T1} reports the statistical information about our fits under various conditions.
  As a general statement, the fit quality is always either reasonable or very good as clear from the last
  row in Table \ref{T1}. 
  
   As a first remark, neglecting to account for photon vacuum polarization effects  does not end up 
   with a dramatic failure~; however, there is a general improvement while introducing the corresponding
function. The negligible degradation observed for CLEO data is entirely produced by the value found for
the CLEO mass shift parameter (which contributes to the $\chi^2$, as explained above)
will be discussed below. 
This statement clearly  follows from comparing the two data columns named ``full fit" in Table \ref{T1}.

The gain in $\chi^2$, while including the photon VP, is  13.5 units without any additional parameter freedom
and, in terms of fit probability, one wins 20 \%. 
Therefore, one may conclude that the data description prefers including explicitly the 
photon VP  while fitting the $e^+e^-$ data. Under realistic conditions, the fit
probability is then always of the order 75 \% or better.  

 The fits have always been performed
using the package {\sc minuit} \cite{James:1975dr} and the errors quoted are always the improved
errors returned by the routine {\sc minos}. This has allowed us to check
that the minimum $\chi^2$ was always locally parabolic, which provides symmetric errors.

Tables  \ref{T2a} and \ref{T2b} display the fit parameter values as returned by the fit 
under the conditions defined by the various titles of the data columns. We only 
provide results including the photon VP inside the definition of the pion form
factor for $e^+ e^-$ annihilations. Comparing the various data columns in Tables  \ref{T2a}
and \ref{T2b}, clearly illustrates that the  fit parameter values stay close 
together, and generally widely within their ({\sc minos}) errors. The single exception
is obtained while removing the spacelike data~; in this case, the coefficients
for the subtraction polynomial $\Pi_{\pi \pi}^{W/\gamma}$ differ significantly
from all other cases. 
This could have influenced the predicted values for the $V \ra e^+ e^-$ partial
widths~; however, we have checked that this is not the case numerically.

Among the fit parameter values given in Table \ref{T2a}, the most interesting 
are clearly  the fit values for the scale factors which are nicely consistent with
the corresponding experimental information recalled at the beginning of this Section.

The single exception is the CLEO global invariant mass shift which is found consistent with
zero. Taking into account the way  the 0.9 MeV expected shift has been determined\footnote{
In order to make consistent the $\rho$ parameters derived from fits to the ALEPH and
CLEO data.} \cite{CleoPriv}, this information is interesting. 
It will be rediscussed when examining the fit 
residuals which provides an additional important information.

\vspace{0.5cm}

\indent \indent
The third and fourth data columns in Table \ref{T1} provide further information~: 

{\bf i/} Removing the $\tau$ data from the fit sample, one can construct the {\it predicted}
distributions for the ALEPH and CLEO data which are fully derived  using  our model together with only 
light meson 
partial width decays and the $e^+e^-$ data. The predicted $\chi^2$ distance to CLEO data
is practically unchanged with respect to fitting with them, while  the prediction for the ALEPH data 
is not as good even if it remains 
reasonable{\footnote{The $\chi^2$ distance for the fitted part of the ALEPH spectrum corresponds
to an average  $\chi^2$ per point of 1.28 and thus to an average distance of $1.13 \sigma$
per data point. As a prediction, it is already a good starting point, which is improved
by the global fit to an average distance of $0.90 \sigma$ per data point.}}. 
This indicates that CLEO data are in so nice agreement with predictions
(especially $e^+e^-$ data) that they do not
really constrain the fit~! In this respect, ALEPH data, while introduced in the fit data
set, clearly influence the procedure.

{\bf ii/} Removing only the spacelike data looks a little bit more appealing. The $\chi^2$ distance
of the NA7 and Fermilab data altogether is degraded by $\simeq$ 24 units, 
and the fit probability grows from 74 \% to 93\%, pointing to some slight difficulty in 
accomodating these data sets. However, this result is by no way problematic enough to either
force us to remove the spacelike data or to deeply question their quality.

In order to compute the $\chi^2$ distance of the data samples left out from fit, one had to
choose either the CLEO energy shift or the NA7 and Fermilab scale factors, as they can
no longer be fit. We choose to fix them to their fit value  as given in the first data column in 
Table  \ref{T2a}.

Tables   \ref{T2a} and \ref{T2b}, mostly illustrate that, whatever  the fit conditions
examined, the location of the minimum in the fit parameter space remains practically 
unchanged\footnote{With the exception mentioned above.}.
The results obtained by removing $\tau$ data from the fit sample, those
by removing the spacelike data, those corresponding to removing the photon VP function (not shown)
\ldots are consistent with each other.
Let us note  that the fit parameter $c_1$ in the $\Pi_{\pi \pi}^\rho(s)$ function
({\it i.e.} essentially the $\rho$ self--mass function) has been fixed to zero, as it was not found to
sensitively affect the fits in the energy range we are fitting.
The various fit conditions only affect the fit quality which varies from  good to very good, while including
the photon VP.

\begin{table}[!htb]
\begin{center}
\begin{tabular}{|| c  | c  | c | c | c ||}
\hline
\hline
\hhhu Parameter  & Full Fit &  No $\tau$ &  No Spacelike &  No $\rho$ mass shift \\ 
\hline
\hline
\multicolumn{5}{|c|}{Subtraction Polynomial : $\Pi_{\pi\pi}^\rho(s)$} \hhhb \\
\hline
$c_1$ \hhhb & \bf{0} & \bf{0} & \bf{0}& \bf{0} \\
\hline
$c_2$\hhhb & $-0.467 \pm 0.013$ &  $-0.463 \pm 0.014$ &  $-0.472 \pm 0.013$ &  $-0.470 \pm 0.013$\\
\hline
\hline
\multicolumn{5}{|c|}{Subtraction Polynomial : $\epsilon_2(s)$}  \hhhb \\
\hline
$c_1$ \hhhb & $-0.071 \pm 0.003$ &  $-0.071 \pm 0.003$ &  $-0.072 \pm 0.003$ &  $-0.072 \pm 0.003$\\
\hline
$c_2$\hhhb & ~~~$0.045 \pm 0.004 $ & ~~~$0.045 \pm 0.004 $ & ~~~$0.046 \pm 0.004 $ &~~~$0.045 \pm 0.004 $\\
\hline
\hline
\multicolumn{5}{|c|}{Subtraction Polynomial : $\epsilon_1(s)$}  \hhhb \\
\hline
$c_1$ \hhhb & $-0.017 \pm 0.001$ &  $-0.017 \pm 0.001$ &  $-0.017 \pm 0.001$ &  $-0.017 \pm 0.001$\\
\hline
~~$c_2$\hhhb & ~~~$0.020\pm 0.001$ &  ~~~$0.020 \pm 0.001$ & ~~~$0.020 \pm 0.001$ & ~~~$0.020 \pm 0.001$ \\
\hline
\hline
\multicolumn{5}{|c|}{Subtraction  Polynomial $\Pi_{\pi\pi}^{W/\gamma}(s) $ } \hhhb \\
\hline
$c_1$ \hhhb & $0.918 \pm 0.061$ & $0.944 \pm 0.068$ & $0.727 \pm 0.074$ & $0.915 \pm 0.060$\\
\hline
$c_2$\hhhb & $0.433 \pm 0.106$ & $0.361 \pm 0.115$  & $0.831 \pm 0.145$ & $0.440 \pm 0.105$\\
\hline
\hline
\end{tabular}
\end{center}
 
\caption{
\label{T2b}
Parameter values under various strategies (cont'd).
Boldface parameters are not allowed to vary.
Each subtraction polynomial is supposed to be written $c_1 s+ c_2 s^2$. 
}
 
\end{table}

\vspace{0.5cm}
\indent \indent
Figure \ref{NskFig} shows the fit with the $e^+e^-$
data in the timelike region superimposed. The global scale factor effects are accounted for.
In the $\phi$ mass region, the lineshape  is a prediction essentially derived
from the phase and branching fraction of the $\phi \rightarrow \pi^+ \pi^-$ decay  mode
as measured by the SND Collaboration \cite{SNDPhi}. Information on the full (local)
invariant mass spectrum  (when available) would certainly improve this prediction.

Likewise, Figure  \ref{tauFig} shows the fit function and the ALEPH \cite{Aleph} 
and CLEO \cite{Cleo} data superimposed.  One may note that the highest data point from CLEO
data lies at $\simeq 2 \sigma$ of the fitting curve. Actually, the lineshape of the CLEO
data in the neighborhood of the maximum raises some difficulty while fitting, as will be seen
shortly  with the fit residuals.

Leftside Figure  \ref{NA7Fig} shows the spacelike data \cite{NA7,fermilab2}
together with the fit function. One may note a small, but systematic upwards shift
of the fit compared with the NA7 data which certainly explains the jump in
the fit probability while removing this data sample from the fit.

One may conclude from Figures \ref{NskFig}, \ref{tauFig},  \ref{NA7Fig} and the fit probabilities
that the agreement of the data with the model functions is good and that no obvious drawback
shows up.

\vspace{0.5cm}
\indent \indent
In order to refine this statement, we have had  a closer look at the fit residual plots.
The fit residuals for the new Novosibirsk data are shown in Figure \ref{NskRes}. The plotted experimental
errors do not take into account the bin--to--bin correlations generated by the global scale error
common to all data sets.
As the difference between our pion form factor functions $F_\pi^e(s)$ and $F_\pi^\tau(s)$
essentially lies in the $\rho^0-\omg-\phi$ mixing scheme produced by breaking Isospin Symmetry,
this Figure can be compared with Figure 1 in \cite{Davier2007} or Figure 9 in \cite{DavierPrevious3} 
where a systematic $s$--dependent effect was pointing towards a consistency problem between 
$e^+e^-$ and $\tau$ data~; such an effect is no longer observed, pointing to a presently
more adequate manner of performing the breaking of Isospin Symmetry.
One may possibly note that the dispersion
of the residuals is very small everywhere for the 1998 CMD2 data \cite{CMD2-1998-1,CMD2-1998-2},
while it is  larger for the 1995 CMD2 \cite{CMD2-1995corr} and SND \cite{SND-1998} data which 
additionally, are moved in opposite directions in  the $\omg$ mass region. This indicates
that our fit parameter values are dominated by the 1998 CMD2 data.

The residual distributions for $\tau$ data --the upper plots in Figure \ref{TauRes}--
look more interesting. The  arrows indicate the  limit of the fit regions.
The errors plotted are certainly underestimated, as the bin--to--bin correlations are not 
accounted for in the drawings~; moreover, the errors produced by identifying invariant 
mass coordinate and central bin value also are not considered.

One observes now a small but clear  $s$--dependent structure above  the $\rho$ peak location
(more precisely above $s \simeq 850$ MeV) which certainly reflects the influence of the 
unaccounted for higher mass vector mesons. One has also examined the effect of
removing the parameter $\delta m^2$ by  fixing it identically to zero while fitting.
The $\tau$ data residuals are given by the lower plots in Figure  \ref{TauRes}.
One clearly observes the rise of a structure at the $\rho$ peak location in
ALEPH data which is therefore a clear signal of a $\rho^0-\rho^\pm$ mass difference.
The mass shift observed is~:   
\be
\displaystyle
m_{\rho^0}^2-m_{\rho^\pm}^2 = -\delta m^2=(0.27 \pm 0.10)~ 10^{-2} ~~~{\rm GeV}^2 
\Longleftrightarrow m_{\rho^0}-m_{\rho^\pm} = 1.73 \pm 0.60 ~~~{\rm MeV}
\label{massdiff}
\ee
in good agreement with several other reported values \cite{RPP2006}. This question
will be discussed with more details in the Subsection devoted the $\rho$ meson parameters.

In contrast,  
 the shape of the CLEO residual distribution rather indicates
 a systematic effect in CLEO data  located only in the $\rho$ peak region. 
 The global CLEO energy shift of 0.9 MeV serves to recover from the disagreement 
with ALEPH data. However, these plots
clearly show that the problem of systematics is not global but local and that there is no
evidence for a significant global invariant mass shift within the CLEO data in our fitted range. As
this residual behaviour is also observed in the standalone fit performed in \cite{Cleo}
(see Figure 10 therein), it should not follow from the constraints specific to our
model.

Actually, there is a correlation between $\delta m^2$ and the CLEO energy shift $\varepsilon$
which vanishes when performing a simultaneous fit of ALEPH and CLEO data. In order to check 
this statement, we have removed ALEPH data from the fit data  set. Then, fixing  
$\delta m^2 \equiv 0$, we get $\varepsilon=1.57 \pm 0.40$  MeV and, conversely, 
fixing $\varepsilon\equiv 0$ results in $~-\delta m^2= (0.36 \pm 0.08)~ 10^{-2}$ GeV$^2$,
with no change in the $\chi^2$ value and always the same residual shape as shown in Figure 
\ref{TauRes}. Therefore, the value for $\delta m^2$ is set
by the ALEPH data and should be confirmed by forthcoming data sets. Correspondingly,
it is the use of ALEPH data which indicates that the CLEO energy shift could well be
consistent with zero.

All the reported exercises also show, as clear from Tables    \ref{T2a} and \ref{T2b},
that the fit parameter values are stable (with the exception already mentioned). 
This means that our model is overconstrained and that, practically, only the fit quality 
({\it i.e.} the height of the minimum) is affected by the various conditions we have imagined. 
We were also aware of possible correlations between the subtraction polynomials. 
Looking at the fit error covariance matrix, we did not observe strong correlations
between parameters belonging to different polynomials, which seems to indicate that they are
indeed independent.

Therefore, one may consider that the description of all form factor data supports our mixing model, 
as reflected by the statistical fit qualities reported in Table \ref{T1} under various conditions.

Finally, rightside Figure  \ref{NA7Fig} shows the {\em predicted} phase of the $I=1$ part of the
pion form factor together with the measured $P_{11}$ ($\pi \pi$) phase data
from \cite{Ochs,Protopescu}. Clearly, the description is good, keeping in mind that
some contributions have not been included, especially the exchange of a spacelike $\rho$.
Therefore, this Figure indicates that the neglected diagrams should contribute not  more than
 a few degrees to the phase.
\vspace{0.3cm}
\indent \indent

Therefore,  the description of all form factor data can be considered
as satisfactory and  provides a solid ground to our main assumptions~: 

{\bf j/} The bulk of Isospin Symmetry effects which create the difference between
$e^+e^-$ and $\tau$  form factor lineshapes is a  $\rho^0-\omg-\phi$ mixing
scheme of dynamical ({\it i.e.} $s$--dependent) structure. 

{\bf jj/} An appropriate subset of meson partial width decays and the  $e^+e^-$ 
form factor data  mostly suffice to set up a $\rho^0-\omg-\phi$ mixing
scheme able to derive the $\tau$ spectrum with good precision.

{\bf jjj/} The effects of   higher mass vector mesons in the mass region below
1.0 GeV, even if somewhat visible on the upper wing of the $\rho$ peak,  are 
negligible.

{\bf jv/} The (observed) $\rho^0-\rho^\pm$ mass shift is very small
and of negligible effect. New $\tau$ data may confirm its relevance,
as this follows only from ALEPH  data. 
\vspace{0.5cm}

\indent \indent
In view of the residuals shown in Figure \ref{NskRes} and, even more clearly
in Figure \ref{TauRes}, the statement in {\bf jjj/}  can be precisely commented upon.
Because of their statistical accuracy, the Novosibirsk data do not exhibit any undoubtful
effect of higher mass resonances within the whole fitted range~; one may possibly
guess a dip (of small significance) in the region between 850 and 900 MeV. In the CLEO
data (right plots in Figure \ref{TauRes}), the residual structure is unclear
in the region between $\simeq 700$ and $\simeq 950$  MeV, while in the region above $\simeq 1$ GeV,
there is undoubtfully a missing structure which can reasonably be attributed
to the $\rho(1450)$ meson low mass tail. 
If one relies on the ALEPH data (left plots in Figure \ref{TauRes}),
one is instead tempted to state that higher mass vector meson effects have influence down to
 $\simeq 850$  MeV, ({\it i.e.} deep inside the high mass wing of the $\rho(770)$
meson. Additional $\tau$ data sets and large  statistics $e^+ e^-$ data sets
 (collected through the radiative return method)  are expected in a near future 
from B factories and from DAPHNE, hopefully with controlled systematics.
One may imagine that they should allow to clarify the situation in the region from
the $\omg$  to $\varphi$ peak.

\subsection{Light Meson Decays}
\indent \indent As a preliminary remark, when fitting partial widths 
(actually coupling constants), the recommended data
used are the  partial widths 
taken from the RPP \cite{RPP2006}, when available. If not, they are derived
from the branching ratios and the full widths. Sometimes, this procedure reveals
a surprising information. For instance, for $\eta \ra \gamma \gamma$, the ratio
of the ``fit" partial width error to the corresponding central value is $0.026/0.51=0.05$,
while the corresponding information derived from the quoted ``fit" branching fraction is 
$0.26/39.98=0.007$, which might look somewhat optimistic. 

The numerical estimates of branching fractions have been calculated using
the information returned by the {\sc minos} program  and take into account
the parameter error covariance matrix  in the standard way
(as recalled in Section 7.3 of \cite{su2}, for instance). This is mandatory as some error correlation
coefficients are very large, namely those among the two fit parameters hidden inside $\epsilon_1(s)$,
or inside $\epsilon_2(s)$  are about 95\%. Most  other
error correlation matrix elements  are below the 10 \% level. We therefore consider that our
error estimates are accurate.

On the other hand, all partial width results
we compute have been derived using the accepted values for all vector 
and pseudoscalar meson masses \cite{RPP2006}. In order to produce the branching
ratios as given in Table \ref{T3}, we have also divided these partial widths 
by the accepted total widths reported in the latest issue of the Review of Particle
Properties \cite{RPP2006}. The errors on masses and widths have been taken into account in the computer
code used in order to derive the  reconstructed branching ratios.

\subsubsection{Radiative Decays Of Light Mesons}
\indent \indent The fit values for the branching fractions of light mesons radiative decays
are displayed on top of Table \ref{T3}.  Most decay modes involving vector mesons are in 
nice correspondence with their recommended values \cite{RPP2006}. 

The value returned for the $\omg \ra \pi^0 \gamma $ branching ratio is about 3 $\sigma$
from the presently recommended value \cite{RPP2006}, but is in good agreement with the former
recommended value ($(8.5 \pm 0.5)~10^{-2}$ as well as the value found
in the fit of the $e^+ e^- \ra \pi^0 \gamma$ 
reported in \cite{pi0gam}  ($(8.39 \pm 0.25)~10^{-2}$). 
This indicates that the recommended central value for 
this decay mode can be questioned.
 
 On the other hand, as already commented upon, the branching fraction
 found for  $\omg \ra \eta \gamma $, is in much better agreement with the average value 
 proposed by the PDG in \cite{RPP2006} ($(6.3 \pm 1.3)~10^{-4}$) than their so--called fit 
 value reported in  Table \ref{T3}~; this result is also in perfect agreement  with the
 Crystal Barrel \cite{AbeleOmg} measurement ($(6.6 \pm 1.7)~10^{-4}$) as well as the
 measurement obtained  in a fit to the $e^+ e^- \ra \eta \gamma$ cross section 
 \cite{pi0gam} ($(6.56 \pm 2.5)~10^{-4}$).  We also consider confidently
 our result for this decay mode.
 
 The ratio~:
  \be 
 \displaystyle \frac{\Gamma(\omg \ra \eta \gamma) }{\Gamma(\omg \ra \pi^0 \gamma) }
 = (0.802 \pm 0.011) ~10^{-2}
   \label{omgrap1}
 \ee
depends only weakly on  the mass and width definitions of the $\omg$ meson and
is in agreement with all reported direct measurements in the RPP \cite{RPP2006}.
This also gives support to both fit results.

 The contribution of the $\eta^\prime \ra \rho^0 \gamma$  mode to the $\chi^2$ is 
 2.23, while all others are smaller or of  order 1. This may indicate 
 that   the box anomaly \cite{box,Abele} shows up and might have been accounted for.

 The only difficult point of the model is the $\simeq 1.9 \sigma$ departure of the partial
 width for $\eta \ra \gamma \gamma$  from the expected value commented upon at the beginning
 of this Section. Whether this could be due to our assuming that the pion decay constant is not
 affected by Isospin Symmetry breaking effects is an open possibility. 
 Instead, the partial width for $\eta^\prime \ra \gamma \gamma$ fits
 nicely its expected value, possibly because of its larger experimental uncertainty\footnote{
 In order to test this assumption, we have left $f_{\pi^0}$ free in our fits 
 and, for $f_{\pi^0}= 87.9 \pm 2.4$ MeV, we have reached a probability slightly above
 80 \%,  with $\Gamma(\eta \ra \gamma \gamma)$ at 0.55 $\sigma$ from its recommended value
  \cite{RPP2006} and $\Gamma(\eta^\prime \ra \gamma \gamma)$ at only 0.37 $\sigma$.
  This has to be compared with the reported value extracted from $\Gamma(\pi^0 \ra \gamma \gamma)$
  which provides $f_{\pi^0}=91.92 \pm 3.54$ MeV.}. 
 
 Finally, we should note that our model gives a precise indirect measurement
 of $f_K/f_\pi$~:
 \be 
 \displaystyle \left[ \frac{f_K}{f_\pi}\right ]^2 = 
 1.503 \pm 0.010_{stat} \pm 0.002_{model} ~~~\Longrightarrow ~~~\frac{f_K}{f_\pi}=1.226 \pm 0.004_{stat} \pm
 0.001_{model}
  \label{fkfpi}
 \ee
\noindent where the second quoted uncertainty reflects details of the model
together with the effects of including  the spacelike data in the fit. This is in balance with the 
 corresponding quantity which can be derived from the reported world average data \cite{RPP2006} as
 $f_K/f_\pi= 1.223 \pm 0.010$, assuming that the errors on $f_\pi$ and $f_K$ are uncorrelated.
 
\subsubsection{Leptonic Decays Of Light Vector Mesons}
\indent \indent 
Table \ref{T3} indicates that our model nicely accomodates the
$\omg \ra e^+ e^-$ and $\phi \ra e^+ e^-$ partial widths giving 
values which coincide with their recommended values \cite{RPP2006}.

Our result for $\rho^0 \ra e^+ e^-$ is derived from the same data which underly the
other proposed values \cite{RPP2006} and has been obtained with a careful account of all statistical 
and systematic reported errors. Therefore, this value can be confidently considered~; 
one should note that it exhibits a $\simeq 10 \sigma$ distance 
to the presently accepted branching fraction \cite{RPP2006}. 
A more straightforward information
coming out from our fits is the corresponding partial width~:

\be
\Gamma(\rho \ra e^+ e^-) = (8.34 \pm 0.10 \pm 0.31) 10^{-3}~~{\rm MeV}
\label{rhowidth}
\ee
 \noindent where the first error merges all statistical and systematic uncertainties
 commented upon in the body of the text~; the second error takes into account the real
 uncertainty affecting the $\rho$ mass used in order to derive the partial width
 from the coupling. It has conservatively been fixed to 10 MeV for reasons
 which will become clearer shortly. The corresponding partial width as given in
 \cite{RPP2006} is $ (7.02 \pm 0.11 ) 10^{-3}~$ MeV.

\subsubsection{The $\omg / \phi \ra \pi^+ \pi^-$ Decays}
\indent \indent 
The value found for the $\phi \ra \pi^+ \pi^-$ partial width compares well with
its measured value \cite{SNDPhi}. Actually, one may suspect that this
datum  prominently influences some of our free parameters, 
certainly those in the expression for $\epsilon_1(s)$. The phase of the corresponding coupling 
constant being close enough ($1.4 \sigma$) to expectation  \cite{SNDPhi} might indicate that
the data (modulus and phase) for this mode  carry small systematic uncertainties.
 
The branching fraction we get for the $\omg  \ra \pi^+ \pi^-$ mode is more appealing.
It is derived from all data involved in this measurement with a precise account of
all systematic uncertainties. Additionally, the quality of the measurement we propose
probably does not suffer from significant model uncertainties, as the $\rho-\omg$ interference region
is quite well described (see the insets in Figure \ref{NskFig}). Therefore, our 
conclusion for this decay mode is either of~: 
\be
{\rm Br}(\omg \ra \pi^+ \pi^-)= (1.13 \pm 0.08) \% ~~~~~~~~, ~~~~~
 \Gamma(\omg \ra \pi^+ \pi^-) = (9.59 \pm 0.80)~ 10^{-2}~~{\rm MeV}
\label{omgwidth}
\ee
using the recommended value for  width and the mass of the $\omega$ meson \cite{RPP2006}.

This new datum may influence the global fit of all the $\omega$ decay modes in isolation. 
This is of concern for our purpose, as one has noticed that the disagreements observed 
between the PDG recommended values \cite{RPP2006} and  our results for 
the $VP\gamma$ modes refer mostly to the $\omg \ra  (\eta/\pi^0) \gamma$ branching ratios.
Along this line, our fit solution provides~:
\be
\displaystyle \frac{\Gamma(\omg \ra \pi^+ \pi^-)} {\Gamma(\omg \ra \pi^0 \gamma)}=
0.14 \pm  0.01
\label{omgrap}
\ee
to be compared with  the single existing measurement by KLOE \cite{KLOEmrho} which provided $0.20 \pm 0.04$
and a $1.5 \sigma$ distance.
Therefore, our surprising estimate of the two pion mode together with radiative decays is in
accord with experimental expections.

Finally, the unfitted Orsay phase for the coupling $\omg  \ra \pi^+ \pi^-$ is found 
close to its expected value from a 
standalone fit to the so--called old timelike data \cite{ffVeryOld}, while our fit for the phase of the
$\phi \ra \pi^+ \pi^-$ coupling is in good agreement with its measured value \cite{SNDPhi}.

\subsubsection{The $\phi \ra K \overline{K}$ Decays}
\indent \indent 
As explained in the body of the text, we have been led to leave 
both $\phi \ra K \overline{K}$ decay widths outside our fit 
procedure, as there is some uncertainty with possible factors,
like Coulomb corrections, which may affect the usual
coupling constant contributions to both partial
widths\cite{BGPter,Fischbach,Oliver}.

Therefore, the values reported in Table \ref{T3} are predictions
only influenced by the other decay modes and without any additional 
correction factor to each of the $\phi \ra K \overline{K}$ branching ratios. 
 The numerical values found for these branching ratios
clearly illustrate that our model is overconstrained and provides precise
values for the coupling constants of both $\phi \ra K \overline{K}$ modes.

The ratio of the prediction to the recommended central values is 1.022 for the 
charged mode ({\it i.e.} a $1.8  ~\sigma$ distance) and 0.97 for the neutral decay mode
({\it i.e.} a $2.0 ~\sigma$ distance). Taking into account the model uncertainty 
reported in Table \ref{T3}, the agreement could be considered as satisfactory.

Now, if correction factors have to be applied, they are expected to improve the prediction
for the rates. Therefore, they should be of the order 0.976  and  1.031 
for, respectively, the charged and neutral decay widths.
This clearly invalidates the traditional 1.042 correction factor proposed
in order to account for Coulomb interaction among the charged kaons\footnote{
See \cite{BGPter} for a detailed account of the usual way to deal with Coulomb corrections
and Isospin Symmetry breaking effects in  $\phi$ decays. }.
 
 Correcting both modes as argued in \cite{Fischbach}, even if able to provide a
 good account for the ratio $\Gamma(\phi \ra K^+ K^-)/\Gamma(\phi \ra K^0 \overline{K}^0)$,
 does not allow a good account of both modes separately, as the corrections proposed
 turn out to increase the expected rate for both modes. 
 
 Within the framework of our model, if correction factors have to be applied,
 they should not increase the charged decay mode by more than $\simeq 1\%$. There is more freedom
 with the neutral decay mode. Therefore, in order to fix one's ideas, one has let
 a correction factor for only the neutral decay mode to vary. In this 
 case, of course, the correction factor ($= 1.047 \pm 0.024$)                                                                        
 is found such that the neutral mode exactly coincides
 with its measured value, which could be expected beforehand. However,
 more interesting  is that the $\chi^2$ contribution of the charged decay mode
 (which does not explicitely depend on this factor) is only 0.3 (a $0.5 ~\sigma$ effect). 
 This indeed confirms that
 only the predicted neutral decay width might have to be corrected significantly.
 Taking into account that systematic effects are harder to estimate for the
 charged mode than for the neutral one (see footnote 16 in \cite{su2}), this may look
 a physical effect. Whether the ``mixed isoscalar and isovector source" scheme
 of \cite{Oliver} can account for such an effect would be interesting to explore.
 
  As a summary, our analysis tends to disfavor a significant correction factor
 to the  $\phi \ra K^+ K^-$  decay width (above the 1.01 level). It would rather favor  
 a significant correction factor for only the neutral mode  $\phi \ra K^0 \overline{K}^0$
 (which could be as large as 1.047 for the rate). If the traditional scheme of Coulomb corrections
 should really apply, both measured widths for $\phi \ra K \overline{K}$ are hard to 
 understand, as already stated in \cite{BGPter,su2}.

 \subsubsection{What Are The $\rho$ Parameters?}
\indent \indent 
For objects as broad as the  $\rho$ (or $K^*$) meson, the definition for mass and width
(from experimentally accessible information) is not a trivial matter \cite{Tony}, 
 and no physically motivated uniqueness statement can be formulated. 
 Having defined in our model the $\rho^0$ and 
  $\rho^\pm$ propagators as analytic functions (or rather meromorphic 
  functions on a 2--sheeted Riemann surface with branch point at threshold), one has at
  disposal the poles of the propagators. This has been shown to provide 
  the most stable definition of the mass
  and width  \cite{Tony}. If one assumes that Analyticity of S--matrix elements is 
  a basic principle, this is also the most model independent definition. Indeed, 
  whatever are the working assumptions, the pole basically tells where the peak is 
  really located  and how wide is the invariant mass distribution around the peak
  (typically, close to the full width at half maximum). Obviously, any given model cannot
  successfully describe the relevant data without having the pole located at the place
  requested by the data. The specific character of a given model is basically concentrated
  in the regular part of the Laurent expansion of the amplitude. To be complete, departures
  from this statement may exist but is (parameter, model)  freedom is  essentially governed by the
   magnitude of the experimental error bars. 
  
  However, based on the expectation that the mass difference between  charged and neutral 
  $\rho$ mesons is only of the order a few MeV's at most ({\it i.e.} 
  $\delta m_\rho \simeq {\rm a~few~}10^{-3} m_\rho$),
  one may guess that this difference could be less sensitive to mass definitions.
  
  
  Our final results for the complex $s$ locations of the $\rho$ meson poles can be derived
  from our fit parameter values. Sampling them by taking into account the parameter error
  covariance matrix, one gets~:
  \be
  \left \{
  \begin{array}{lll}
  s_{\rho^0}   =  (0.5782 -i~ 0.1099) \pm (0.9 +i~0.5)~10^{-3}~~~~~~({\rm GeV}^2)\\[0.5cm]
  s_{\rho^\pm} =  (0.5760 -i~  0.1095) \pm (1.0 +i~0.5)~10^{-3}~~~~~~({\rm GeV}^2)\\[0.5cm]
  s_{\rho^0} -  s_{\rho^\pm} = (2.26 -i~0.38) ~10^{-3} \pm (0.83 + i~0.14) ~10^{-3}~~~~~~({\rm GeV}^2)\\
  \end{array}
  \right .
  \label{massrho1}
  \ee
  with uncertainties folding all reported statistical and systematic errors.
  
  In order to compare with related information available in the literature, one has to relate 
  the $\rho$ pole locations with the usual $M_\rho$ and $\Gamma_\rho$.
  
  Defining as \cite{Bernicha} $s_R=M_R^2-iM_R \Gamma_R$, one gets~:
  \be
  \left \{
  \begin{array}{lll}
  M_{\rho^0} 	= 760.4 \pm 0.6 ~~~{\rm MeV}~~~,& \Gamma_{\rho^0}= 144.6 \pm 0.6 ~~~{\rm MeV}\\[0.5cm]
  M_{\rho^\pm} 	= 758.9 \pm 0.6 ~~~{\rm MeV}~~~,& \Gamma_{\rho^\pm} = 144.3 \pm 0.5 ~~~{\rm MeV}\\[0.5cm]
  M_{\rho^0} -M_{\rho^\pm}=1.51 \pm 0.53~~~{\rm MeV} ~~~,& \Gamma_{\rho^0}-\Gamma_{\rho^\pm}=
 0.22 \pm 0.08 ~~~{\rm MeV}
 \end{array}
  \right .
  \label{massrho2}
  \ee
 which for $\rho^0$ are slightly larger than those found by \cite{Bernicha} using only the so--called 
 old timelike data. One may note that the mass difference is affected by a smaller uncertainty than
 masses separately~; this effect is even much more pronounced for the width difference. 
 This error shrinking reflects the correlations contained in the parameter error
 covariance matrix of our fit.
 
 One may also choose \cite{Melikhov2} $s_R=(M_R-i\Gamma_R/2)^2$ and obtain slightly different 
 values (not provided). Defining the mass by the location of the maximum of the distribution and the width
 by the full width at half maximum cannot be derived easily from Eqs. (\ref{massrho1})~; they are~:
 
   \be
  \left \{
  \begin{array}{lll}
  M_{\rho^0} 	= 762.1 \pm 0.6 ~~~{\rm MeV}~~~,& \Gamma_{\rho^0}= 144.5 \pm 0.6 ~~~{\rm MeV}\\[0.5cm]
  M_{\rho^\pm} 	= 760.8 \pm 0.6 ~~~{\rm MeV}~~~,& \Gamma_{\rho^\pm} = 144.5 \pm 0.5 ~~~{\rm MeV}\\[0.5cm]
  M_{\rho^0} -M_{\rho^\pm}=1.22 \pm 0.53~~~{\rm MeV}~~~,& \Gamma_{\rho^0}-\Gamma_{\rho^\pm}=0.02 \pm 0.08
 \end{array}
  \right .
  \label{massrho2b}
  \ee
One may consider these values as they are in consistency with the way the mass and width for
objects like the $\omg$ and $\phi$ mesons are usually defined \cite{RPP2006}. The difference 
between these results and those in Eqs. (\ref{massrho2}) could be attributed to the influence
of the regular part of the invariant mass distribution which distorts a little bit
the distribution lineshape. Finally, in view of Eqs. (\ref{massrho2}) and (\ref{massrho2b}),
one cannot be really conclusive about the sign of the width difference central value, as it sensitively
varies with parameter definitions (see also footnote \ref{rpr0mass}).

\vspace{0.5cm}

\indent \indent
Our mass difference values can be compared with results from other reactions available in the literature \cite{RPP2006}.
Limiting oneself to the most recent estimates (all with large statistics), one has~:
  \be
  \left \{
  \begin{array}{lll}
m_{\rho^0}-m_{\rho^\pm}&=0.4 \pm 0.7 \pm 0.6 ~~{\rm MeV} & {\rm KLOE~~2003~\cite{KLOEmrho},~~1980~ Kevents}
\\[0.5cm]
m_{\rho^0}-m_{\rho^\pm}&=1.3 \pm 1.1 \pm 2.0 ~~{\rm MeV} & {\rm SND~~2002 ~\cite{SNDmrho}, ~~500~ Kevents}
\\[0.5cm]
m_{\rho^0}-m_{\rho^\pm}&=1.6 \pm 0.6 \pm 1.7 ~~{\rm MeV} & {\rm Crystal  ~Barrel  ~~1999 ~\cite{CBARmrho},~~ 600~ Kevents}
 \end{array}
  \right .
  \label{masscompar}
  \ee
which compare satisfactorily with our fit results in either  of Eqs. (\ref{massrho2}) or Eqs. 
(\ref{massrho2b}). These experiments  analyze the $\pi^+ \pi^- \pi^0$ final state produced in $e^+e^-$ annihilations
at the $\phi$ mass \cite{KLOEmrho,SNDmrho} or in $p\overline{p}$ annihilations at rest \cite{CBARmrho} using standard
varying width Breit--Wigner shapes for both charged and neutral $\rho$ meson distributions.

The ALEPH Collaboration has also  performed a global fit \cite{Aleph} of the $e^+e^-$ Novosibirsk data 
together with  the ALEPH and CLEO $\tau$ data sets (as we did) and gets~:
  \be
  \begin{array}{lll}
m_{\rho^0}-m_{\rho^\pm}&=-2.4  \pm 0.8 ~~{\rm MeV} & {\rm ALEPH~~2005~\cite{Aleph} }
 \end{array}
  \label{massAleph}
  \ee
significantly different from our findings. However, their fit residuals show a $s$--dependence
below $s \simeq 1$ GeV$^2$ which is absent from our data\footnote{Ref. \cite{Aleph} does not provide,
strictly speaking, the fit residuals, but this can be guessed from Figure 67 therein.}.

The fit results from other experiments shown in Eqs. (\ref{masscompar}) are 
consistent with either sign for $m_{\rho^0}-m_{\rho^\pm}$~; our own results
favor $m_{\rho^0}-m_{\rho^\pm} >0$ (with resp.
$2.9~\sigma$ and $2.3~\sigma$ significance), 
while naively one may expect
the opposite. However, Bijnens and Gosdzinsky \cite{Bijnens}, analyzing within the ChPT framework
 all contributions to this mass difference, concluded that~:
  \be
  -0.4 {\rm ~MeV~} < m_{\rho^0}-m_{\rho^\pm} < +0.7{\rm ~MeV~} 
  \label{bijnens}
  \ee
All measurements given in Eqs. (\ref{masscompar}) are consistent with this mass interval. 
The ALEPH mass difference is at a $2.5 ~\sigma$ distance from the lower bound of 
Eq. (\ref{bijnens}). Concerning
our results, our largest estimate of the mass difference ($1.51 \pm 0.53$) is only at 
$1.5 ~\sigma$ from the upper bound, while our smallest estimate ($1.22 \pm 0.53$) is at
$1.0 ~\sigma$  from this upper bound\footnote{Comparing with the upper limit is the most favorable 
case for our results, while comparing with the lower limit is the most favorable case for ALEPH.}. 
Therefore, awaiting new measurements which may confirm
the ALEPH spectrum at the peak location, one may conclude that our fit results 
are in good agreement with ChPT expectations. From a statistical point of view, 
this agreement has even been marginally
improved compared with the ALEPH best fit result.
\vspace{0.5cm}

\indent \indent

If  one decides to
parametrize the distributions with varying width Breit--Wigner shapes, one would recover
more traditional mass values as tabulated in \cite{RPP2006}.
However, their model dependence (not only their definition dependence) should be stressed.
Because of having decay and pion form factor data intricated, it is not easy to 
perform this check within the present context. However, a good approximation of 
using varying width Breit--Wigner expressions is to define the $\rho$ masses by~:
  \be
  {\rm  Re}(D^{-1}_{\rho}(M_{\rho}^2)) =0
  \label{rhodef}
  \ee
which, by the way, is fulfilled by the standard Gounaris--Sakurai propagator 
\cite{Gounaris,Aleph}. In this case we get~:
 \be
  \begin{array}{lll}
 M_{\rho^0} 	= 774.8 \pm 0.6 ~~{\rm MeV}~~,&  M_{\rho^\pm} = 773.3 \pm 0.7~~{\rm MeV}~~,
 & M_{\rho^0} - M_{\rho^\pm} 	= 1.48 \pm 0.50 ~~{\rm MeV}
 \end{array}
  \label{rhodef2}
  \ee
which can be compared with  standard values for masses and provide a mass diffference 
in good agreement
with our results above. This definition of  $\rho$ meson parameters is very close
to the corresponding ones following from using Gounaris--Sakurai expressions. Then,
comparing Eqs. (\ref{rhodef2}) with the corresponding ALEPH results \cite{Aleph}
 is a way to exhibit the effect of the additional isospin breaking mechanism we propose.

 Going a step further along the same line, one may also choose to define masses by 
 the HK mass values as they come  from the Lagrangian\footnote{
 This has a clear physical meaning~: It is the mass of the $\rho$ meson while working
 at tree level (when possible, as maybe farther  inside the spacelike region than
 we have gone). However, one should keep in mind that, numerically, they are derived
 from fits using expressions containing the unavoidable self--mass corrections.}
  and our fitted  parameters. Then one gets~:
  \be
  \begin{array}{lll}
  M_{\rho^0} = 782.1 \pm 2.1 ~~{\rm MeV}~~,~~M_{\rho^\pm} = 780.4 \pm 2.2~~{\rm MeV}~~,
  ~~M_{\rho^0}-M_{\rho^\pm} =1.74 \pm 0.60~~{\rm MeV}
 \end{array}
  \label{massrho3}
  \ee
  
  This result is also interesting. Indeed, as stated above (see Footnote \ref{rpr0mass}),
  a reasonable breaking of Isospin Symmetry at Lagrangian level, while producing
  a (HK)  mass difference between the $\rho^0$ and $\rho^\pm$ mesons returns
  $m_{\rho^0}=m_{\omg}$. As one may think that the HK $\omg$ mass is close
  to the tabulated value \cite{RPP2006} ($m_{\omg}=782.65 \pm 0.12$ MeV), it may have
  a meaning to find that the HK mass for the $\rho^0$  is consistent with the accepted $\omg$
  mass. For this purpose, it should be noted that the $\omg$ mass value used in our fits was
  fixed at this accepted value, and then cannot directly influence the HK value for $m_{\rho^0}$.
 
  Defining the width using Eq. (\ref{DD12}) and using\footnote{We have averaged the pion masses
  used for the charged $\rho$ width computation.} the HK $\rho$ masses given in Eqs. (\ref{rhodef2})
 one derives the following information for the $\rho$ widths\footnote{
 We tried to decouple $g_{\rho^0 \pi \pi}$ from $g_{\rho^\pm \pi \pi}$ by defining
 $g_{\rho^\pm \pi \pi} = g_{\rho^0 \pi \pi} +\delta g$ with $\delta g$ submitted to fit.
 The minimization procedure returned $\delta g =-0.018 \pm 0.014$ and an improvement of the 
 $\chi^2$ of the order 0.5, quite unsignificant.}~:
  \be
  \begin{array}{lll}
   \Gamma_{\rho^0} = 174.3 \pm 2.0 ~~~{\rm MeV}~~,&  \Gamma_{\rho^\pm} = 175.0 \pm 2.0~~{\rm MeV}~~,
 & \Gamma_{\rho^0} - \Gamma_{\rho^\pm} 	= -0.7 \pm 0.2 ~~{\rm MeV}.
 \end{array}
  \label{massrho4}
  \ee
which are substantially larger than all other definitions
and with a difference going in the opposite way compared with above~; however  this follows from
the most usually employed formula for the two--body decay widths of the $\omg$ and $\phi$ mesons.
One may argue that the larger mass and width exhibited by Eqs. (\ref{massrho3}) and (\ref{massrho4})
compared with usual is related with our having the HLS parameter $a$  significantly different 
from 2 (a $3~\sigma$ effect).

In summary, as far as $\rho$ meson parameters are concerned, we consider that the most
relevant information are the $s$ location of the poles as given by
Eqs. (\ref{massrho1})~; definitions of the mass and width using $s_R=M_R^2-iM_R \Gamma_R$,
or $s_R=(M_R-i\Gamma_R/2)^2$, or something else, can be derived algebraically and some have been given.

\subsection{A Few More Comments On The Model}
\indent \indent
In order to justify the change from ideal to physical fields, one should check that
the functions in the non--diagonal elements  of the $R$ matrix in Eq. (\ref{Mass7}) are
small compared to 1 in the relevant invariant mass range. For this purpose, we have had 
a closer look at the functions~:
\be
\left \{
\begin{array}{lll}
F_{\rho\omg}(s) = & \displaystyle \frac{\epsilon_1(s)}{\Pi_{\pi \pi}(s)-\epsilon_2(s)}\\[0.5cm]
F_{\rho\phi}(s) = & \displaystyle   
\frac{\mu \epsilon_1 (s)}{(1-z_V) m^2 + \Pi_{\pi \pi}(s) -\mu^2 \epsilon_2(s)}\\[0.5cm]
F_{\omg\phi}(s) = & \displaystyle\frac{\mu \epsilon_2(s) }{ (1-z_V) m^2 + (1-\mu^2) \epsilon_2(s)}
\end{array}
\right .
\label{mixingAngles}
\ee
computed with the fit parameter values. These are the entries of the $R(s)$ matrix which defines
our transformation from ideal (bare) fields to physical fields.

Figure \ref{MixAng} shows the real and imaginary parts of these functions.
They are all small compared to 1 all along the physical region~:
$|F_{\rho\omg}(s)| \simeq {\cal O}(10^{-2})$, $|F_{\rho\phi}(s)| \simeq {\cal O}(10^{-2})$
and $|F_{\omg\phi}(s)| \simeq {\cal O}(10^{-1})$. As expected $F_{\omg\phi}(s)$, which
does not vanish in the Isospin Symmetry limit, is larger and one observes an order of magnitude
difference. 

$F_{\rho\omg}(s)$ represents the traditional  
$\rho-\omg$ mixing and  its behaviour translates in our modelling the known large (Orsay) phase
by the quasi--vanishing of its real part around $m_\omg^2$. It exhibits around 
$\sqrt{s} \simeq  0.3 $ GeV, the two--pion threshold,  an unexpected behaviour which is
actually too small to influence numerically the pion form factor.
Whether this local effect should be considered seriously is unclear, taking into account 
the approximations done in order to work out the model\footnote{This goes along
with the remark that one would have preferred a solution for  $F_{\rho\omg}(s)$ 
which vanishes at $s=0$ as $F_{\rho\phi}(s)$  and $F_{\omg\phi}(s)$ do, even for non identically 
vanishing of $\epsilon_1$.
This could also be a consequence of working at first order in $\epsilon_1$ and $\epsilon_2$.
However, the neglecting of the anomalous loop effects may  play some role near $s=0$.
}.
 The important point here is that, even
if narrow, the amplitude does not exceed a 10 \% level there and a few percents
all along the physical region (namely the $\rho/\omg$ and $\phi$  peak locations). 
 
$F_{\rho\phi}(s)$  is of the same order of magnitude than $F_{\rho\omg}(s)$ 
but much smoother all along the physical region~; 
its real and imaginary parts around the $\phi$ meson mass are comparable ($\simeq$ 1 \%).

$F_{\omg\phi}(s)$ is more interesting as it represents what is traditionally
attributed to
 a constant mixing angle of a few degrees \cite{rad}. It is indeed what is 
exhibited, as this function is real at  $m_\omg^2$ and close to real at $m_\phi^2$. 
However, the numerical values of the mixing angle vary significantly~: at the $\omg$
mass the angle is 0.45$^\circ$, while it is 4.64$^\circ$ at the $\phi$ mass.
This tends to indicate that the notion of mixing angle has somewhat to be readdressed.

As explained above, the (not too small) magnitude of 
$F_{\omg\phi}(s)$ could have been inferred from the HLS model, as the transitions
between $\omg_I$ and $\phi_I$ follow from (kaon, $K^*K$ and $K^* \overline{K}^*$) loop effects 
and not from supplementing them with Isospin Symmetry breaking effects.

\vspace{1.cm}

\indent \indent
Figure \ref{CVCFig} emphasizes the important features of our model for the pion form factor.
The upper plot shows the function~:
\be
H(s)= \displaystyle
\frac{|F_\pi^e(s)|^2-|F_\pi^\tau(s)|^2}{|F_\pi^\tau(s)|^2}
\label{CVC1}  
\ee
which summarizes the breaking of Isospin Symmetry all along the physical region.
The strong effects at the $\omg$ and $\phi$ mass locations could have been expected.
However, one clearly sees  a non--zero ``background" contribution extending 
down to threshold (and even below) and beyond the $\phi$ mass. This simply illustrates that our
$\rho-\omg-\phi$ mixing scheme is really invariant mass dependent. This is why it can
suppress the unwanted effects exhibited by residuals in more standard approaches
(see for instance \cite{Davier2007}, \cite{DavierPrevious2} or \cite{DavierPrevious3}).

The lower plot shows instead~:
\be
H_1(s)= \displaystyle
\frac{|F_{\pi,I=1}^e(s)|^2-|F_\pi^\tau(s)|^2}{|F_\pi^\tau(s)|^2}
\label{CVC2}  
\ee
where $F_{\pi,I=1}^e(s)$ is identified with the $\rho^0$ part of the pion form factor,
as traditionally done. It clearly exhibits  that the $\rho^0$ and
 $\rho^\pm$ mesons are different kinds of objects in our modelling. 
 We indeed  observe\footnote{The local minimum
 just above threshold, which reflects the structure of $F_{\rho\omg}(s)$
 we discussed, may be an artifact of the model. Its numerical
 value at this location makes it however totally invisible.}  an effect of several percents
 and functionally  $s$--dependent.

 Stated otherwise, the $\rho^\pm$ is indeed a
 pure isospin 1 meson, while the $\rho^0$ meson is actually a mixture of 
 isospin 1 ($\rho^0_I$) and isospin 0 states ($\omg_I$ and $\phi_I$). A real
 extraction of the isospin 1 component of $F_\pi^e(s)$ should isolate the isospin
1 part of the $\rho$, $\omg$, and $\phi$ amplitudes, which is exactly what our
model does, by construction. This allows us to agree with the analysis by K. 
Maltman \cite{Maltman2,Maltman3} concluding that the $\rho$ part of the pion
form factor in $e^+e^-$ data does not behave  as being isospin 1~; however,
this does not invalidate the $e^+e^-$ data.

Therefore, the single departure from CVC one observes
is simply the tiny shift between the $\rho^0$ and $\rho^\pm$ masses which follows
from ALEPH data.

It also follows from our model that using $\tau$ data in order to reconstruct
the equivalent $e^+e^-$ spectrum is not straightforward and requires a 
non--trivial physics input  as shown by both plots in Figure \ref{CVCFig}.

\vspace{0.5cm}
\indent \indent
One is tempted to think to extending the model in order to include higher mass vector meson 
nonets. In order to do it properly within the HLS context, one may change in
the covariant derivative (see Eqs. (\ref{AA7})) the (presently) single vector 
term $g V$  to $\sum g_i V_i$. This induces further transition amplitudes
 like, for instance, $ \rho^0(770) \longleftrightarrow \rho^0(1450)$ 
 which may sharply complicate the model. It is not easy to have a feeling
for  the numerical magnitude of the inter--nonet transitions as, now, pion and kaon loops
  contribute. Moreover, the  higher mass vector mesons are in the region where
  all the thresholds of the $VP$ channels are open. Therefore, one may also have
  to include the corresponding loops explicitly. Whether the problem will not become
  numerically untractable is therefore unclear.
 
 For this pupose, one has certainly noticed that the problem we have examined
 is highly non--linear in the parameters. This makes the search for solutions
 highly dependent on getting a starting point (in the parameter phase--space) reasonably close
 to the final solution that the fit procedure may succeed. In the present case, 
 it was already a non--trivial (and highly time consuming) task.
  
 Finally, as clear from the downmost Figure \ref{CVCFig},
  our model predicts a few percent effect for isospin breaking
 in the resonance region as well as in the low energy region where ChPT applies.
 Although our pion form factor fit results are good down to the lowest energy points 
 in $e^+e^-$ data (about $360$ MeV), 
 it would be interesting to compare with ChPT estimates in this region.
 Examining this question, the authors of Ref. \cite{RhoOmg6} argued that 
 a reliable answer may call for a $\cal O$$(p^6)$ estimate still missing.

\section{Conclusion}
\label{summary}
\indent \indent
Within the context of the HLS model, as for other models, at tree
level the so--called ideal fields ($\rho_I$, $\omg_I$, $\phi_I$) are mass
eigenstates. This simple picture vanishes at one--loop order. In this
case,  kaon loops generate non--zero amplitudes allowing $\omg_I \longleftrightarrow \phi_I$ 
transitions. Breaking of Isospin Symmetry in the pseudoscalar sector  generates 
a mass difference between kaons and, besides,  the transition amplitudes 
$\omg_I \longleftrightarrow \rho_I$  and $\phi_I \longleftrightarrow \rho_I$,
even if of small magnitude, are no longer vanishing. Additionally, these amplitudes 
exhibit a dependence on $s$, the square of the momentum flowing through the
vector meson lines. As the physical fields $\rho$, $\omg$, $\phi$
 are expected to be eigenstates of the squared mass matrix,
this unavoidably leads to define them as linear combinations
of their ideal partners. However, as the transition amplitudes
are $s$--dependent, it is clear that these combinations should also be
$s$--dependent.

We substantiate these considerings starting from the  HLS Lagrangian,
modified by including in the squared mass matrix of the neutral vector
mesons all  self--energies and transition amplitudes.
Making the assumption that the physical neutral vector fields should be
eigenstates of the loop modified squared mass matrix of the (ideal)
neutral vector meson, we solve the eigenvalue problem perturbatively.
This leads to physical vector meson fields expressed as linear
combinations of their ideal partners with definite $s$--dependent
coefficients, which are actually analytic -- or, rather, meromorphic -- functions of $s$.
Of course, this algebra leaves unchanged the charged and/or open strangeness
sector, as the starting fields acquire a running mass but no
transition from one to another meson field.

The main mechanism producing the vector meson field mixing 
is the occurence of neutral and charged kaon loops in transitions
between the ideal neutral vector meson fields. We have also shown that
the anomalous HLS--FKTUY sector and the Yang--Mills lagrangian piece
provide as supplementing mechanisms the  $K^*K$ and $K^*\overline{K}^*$
loops, occuring in transitions in nice correspondence with kaon
loops. Within this framework, the $\omg \phi$ mixing has been shown to
proceed from quantum (loop) effects, while the $\rho^0\omg $ and $ \rho^0\phi$
mixings follow from Isospin Symmetry breaking effects and vanish when
this symmetry is restored. From a numerical point of view,
the bulk of the effect is carried more by the subtraction polynomials
than the loop expression themselves, as these are logarithmic
functions of small amplitude in the physics region of interest.

The vector meson mixing allows to dynamically generate isospin violating couplings
$\phi \pi^+ \pi^-$ and $\omg \pi^+ \pi^-$ at the (modified HLS)
Lagrangian level. With this at hand, we have been able to construct
the pion form factor expression at one--loop order modified in
order to account for Isospin Symmetry breaking through only
a dynamically generated $\rho_I-\omg_I-\phi_I$ mixing scheme.
{\it A priori}, this fully affects the $e^+e^-$ physics but, by no way,
the $\tau$ physics~; in this sector, as a second order refinement,
we have been led to accept shifting with respect to each other the Higgs--Kibble 
masses of the $\rho^0$ and $\rho^\pm$ mesons. This provides a tiny effect,
nevertheless clearly visible in ALEPH $\tau$ data, but not  obvious 
in CLEO data. New data expected from B-factories may confirm
the need for this mass shift. As a final result, we end up with
structureless residuals in the fitted regions, which confirm
that our dynamical mixing scheme is appropriate.

The mixing properties have been introduced in the anomalous decay amplitudes
$V \ra P \gamma$ and $P \ra V \gamma$. These processes actually represent
our main lever arm  while defining numerically our isospin breaking scheme.

Beside a good description of the $V \ra P \gamma$ and $P \ra V \gamma$
decay data, this allows a  very good simultaneous description
of all pion form factor data from the close spacelike region to the
$\phi$ mass, in $e^+e^-$ annihilations as well as in $\tau$  decays.
The physical  ground of this result can be traced back to the fact
that the $\rho^0$ meson is a mixture of isospin 0 and 1 states (as
the $\phi$ and $\omg$ meson), in contrast to 
the $\rho^\pm$ meson which is purely isospin 1. Actually,
extracting the isospin 1 part of the pion form factor in $e^+e^-$
annihilations requires to split up the $\rho^0$, $\phi$ and $\omg$
contributions. This is done automatically by our model, and one can
claim that such a splitting cannot be done without some model.

The net result of this model is to prove that the lineshapes for the pion
form factor in $e^+e^-$ annihilations  and in $\tau$  decays are
perfectly consistent with each other, without any further breaking of CVC
than a possible tiny $\rho^0-\rho^\pm$ mass difference. In a further step, 
one may include the data from KLOE as well as data expected to come
from BaBar and Belle concerning the $\tau$ spectrum on the one hand and the $e^+ e^-$ 
Initial State Radiation samples on the other hand. However, our comparison 
of $e^+e^-$ and $\tau$ data does not seem to leave room for any kind of new physics.

\vspace{0.5cm}
\indent \indent
On the other hand, we have shown that this model allows a good account of all
decays of the form  $V \ra P \gamma$ and $P \ra V \gamma$. The case of
the $\omg \ra \eta \gamma$ and $\omg \ra \pi^0 \gamma$ partial widths, where some disagreement
is observed with the so--called ``fit" values proposed by the PDG, has been discussed and
we have argued that the real situation is somewhat unclear.

Our dealing with the pion form factor data has led us to propose
improved values for data sharply related with the $e^+e^- \ra \pi^+ \pi^-$
annihilation process, namely the $\rho^0 \ra e^+e^-$ and  $\omg \ra \pi^+ \pi^-$
partial width decays. In both cases, we find that the reference 
values should be significantly modified and we propose for these 
new reference data.

Finally, we have briefly commented upon the mass and width of the $\rho^0$ 
and $\rho^\pm$ mesons and argued that the best motivated definition should rely
on the pole position in the complex $s$--plane and related definitions.

\vspace{0.5cm}
\indent \indent
Having shown that  $e^+e^-$ data and $\tau$ data are perfectly consistent with each other
provided one uses  an appropriate model of Isospin Symmetry breaking, we can conclude that there is no reason
to question the $e^+e^-$ data. This result is important as these data
serve to estimate numerically the hadronic
photon vacuum polarization used in order to  predict  the value of the muon anomalous 
magnetic moment $g-2$ . Therefore our model indirectly confirms the 3.3 $\sigma$ discrepancy
between the BNL direct measurement of the muon anomalous magnetic moment 
and its theoretical estimate.

\section*{Acknowledgements}
\indent \indent
We gratefully acknowledge S. Eidelman (Budker Institute, Novosibirsk) 
for information on the Novosibirsk data and their systematics and for several
useful and friendly discussions and comments. We also benefited from useful information 
in order to deal in the most appropriate way with the ALEPH data
from A. H\"{o}cker   (CERN, Geneva) and with the CLEO
data from J. Urheim (Indiana University, Bloomington) who also made
valuable remarks. We also acknowledge
B. Pietrzyk (LAPP, Annecy) and A. Vainshtein (University of Minnesota, Minneapolis)
 for interesting discussions and comments.
M. Davier (LAL, Orsay) has drawn our attention on the 
photon vacuum polarization effects, we first neglected~; 
we are indebted to him of having provided us with his code for calculating 
the hadronic photon vacuum polarization in the timelike region
before publication. We finally are thankful to H. Burkhardt
(CERN, Geneva) for having provided us with the hadronic photon vacuum polarization
in the close spacelike region. Both pieces of information have 
been quite useful to the present work.\\

\begin{table}[ph]
\vspace*{-2.5cm}
\begin{center}
\begin{tabular}{|| c  | c  | c ||}
\hline
\hline
\hhhc
\hhhd Decay Mode & Fit  Value  & PDG/Reference\\
\hline
\hline
$\rho \ra \pi^0 \gamma~ [\times 10^{4}]$ \hhhv   & $5.17 \pm 0.04 $    	& $6.0 \pm 0.8$ \\
\hline
$\rho \ra \pi^\pm \gamma~ [\times 10^{4}]$\hhhv  & $ 5.03 \pm 0.03$    	& $4.5 \pm 0.5$ \\
\hline
$\rho \ra \eta \gamma~ [\times 10^{4}]$ \hhhv    & $ 3.05 \pm 0.04$ 	& $2.95 \pm 0.30$ \\
\hline
$\eta' \ra \rho \gamma~ [\times 10^{2}]$ \hhhv   & $33.3 \pm 1.0 $ 	& $29.4 \pm 0.9$ \\
\hline
\hline
$ K^{*\pm} \ra K^\pm \gamma  [\times 10^{4}]$\hhhv& $ 9.8 \pm 0.9$ 	& $9.9 \pm 0.9$ \\
\hline
$ K^{*0} \ra K^0 \gamma  [\times 10^{3}]$\hhhv	& $ 2.26 \pm 0.02$ 	& $2.31 \pm 0.20$ \\
\hline
\hline
$\omg \ra \pi^0 \gamma~ [\times 10^{2}]$ \hhhv	& $8.23 \pm 0.04 $ 	& $8.9^{+0.27}_{-0.23}$ $(*)$\\
\hline
$\omg \ra \eta \gamma~ [\times 10^{4}]$ \hhhv	& $ 6.60 \pm 0.09$ 	& $4.9 \pm 0.5$ $(*)$ \\
\hline
$\eta' \ra \omg \gamma~ [\times 10^{2}]$ \hhhv	& $3.14 \pm 0.10 $ 	& $3.03 \pm 0.31$ \\
\hline
\hline
$\phi \ra \pi^0 \gamma~ [\times 10^{3}]$ \hhhv	& $ 1.24 \pm 0.07$ 	& $1.25 \pm 0.07$ \\
\hline
$\phi \ra \eta \gamma~ [\times 10^{2}]$ \hhhv& $ 1.292 \pm 0.025$ 	& $1.301 \pm 0.024$ \\
\hline
$\phi \ra \eta' \gamma~ [\times 10^{4}]$ \hhhv& $0.60 \pm 0.02 $ 	& $0.62  \pm 0.07$ \\
\hline
\hline
$\eta \ra \gamma \gamma~ [\times 10^{2}]$ \hhhv& $35.50 \pm 0.56 $ 	& $39.38 \pm 0.26$ \\
\hline
$\eta' \ra \gamma \gamma~ [\times 10^{2}]$ \hhhv& $2.10 \pm 0.06 $ 	& $2.12 \pm 0.14$ \\
\hline
\hline
$\rho \ra e^+ e^- ~ [\times 10^{5}]$ \hhhv& $5.56 \pm 0.06 $ 		& $4.70 \pm 0.08$ $(**)$\\
\hline
$\omg \ra e^+ e^- ~ [\times 10^{5}]$ \hhhv& $7.15 \pm 0.13 $ 		& $7.18 \pm 0.12$ \\
\hline
$\phi \ra ~ e^+ e^- [\times 10^{4}]$ \hhhv& $ 2.98 \pm 0.05$ 		& $2.97 \pm 0.04$ \\
\hline
\hline
$\omg \ra \pi^+ \pi^-  [\times 10^{2}]$ \hhhv	&$1.13 \pm 0.08$	& $1.70 \pm 0.27$ $(**)$\\
\hline
$g_{\omg \pi^+ \pi^- }$ phase [degr]\hhhv		&$101.2 \pm 1.6$ 	& $104.7 \pm 4.1$ $(**)$ ~~ \cite{ffVeryOld}\\
\hline
\hline
$\phi \ra \pi^+ \pi^-  [\times 10^{5}]$ \hhhv	&$7.14 \pm 1.7 $ 	& $7.3 \pm 1.3$ \\
\hline
$g_{\phi \pi^+ \pi^- }$ phase [degr]\hhhv		&  $-27.0 \pm 0.5 $ 	& $-34 \pm 5$ ~~ \cite{SNDPhi}\\
\hline
\hline
$\phi \ra K^+ K^-  [\times 10^{2}]$ \hhhv	& $50.3 \pm  1.0$ 	&  $49.2 \pm 0.6$ $(**)$\\
\hline
$\phi \ra K^0_S K^0_L  [\times 10^{2}]$ \hhhv	&$33.0 \pm 0.7$		& $34.0 \pm 0.5$ $(**)$\\
\hline
\hline
\end{tabular}
\end{center}
\caption{
\label{T3}
Reconstructed Branching fractions for radiative and leptonic decays
using any of the  various fit strategies. The reported values are the mean value
and the rms of the simulated distributions. The last column displays the recommended
branching ratios  \cite{RPP2006}. The symbol $(*)$ indicates data commented upon in the text,  
$(**)$  indicates data which are not introduced in the fit procedure. 
}
\end{table}

\newpage
\section*{\Large{Appendices}}
\appendix
\section{The Full HLS Non--Anomalous Lagrangian}
\label{HLSLagrangian}
\label{AA}

\indent \indent
The construction of the non--anomalous Lagrangian of the Hidden Local Symmetry (HLS) Model has 
been presented in great detail several times by its authors (see for  instance \cite{HLSOrigin} 
or more recently \cite{HLSRef}). Let us simply outline the main steps of the construction 
procedure.

The HLS model allows to produce a theory with vector mesons as gauge bosons of a hidden local 
symmetry. These acquire a mass because of the spontaneous breakdown of a global chiral symmetry
$G_{{\rm global}} = U(3)_L \otimes U(3)_R$. The chiral Lagrangian is written~:
\be
{\cal L}_{{\rm chiral}}= \displaystyle \frac{f_\pi^2}{4} {\rm Tr} [\pa_\mu U \pa^\mu U]
\label{AA1}
\ee
\noindent where $U(x)=\exp{[2 i P(x)/f_\pi}]$~; here $ f_\pi$ is identified with the pion decay constant 
($ f_\pi=92.42$ MeV) and $P$ is the matrix of pseudoscalar mesons (the Goldstone bosons
associated with the spontaneous breakdown of $G_{{\rm global}}$). This matrix~:

\be
P= P_8 +P_0=\frac{1}{\sqrt{2}}
  \left( \begin{array}{ccc}
          \displaystyle  \frac{1}{\sqrt{2}}\pi^0+\frac{1}{\sqrt{6}}\eta_8+
            \frac{1}{\sqrt{3}}\eta_0&\displaystyle \pi^+ & \displaystyle  K^+ \\
            \displaystyle \pi^-  & \displaystyle -\frac{1}{\sqrt{2}}\pi^0+\frac{1}{\sqrt{6}}\eta_8
            +\frac{1}{\sqrt{3}}\eta_0  &  \displaystyle K^0 \\
           \displaystyle  K^-             &  \displaystyle \overline{K}^0  &\displaystyle 
             -\sqrt{\frac{2}{3}}\eta_8 +\frac{1}{\sqrt{3}}\eta_0 \\
         \end{array} 
  \right),
\label{AA2}  
\ee
\noindent contains a singlet term besides the octet term~;  appropriate combinations of
$\eta_8$ and $\eta_0$ correspond to the physical pseudoscalar fields $\eta$ and $\eta^\prime$.
Here and throughout this paper  we restrict ourselves to three flavours. 

However, besides the global symmetry $G_{{\rm global}}$, the chiral Lagrangian
possesses a local symmetry
$H_{{\rm local}} = SU(3)_V$ which is included in the HLS approach by splitting up $U$ as~:

\be
U(x)=\xi_L^\dagger \xi_R
\label{AA3}  
\ee
\noindent where the $\xi$ fields undergo the local transformation. These variables are parametrized
as~:
\be
\xi_{R,L} = \displaystyle  e^{i \displaystyle \sigma/f_\sigma} e^{\pm i \displaystyle P/f_\pi}
\label{AA4}  
\ee
 
\noindent and the scalar field $\sigma$ is usually 
eliminated through a gauge choice, and can be considered absorbed into the gauge 
bosons and removed. However, the decay constant $f_\sigma$ goes on appearing in the model
through the HLS fundamental parameter $a=f_\sigma^2/f_\pi^2$. Using this parametrization
Eq. (\ref{AA1}) can be rewritten~:

\be
{\cal L}_{{\rm chiral}}= \displaystyle - \frac{f_\pi^2}{4} 
{\rm Tr} [(\pa_\mu\xi_L\xi_L^\dagger -\pa_\mu\xi_R\xi_R^\dagger)^2]
\label{AA5}  
\ee

This Lagrangian can be gauged for electromagnetism, weak interaction and the hidden local
symmetry by changing the usual derivatives $\pa_\mu$ to covariant derivatives $D_\mu$
\cite{HLSOrigin,HLSRef} and one then gets~:
\be
\left \{
 \begin{array}{ccc}
 {\cal L}_{HLS}= &{\cal L}_A + a {\cal L}_V \\[0.5cm]
 {\cal L}_A = &\displaystyle  -\frac{f_\pi^2}{4} {\rm Tr} [(D_\mu\xi_L\xi_L^\dagger -D_\mu\xi_R\xi_R^\dagger)^2] &
 \displaystyle \equiv -\frac{f_\pi^2}{4} {\rm Tr} [L-R]^2\\[0.5cm]
 {\cal L}_V = &\displaystyle  -\frac{f_\pi^2}{4} {\rm Tr} [(D_\mu\xi_L\xi_L^\dagger +D_\mu\xi_R\xi_R^\dagger)^2] &
 \displaystyle \equiv -\frac{f_\pi^2}{4} {\rm Tr} [L+R]^2\\[0.5cm]
 \end{array} 
 \right .
\label{AA6}  
\ee
\noindent using obvious notations. 

Now let us turn to the covariant derivatives. These are given by\cite{HLSRef}~:
\be
\left \{
\begin{array}{ccc}
D_\mu \xi_L  = \displaystyle  \pa_\mu \xi_L -i g V_\mu \xi_L +i \xi_L {\cal L}_\mu\\[0.5cm]
D_\mu \xi_R  = \displaystyle  \pa_\mu \xi_R -i g V_\mu \xi_R +i \xi_R {\cal R}_\mu
 \end{array} 
 \right .
\label{AA7}  
\ee
\noindent (where we have factored out the universal vector meson coupling constant) with~: 
\be
\left \{
\begin{array}{l}
{\cal L}_\mu =   \displaystyle  e Q A_\mu + \frac{g_2}{\cos{\theta_W}} (T_z -\sin^2{\theta_W})Z_\mu 
+\frac{g_2}{\sqrt{2}} (W^+_\mu T_+ + W^-_\mu T_-)\\[0.5cm]
{\cal R}_\mu =   \displaystyle e Q A_\mu - \frac{g_2}{\cos{\theta_W}} \sin^2{\theta_W} Z_\mu 
 \end{array}  
 \right .
\label{AA8}  
\ee

\indent \indent Eqs. (\ref{AA7}, \ref{AA8}) introduce the  matrix of vector meson fields (the gauge bosons
of the hidden local symmetry) which is~:
\be
V=\frac{1}{\sqrt{2}}
  \left( \begin{array}{ccc}
   \displaystyle (\rho^I+\omega^I)/\sqrt{2}  & \displaystyle \rho^+             &  \displaystyle K^{*+} \\[0.5cm]
    \displaystyle  \rho^-    & \displaystyle  (-\rho^I+\omega^I)/\sqrt{2}    &  \displaystyle  K^{*0} \\[0.5cm]
      \displaystyle        K^{*-}           & \displaystyle  \overline{K}^{*0}  &  \displaystyle  \phi^I   
         \end{array}
  \right) 
\label{AA9}
\ee
\noindent in terms of the so--called ideal field combinations
 (indicated by the superscript $I$) for the neutral vector mesons,  which should be distinguished from the
 physical fields introduced in the main text. $ A_\mu$ is the electromagnetic field and $e$ the unit electric charge,
 $g_2$ and $\theta_W$  are respectively the gauge weak coupling constant and the weak (Weinberg) angle. $Z_\mu$ and $W^\pm_\mu$
 are, of course the weak gauge boson fields. $Q$, the quark charge matrix and the $T$ matrices are SU(3) matrices~: 
 $Q=1/3 ~{\rm{Diag}}(2,-1,-1)$ and $T_z=1/2 ~{\rm{Diag}}(1,-1,-1)$ , while $T_+=(T_-)^\dagger$ with~:
\be
 T_+=
  \left( \begin{array}{ccc}
 	0 &V_{ud} & V_{us}\\
	0 &0 & 0\\
 	0 &0 & 0\\
         \end{array}
  \right) 
\label{AA10}
\ee
\noindent in terms of elements of the Cabibbo-Kobayashi-Maskawa matrix elements. In this work, the $Z_\mu$ terms have not
to be considered~; they have been given for completeness. The HLS Lagrangian given above should be completed with the vector
meson kinetic energy term\cite{HLSRef} but also with the usual free Lagrangian for electromagnetic and weak boson fields.
The leptonic sector  also has to be added~; it is written as per usual~:
\be
\begin{array}{ll}
{\cal L}_{\ell,\nu} &= \displaystyle \sum_{\ell = (e ,\mu ,\tau)} 
\left [q_\ell~ \ell^- \gamma^\mu \ell^+ A_\mu -\frac{g_2}{2 \sqrt{2}}
\overline{\nu}_{\ell} \gamma^\mu(1-\gamma_5)\ell^- W^+ + \cdots \right] 
\end{array}
\label{AA11}
\ee

From a practical point of view, $g_2$ defined above is 
related with the Fermi constant $G_F$ and the W boson mass by~:
\be
g_2 = 2 m_W ~ \sqrt{G_F \sqrt{2}}
\label{AA12}
\ee
and it is useful to note that at the $\tau$ lepton mass scale one has \cite{RPP2006}~:
 $$ g_2=0.629 ~~~~ ({\rm and}~~~~ e=0.30286)~~.$$

\section{The HLS Anomalous Sector}
\label{BB}
\indent \indent
QCD admits a non-abelian anomaly which  explicitly breaks chiral symmetry. This anomaly is well reproduced
by the Wess--Zumino--Witten Lagrangian \cite{WZ,Witten}~; this has been incorporated within the HLS context
by Fujiwara, Kugo, Terao,  Uehara and  Yamawaki along with vector mesons \cite{FKTUY,HLSRef}. In this way,
it becomes possible to provide a framework which allows one to describe most decays of vector mesons, and especially modes like
$\omg \ra \pi^+ \pi^- \pi^0$ and others more relevant in the present context. 

Let us briefly outline the derivation and its assumptions which has been presented in comprehensive reviews
\cite{BKY,Meissner,HLSRef}.  The anomalous action can be cast under the form~:
\be
\begin{array}{ll}
\Gamma&=\Gamma_{\rm WZW}+\Gamma_{\rm FKTUY}\\[0.5cm]
\Gamma_{\rm FKTUY}&=\sum_{i=1}^{4}c_i~\int d^4x\;{\cal L}_i
\end{array}
\label{BB1}
\ee
\noindent where $\Gamma_{\rm WZW}$ is the original WZW Lagrangian. The Lagrangian pieces ${\cal L}_i$
where first given in \cite{FKTUY} and each of them contains $APPP$ and $AAP$ pieces which 
would contribute to the anomalous decays, but are cancelled by $APV$ terms. 
These Lagrangians contain also $VPPP$ and $VVP$ pieces \cite{FKTUY,HLSRef}. A priori, the weighting 
coefficients $c_i$  are arbitrary. However, in order to reconcile this Lagrangian with decay data, especially  
$\omg \ra \pi^+ \pi^- \pi^0$, FKTUY  \cite{FKTUY,HLSRef} finally choose the following combination~:
\be 
{\cal L}^{\rm FKUTY}=-\frac{3g^2}{4\pi^2 f_\pi}\epsilon^{\mu\nu\rho\sigma}
{\rm Tr}[\pa_\mu V_\nu\pa_\rho V_\sigma P]-\frac{1}{2}{\cal L}_{\gamma PPP}~~,
\label{BB2}
\ee   
\noindent which turns out to complement the usual WZW term for $\gamma P P P$ interaction with only
a $VVP$ term. In this model, for instance  the decay $\pi^0 \ra \ggam$ occurs 
solely through $\pi^0 \ra \omg \rho^0$ followed by the (non--anomalous) 
transitions $\omg \ra \gamma$ and $\rho^0 \ra \gamma$ and the partial width is 
identical to the Current Algebra prediction reproduced by ${\cal L}_{\gamma \gamma P}$~: 
\be
{\cal L}_{\gamma \gamma P} =  \displaystyle
-\frac{N_c e^2}{4 \pi^2 f_{\pi}} 
\epsilon^{\mu \nu \rho \sigma}
\partial_{\mu}A_{\nu} \partial_{\rho}A_{\sigma} {\rm Tr} [Q^2P] \\[0.5cm]
\label{BB3}
\ee

The model given by Eq. (\ref{BB2}) with~:
\be
{\cal L}_{\gamma P P P} =    \displaystyle
-\frac{ieN_c}{3 \pi^2 f_\pi^3}
\epsilon^{\mu \nu \rho \sigma}
A_{\mu}  {\rm Tr} [Q \partial_{\nu}P \partial_{\rho}P\partial_{\sigma}P]
\label{BB4}
\ee
($N_c=3$) has been also shown \cite{box} to reproduce perfectly the data
on $\eta/\eta^\prime \ra \pi^+ \pi^- \gamma$, especially the most precise ones  
\cite{Abele}. Indeed, with no free parameter, the distortion of the $\rho$
lineshape is accurately accounted for and this can be considered as the
signature for the box anomaly in experimental data. 

Accounting for the light meson radiative decays of the $AVP$ or $AAP$ forms is then
a $VVP$ coupling followed by one or two $V \ra \gamma$ transition(s).
From a practical point of view, it has been also shown  \cite{mixing} that the 
corresponding couplings can be directly derived from the following Lagrangian 
piece~:
 \be
 \begin{array}{lll}
{\cal L}=C \epsilon^{\mu \nu \rho \sigma}
{\rm Tr} [
\partial_\mu(eQA_\nu + g V_\nu)\partial_\rho (eQA_\sigma + g V_\sigma)P]~~&
,~~~C =\displaystyle -\frac{3}{4 \pi^2 f_\pi}~.
 \end{array} 
\label{BB5}	
\ee

Let us note that in the meson decays we are interested in, the weak boson sector
is irrelevant. Finally, one can find the $VVP$ Lagrangian expanded in \cite{rad},
more precisely in Appendix 1 and 4 for respectively the fully flavor symmetric
case and the SU(3)/U(3) broken case{\footnote{In the Appendix
4, the two breaking parameters $\ell_W$ and $\ell_T$ (denoted $z_T$ in the present paper)
have been found in this reference to fulfill $\ell_W \ell_T^2 =1 $.}

\section{SU(3)/U(3) Symmetry Breaking of the HLS Model}
\label{CC}
\indent \indent The HLS Lagrangian we are interested in is the lowest
order expansion of Eq. (\ref{AA6}) supplemented with Eq. (\ref{BB5}).
However, in order to use it with most real data, one cannot avoid
defining an appropriate symmetry breaking mechanism. Several breaking schemes
have been proposed \cite{BKY,BGP,BGPbis,Heath1} as there is no unique way to implement 
such a mechanism in the HLS model. We will prefer the method proposed in 
\cite{Heath1} which looks to be the simplest that automatically fulfills the hermiticity
requirement. This symmetry breaking scheme turns out to modify the
non--anomalous Lagrangian terms in Eq. (\ref{AA6}) to~:

 \be
\left \{
 \begin{array}{ll}
{\cal L}_A = & \displaystyle   -\frac{f_\pi^2}{4} {\rm Tr} [(L-R) X_A]^2\\[0.5cm]
 {\cal L}_V = & \displaystyle  -\frac{f_\pi^2}{4} {\rm Tr} [L+R) X_V]^2\\[0.5cm]
 \end{array} 
 \right .
\label{CC1}	
\ee
\noindent where the SU(3) symmetry breaking matrices $X_A$ and $X_V$ can be written~:
 \be
\left \{
 \begin{array}{ll}
 X_A=& {\rm Diag}(1,1,\sqrt{z_A})\\[0.5cm]
 X_V=& {\rm Diag}(1,1,\sqrt{z_V})~~~~~.
 \end{array} 
 \right .
\label{CC2}	
\ee

As the parameter $z_A=[f_K/f_\pi]^2$ is fixed here by kaon decay data, it
can hardly be considered as a truly free parameter, even if one allows it to vary
within errors \cite{RPP2006} : $z_A=1.495 \pm 0.031$. As shown in the Lagrangian 
pieces given in the main text, the second breaking parameter, $z_V$,  allows one to
shift the $\phi$ meson mass away from those of the $\rho$ and $\omega$ mesons~;
practically we have more freedom in varying it. The full SU(3) broken HLS Lagrangian 
produced by this mechanism (without $W^\pm$ interaction terms) has been given in 
\cite{Heath1}. Using this mechanism, however, the pseudoscalar kinetic energy term
of the HLS Lagrangian is no longer canonical and a renormalization of the pseuscalar
fields is required  \cite{BKY}~:
 \be
 P^\prime= \displaystyle X_A^{1/2} P  X_A^{1/2}
\label{CC3}	
\ee
\noindent where $P$ and $P^\prime$ stand respectively for the bare and renormalized
pseudoscalar field matrices. With this redefinition, the kinetic energy term of
the SU(3) broken Lagrangian is once again canonical. However, the coupling constants
to kaons have to be changed correspondingly by introducing renormalized fields. 

\vspace{1.cm}

\indent \indent
With this symmetry breaking mechanism, the realm of practical relevance for the HLS model
extends to pions and kaons, as far as pseudoscalar mesons are concerned. In order to
bring $\eta$ and $\eta^\prime$ mesons into the game, one needs first to define them
in terms of the $\eta_8$ and $\eta_0$ fields, second to extend the breaking scheme
from SU(3) to U(3).

We use the one angle traditional mixing expression~:
\begin{equation}
\left[
     \begin{array}{ll}
     \displaystyle \eta   \\[0.5cm]
     \displaystyle \eta'   
     \end{array}
\right]
=
\left[
     \begin{array}{lll}
\displaystyle \cos{\theta_P} & -\displaystyle \sin{\theta_P} \\[0.5cm]
\displaystyle \sin{\theta_P} &
\displaystyle ~~\cos{\theta_P} 
     \end{array}
\right]
\left[
     \begin{array}{ll}
     \pi_8\\[0.5cm]
     \eta_0\\
     \end{array}
\right]
\label{CC4}
\end{equation}

It has been shown \cite{WZWChPT} that the one--angle description was equivalent 
at the first two leading  orders to the two--angle, two--decay constant description in favor since the 
Extended Chiral  Perturbation Theory (EChPT) \cite{leutw,leutwb,feldmann}. 

Now the question is how  Nonet Symmetry Breaking (NSB) can be incorporated within the
(SU(3) broken) HLS Lagrangian already defined. This can be done by means of determinant terms 
\cite{tHooft} which break the $U_A(1)$ symmetry~:
\be
{\cal L}={\cal L}_{\rm HLS}+ \frac{\mu^2f^2_\pi}{12}
\ln\det U \cdot\ln\det U^{\dag}
+\lambda\frac{f^2_\pi}{12}\ln\det \pa_\mu U
\cdot\ln\det \pa^\mu U^{\dag}
\label{CC5}
\ee
\noindent where $U$ is defined by Eqs.(\ref{AA3}) and (\ref{AA4}) after removal of the $\sigma$ field
matrix. This can be rewritten more explicitly~:
\be
{\cal L}={\cal L}_{\rm HLS}+{\cal L}_{\rm HLS}^{\prime}
\equiv{\cal L}_{\rm HLS}+\frac{1}{2}\mu^2\eta_0^2+\frac{1}{2}\lambda
\pa_\mu \eta_0 \pa^\mu \eta_0
\label{CC6}
\ee

Therefore, in this manner, one provides both a mass to the singlet and a modification of the
kinetic singlet term which is thus no longer canonical and, then, calls for a renormalization. 
The exact renormalization relation is given in  \cite{WZWChPT}, where it has also been shown
that, at leading order, this transformation is equivalent to using the HLS Lagrangian
but replacing Eq. (\ref{CC3}) by~:
\be
 P^\prime_8 + xP^\prime_0 = \displaystyle X_A^{1/2} (P_8 +P_0) X_A^{1/2}
\label{CC7}	
\ee
\noindent (with obvious notations). The Nonet Symmetry Breaking (NSB) mechanism introduces a
parameter $x$ which can be related \cite{WZWChPT} with $\lambda$ by~:
\be
x=1-\frac{\lambda}{2}B^2 \simeq \frac{1}{\sqrt{1+\lambda B^2}} 
\Longrightarrow \lambda \simeq 0.20 - 0.25 ,
\label{CC8}	
\ee
\noindent ($B=(2 z_A+1)/3 z_A$) with a precision better than $\simeq 5 \%$.

Therefore one has only to equip  the SU(3) broken HLS Lagrangian with the U(3) broken
renormalization condition given by Eq.(\ref{CC7}). Ref. \cite{WZWChPT} showed that,
at leading order in breaking parameters one recovers the ChPT expectations.

\vspace{1.cm}

\indent \indent
In order to achieve this general presentation of the broken HLS model, we
recall in a few words the breaking procedure of the anomalous Lagrangian.
A priori, the transformation to renormalized fields given by Eq. (\ref{CC7}) 
induces a breaking mechanism into the anomalous HLS Lagrangian given by Eqs.
(\ref{BB2}) and (\ref{BB4}). It has been shown \cite{rad,mixing} that, alone,
this breaking scheme (as well as no breaking at all, both) implies that the coupling 
constant ratio  $G_{K^{*0} K^0 \gam}/G_{K^{*\pm} K^\pm \gam}$ equals 0.5 in sharp
disagreement with experimental data \cite{RPP2006}. Interestingly, the non--relativistic
quark model (NRQM) allows more freedom by exhibiting a dependence of this ratio
upon the ratio of quark magnetic moments $r$ \cite{Morpurgo}~:
\be
\displaystyle \frac{G_{K^{*0} K^0 \gam}}{G_{K^{*\pm} K^\pm \gam}} = -\frac{1+r}{2-r}
\label{CC9}	
\ee

In \cite{rad,mixing}, it has been shown that this effect can be obtained
by mixing a symmetry breaking scheme proposed by Bramon, Grau and Pancheri (BGP) \cite{BGP,BGPbis} 
with some sort of vector field renormalization. In the following Appendix, we show that
the guess expressed in \cite{rad,mixing} that a vector field renormalization should take
place is justified while having to perform the SU(3) symmetry breaking of the Yang--Mills
piece in the HLS Lagrangian.

Numerical analysis implies that these two  mechanisms (namely the BGP mechanism and
the vector field renormalization) are highly correlated\footnote{Actually, as stated
in \cite{mixing}, imposing that the two--photon decay widths of the $\eta$ and $\eta^\prime$
mesons should remain as  given by the original Wess--Zumino--Witten Lagrangian \cite{WZ,Witten}
provides the functional correlation first found numerically.} with the neat result that the
broken VVP Lagrangian in Eq. (\ref{BB5}) becomes~:
 \be
{\cal L}=C \epsilon^{\mu \nu \rho \sigma}
{\rm Tr} [
X_T \partial_\mu(eQA_\nu + g V_\nu)X_T^{-2} \partial_\rho (eQA_\sigma + g V_\sigma)X_T P].
\label{CC10}	
\ee
\noindent with $P$ being  replaced by renormalized fields using Eq. (\ref{CC7}) above
and $V$ being understood as already renormalized (as a consequence of breaking the SU(3) symmetry
of ${\cal L}_{YM}$).

Therefore, the Lagrangian we use in order to account for anomalous decays is{\footnote{
Eq. (\ref{BB4}) might have also to be broken similarly. However, existing data on box anomalies
allow  access only to the limited  sector $\pi^0/\eta/\eta^\prime \ra \pi^+ \pi^- \gam$ 
not affected by more breaking than reflected by Eq. (\ref{CC7})~; it has been
shown \cite{box} that Eq. (\ref{BB4}) as it already stands, suffices for satisfactorily 
accounting for the data. There is therefore no need to go beyond for this term.
}} Eq. (\ref{CC10}) with~:
\be
 X_T=  {\rm Diag}(1,1,\sqrt{z_T})
\label{CC11}	
\ee
\noindent where $z_T$ is a parameter to be fitted. We should stress that this
specific breaking, which allows one to recover Eq. (\ref{CC9}) leaves  
all other couplings of physical interest (AVP and AAP) unchanged. One should note that,
except for (conceptually unavoidable) mixing angles, the model we use introduces only
two parameters $z_A$ and $z_T$ in the anomalous sector, the former being essentially fixed by pure kaon
physics. Taking into account that our broken anomalous Lagrangian aims at accounting for
14 decay modes, the parameter freedom is actually very limited.

\section{The Yang--Mills Term Of The HLS Lagrangian}
\label{YM}
\indent \indent
The HLS Lagrangian  defined in the body of the text should be understood supplemented
with the Yang--Mills piece associated with the vector meson fields  \cite{HLSRef}.
Defining the vector field matrix as given in Eq. (\ref{AA9}) (see Appendix A),
this writes \cite{HLSRef}~:
\begin{equation}
\left \{
\begin{array}{lll}
\displaystyle {\cal L}_{YM} &\displaystyle  =-\frac{1}{2} {\rm Tr} [F_{\mu \nu}F^{\mu \nu}]\\[0.5cm]
F_{\mu \nu} &= \partial_\mu V_\nu-\partial_\nu V_\mu -i g [V_\mu,V_\nu]
\end{array}
\right .
\label{un}
\end{equation}   
 
The square of the abelian part provides the usual kinetic energy term
of the vector meson fields and is certainly canonical while working with
ideal fields. When performing the field transformation defined
by Eqs. (\ref{Mass6}) and (\ref{Mass7}), this piece remains canonical in terms of physical fields, 
as a trivial consequence of  the orthogonality property of the matrix $R$~:  $R(s) \widetilde{R}(s)=1$.

\vspace{0.5cm}
\indent \indent
The mixed abelian--non--abelian term contributes to all vector meson
self--energies. Additionally,  it also  provides transition amplitudes
among the ideal $\rho^0$, $\omg$ and  $\phi$ fields. One can check that
these additional contributions to  the transition amplitudes only involve $K^* \overline{K}^*$ loops
and contribute  in the following way to right--hand sides of Eqs. (\ref{SelfMasses})~:
\begin{equation}
\left \{
\begin{array}{lll}
\Pi_{\omg \phi}(s) = & \cdots + g_{\phi K^* {\overline{K}}^*}g_{\omg K^* \overline{K}^*}
\left [ \Pi_{ K^{*\pm}  \overline{K}^{*\mp}}(s) + \Pi_{ K^{*0}  \overline{K}^{*0}}(s)
\right ] \\[0.5cm]
\Pi_{\rho \omg}(s) = & \cdots + g_{\rho K^* \overline{K}^*}g_{\omg K^* \overline{K}^*}
\left [ \Pi_{ K^{*\pm}  \overline{K}^{*\mp}}(s) - \Pi_{ K^{*0}  \overline{K}^{*0}}(s)
\right ] \\[0.5cm]
\Pi_{\rho \phi}(s) = & \cdots + g_{\rho K^* \overline{K}^*}g_{\phi K^* \overline{K}^*}
\left [ \Pi_{ K^{*\pm}  \overline{K}^{*\mp}}(s) - \Pi_{ K^{*0}  \overline{K}^{*0}}(s)
\right ] \\[0.5cm]
\end{array}
\right .
\label{deux}
\end{equation}   
still using obvious notations. Quite interestingly, one sees that the transition amplitudes
$\Pi_{\omg \phi}(s)$, $\Pi_{\rho \omg}(s)$ and $\Pi_{\rho \phi}(s)$ are  modified
in  such a way that $\omg \phi$ always receives an additional non--vanishing contribution, while
the $\rho \omg$ and $\rho \phi$ transitions receive non--vanishing contributions only
if isospin symmetry is broken ($m_{K^{*\pm}} \ne m_{K^{*0}}$). One  may note the striking
correspondence between the Yang--Mills contributions to the transition amplitudes
and those already given in   Eqs. (\ref{SelfMasses}).

Moreover, taking into account the
threshold value of $K^*  \overline{K}^*$ loops, these are certainly real below 1 GeV  
 and their effects can be considered numerically absorbed in the subtraction polynomials
we already use. Denoting by $M$ the $K^*$ mass, the amputated $K^* \overline{K}^*$ loop fulfills~:

\begin{equation}
\displaystyle {\rm Im}~\Pi(s) =-\frac{1}{48 \pi} \frac{s+3 M^2}{M^2} \frac{(s- 4 M^2)^{3/2}}{s^{1/2}}
\label{trois}
\end{equation}   
above the two--$K^*$ threshold. This function can be algebraically derived from the two--pion loop
(see Appendix E) and undergoes the same minimum number of subtractions as this.

If no SU(3) breaking is implemented inside ${\cal L}_{YM}$, the relevant coupling constants 
are~:
\begin{equation}
\left \{
\begin{array}{lll}
\displaystyle 
G_{\omg K^{*+} K^{*-}} =G_{\rho K^{*+} K^{*-}}=-\sqrt{2} G_{\phi K^{*+} K^{*-}}=-\frac{g}{2}\\[0.5cm]
\displaystyle 
G_{\omg K^{*0} \overline{K}^{*0}} = - G_{\rho K^{*0} \overline{K}^{*0}} =- \sqrt{2} G_{\phi K^{*0} \overline{K}^{*0}}
=-\frac{g}{2}
\end{array}
\right .
\label{quatre}
\end{equation}   

Finally, tadpole terms may also contribute to transition amplitudes~; they are generated
by the non--abelian term squared. One can prove that they follow the same pattern
as shown in Eqs. (\ref{deux}) just above, namely still contributing to $\Pi_{\omg \phi}(s)$, while
contributing to  $\Pi_{\rho \omg}(s)$ and $\Pi_{\rho \phi}(s)$ only if isospin
symmetry is broken.
 
\vspace{0.5cm}
\indent \indent
Another remark is of interest and concerns the flavor SU(3) symmetry breaking of the Yang--Mills piece.
Indeed, as we already perform the flavor SU(3) symmetry breaking of the ${\cal L}_{A}$ and ${\cal L}_{V}$
pieces
of the HLS Lagrangian (see Eqs. (\ref{CC1}) and (\ref{CC2})), we should perform likewise with ${\cal L}_{YM}$.
In consistency with our SU(3) symmetry breaking scheme, this should be done as~:
\begin{equation}
\left \{
\begin{array}{lll}
\displaystyle {\cal L}_{YM} &\displaystyle  
=-\frac{1}{2} {\rm Tr} [F_{\mu \nu} X_{YM} F^{\mu \nu} X_{YM}]\\[0.5cm]
X_{YM} &= {\rm Diag}(1,1,z_{YM})
\end{array}
\right .
\label{cinq}
\end{equation}   
It is easy to check that, in order to restore the canonical form of the vector meson kinetic energy term,
one has to renormalize the vector  fields and define~:
\begin{equation}
V_\mu = X_{YM}^{-1/2}V_\mu^{Ren} X_{YM}^{-1/2}
\label{six}
\end{equation}   
This change of fields should be propagated to the anomalous Lagrangian. Doing this way and
introducing the breaking procedure proposed  by Bramon, Grau and Pancheri \cite{BGPbis},
one indeed ends up with Eq. (\ref{CC10}) in Appendix \ref{CC} (for this purpose, see the relevant discussions
in \cite{rad,mixing}) with~:
\begin{equation}
X_T = X_{YM}^{-1/2}~~~~.
\label{sept}
\end{equation}   

In this renormalization, Eqs. (\ref{quatre})
remain valid with replacing there each $G_{\phi K^{*} \overline{K}^{*}}$ by $G_{\phi K^{*} \overline{K}^{*}}/z_T$.
However,
it is interesting to note that the single piece of information  in our global data set which is really sensitive
to this renormalization is the  (charged and neutral) $K^*$ radiative decay.
 Actually, the 
vector field renormalization also affects the ideal $\phi$ meson decay to the $K \overline{K}$ and   $e^+ e^-$  modes~;
because of mixing among ideal fields, this also affects $\omg$ and $\rho$ decays to $e^+ e^-$. However, one
can easily convince oneself that the influence of this renormalization is absorbed numerically 
in the fitted value for the parameter $z_V$. 

\section{Radiative and Leptonic Coupling Constants}
\label{DD} 
 \indent \indent In order to simplify the main text we prefer to list here the 
coupling constants entering decay widths expression which will be treated in this paper.
 Most of them can be derived trivially from expressions already given in Appendix E in \cite{mixing}.
 \subsection{Radiative Decays}
 \indent \indent Starting from the Lagrangian in Eq. (\ref{CC10}), and using the breaking 
procedure as defined by Eq. (\ref{CC7}), one can compute the coupling constants for all 
radiative decays of relevance.  Let us define~:

\begin{equation}
G=\displaystyle 
-\frac{3eg}{8 \pi^2 f_\pi}~~~,~~~ 
G'=\displaystyle -\frac{3eg}{8 \pi^2 f_K}~~~, 
~~~Z= \left [\frac{f_\pi}{f_K} \right]^2=\frac{1}{z_A}~~~,
~~~ \delta_P=\theta_P-\theta_0 ~~~~~(\tan{\theta_0= 1/\sqrt{2}}) .
\label{DD1}
\end{equation}

Some $VP\gamma$ coupling constants are not affected by
 Isospin Symmetry breaking~:
\begin{equation}
\left \{
\begin{array}{lll}
G_{\rho^{\pm} \pi^{\pm} \gamma}=& &\displaystyle \frac{1}{3} G \\[0.3cm]
G_{K^{*0} K^0 \gamma}=&- & 
\displaystyle \frac{G'}{3} \sqrt{z_T} (1+\frac{1}{z_T})  \\[0.3cm]
G_{K^{*\pm} K^{\pm} \gamma}=& &
\displaystyle \frac{G'}{3} \sqrt{z_T} (2-\frac{1}{z_T}) ~~. \\[0.3cm]
 \end{array}
\right .
\label{DD2}
\end{equation}

The $\rho_I P\gamma$ coupling constants are~:
\begin{equation}
\left \{
\begin{array}{lll}
G_{\rho_I \pi^0 \gamma}=& &\displaystyle \frac{1}{3} G \\[0.3cm]
G_{\rho_I \eta \gamma}=& &\displaystyle \frac{1}{3} G
\left[\sqrt{2}(1-x)\cos{\delta_P}-(2x+1)\sin{\delta_P}\right]\\[0.3cm]
G_{\rho_I \eta' \gamma}=& &\displaystyle \frac{1}{3} G
\left[\sqrt{2}(1-x)\sin{\delta_P}+(2x+1)\cos{\delta_P}\right]~~.
\end{array}
\right .
\label{DD3}
\end{equation}

 The other single photon radiative modes provide the following coupling
 constants~:
\begin{equation}
\left \{
\begin{array}{lll}
G_{\omega_I \pi^0 \gamma}=& \displaystyle G \\[0.3cm]

G_{\phi_I \pi^0 \gamma}=& \displaystyle 0 \\[0.3cm]

G_{\omega_I \eta \gamma}=&   \displaystyle \frac{1}{9} G  \left [
\sqrt{2} (1-x) \cos{\delta_P} -(1+2x) \sin{\delta_P} \right ]\\[0.3cm]

G_{\omega_I \eta' \gamma}=& \displaystyle \frac{1}{9} G \left [
(1+2x)\cos{\delta_P} +\sqrt{2}(1-x)\sin{\delta_P} \right ]\\[0.3cm]

G_{\phi_I \eta \gamma}=& \displaystyle \frac{2}{9} G \left [
Z(2+x)\cos{\delta_P}  - \sqrt{2}Z(1-x)\sin{\delta_P} \right ]\\[0.3cm]

G_{\phi_I \eta' \gamma}=&\displaystyle \frac{2}{9}G \left [
\sqrt{2}Z(1-x)\cos{\delta_P} + Z(2+x)\sin{\delta_P} \right ]~~.
\end{array}
\right .
\label{DD4}
\end{equation}

In order to go from ideal field couplings to physical vector field couplings,
one has to use linear combinations of the couplings in Eqs. (\ref{DD3}-\ref{DD4})
weighted by elements of the transformation matrix $R(s)$ given in the body of the
paper. 

\subsection{ $P\gamma \gamma$ and $V-\gamma$ Modes}

\indent \indent
The 2--photon decay modes are not affected by Isospin
Symmetry breaking in the vector sector and keep their usual form within
the HLS model \cite{rad,mixing,WZWChPT}~:
\begin{equation}
\left \{
\begin{array}{lll}
G_{\eta \gamma \gamma} = && 
-\displaystyle \frac{\alpha_{em}}{\pi \sqrt{3} f_{\pi}}
\left [ \frac{5-2Z}{3}\cos{\theta_P}-\sqrt{2} 
\frac{5+Z}{3}x \sin{\theta_P} \right ]\\[0.3cm] 
G_{\eta' \gamma \gamma} = && 
-\displaystyle \frac{\alpha_{em}}{\pi \sqrt{3} f_{\pi}}
\left [ \frac{5-2Z}{3}\sin{\theta_P} 
+ \sqrt{2} \frac{5+Z}{3}x \cos{\theta_P} \right ]\\[0.3cm] 
G_{\pi^0 \gamma \gamma} = && -\displaystyle  \frac{\alpha_{em}}{\pi  f_{\pi}}~~.
\end{array}
\right .
\label{DD5}
\end{equation}
As stated in the text, we actually replace this last coupling by using the world average value 
for $f_{\pi}$ as given in the RPP \cite{RPP2006}.

Finally, in the non--anomalous sector, the  leptonic decay widths of vector mesons depend
on the HLS $V-\gamma$ couplings. For the  ideal combinations, we have~:
\begin{equation}
\left \{
\begin{array}{lll}
f_{\rho_I \gamma} = & \displaystyle a f_{\pi}^2 g \\[0.3cm] 
f_{\omega_I \gamma} = & \displaystyle \frac{f_{\rho_I \gamma}}{3}
\\[0.3cm] 
f_{\phi_I\gamma} = & \displaystyle  -\frac{f_{\rho_I \gamma}}{3}
\sqrt{2} z_V ~~.
\end{array}
\right.
\label{DD6}
\end{equation}

It was shown in \cite{WZWChPT} that the pseudoscalar mixing angle is not a free
parameter, but is related with the SU(3) breaking  parameter $Z(=1/z_A)$ and the Nonet 
Symmetry breaking  parameter $x$ by~:
\begin{equation}
\tan{\theta_P} = \displaystyle \sqrt{2} \frac{Z-1}{2Z+1} x
\label{DD6b}
\end{equation}
with a very good accuracy. This relation is used in our fits as a constraint.

\subsection{Partial widths}
\label{PartialWidths}
\indent \indent
We list for completeness in this Section the expressions for the partial widths
in terms of the coupling constants for the various cases which are examined 
in the text.

The two--photon partial widths are~:
\be
\Gamma(P \ra \gamma \gamma)= \displaystyle 
\frac{m_P^3}{64 \pi} |G_{P\gamma \gamma}|^2~~~,~~P=~\pi^0,~\eta,~\eta'~~.
\label{DD7}
\ee

The leptonic partial widths are~: 
\be
\Gamma(V \ra e^+ e^-)=\displaystyle \frac{4\pi\alpha^2}{3 m_V^3} |f_{V\gamma}|^2~~.
\label{DD8}
\ee

The radiative widths are~:
\be
\Gamma(V \ra P \gamma)= \displaystyle \frac{1}{96 \pi}
 \left[ \frac{m_V^2-m_P^2}{m_V} \right]^3 |G_{VP\gamma}|^2~~,
\label{DD9}
\ee
where $V$ is either of $\rho^0$, $\omg$, $\phi$ and
$P$ is either of $\pi^0$, $\eta$, $\eta'$, and~:
\be
\Gamma(P \ra V \gamma)= \displaystyle \frac{1}{32 \pi}
 \left[ \frac{m_P^2-m_V^2}{m_P} \right]^3 |G_{VP\gamma}|^2~~.
\label{DD10}
\ee

The decay width for a vector meson decaying to $V+P$ is~:
 
\be
\Gamma(V' \ra V P)= \displaystyle \frac{1}{96 \pi}
 \left[ 
 \frac{\sqrt{[m_{V'}^2-(m_V+m_P)^2][m_{V'}^2-(m_V-m_P)^2]}}{m_{V'}}
 \right]^3 |G_{V'VP}|^2~~.
\label{DD11}
\ee

Finally, the partial width for a vector meson decaying into two
pseudoscalar mesons of equal masses is~:
\be
\Gamma(V \ra P P)=\displaystyle \frac{1}{48 \pi}
 \frac{[m_V^2-4m_P^2]^{3/2}}{m_V^2} |G_{VPP}|^2~~.
\label{DD12}
\ee

\section{The Loop Functions}
\label{EE} 
 \indent \indent  The loop functions can be written quite generally as~:
\be
\displaystyle \Pi(s)=f(s) K(s) + P(s)
\label{EE1}
\ee
where $f(s)$ is a polynomial $Q(s)$ divided by some power of $s$. 
The degree of the polynomial $P(s)$ is fixed always at second degree
and we require $P(0)=0$. we have~:
\be
\left \{
\begin{array}{ll}
{\rm{Im}} K(s)=-(s-s_c)^{1/2} (s-s_0)^{1/2} ~~~, ~~~~ (s \ge s_0)  &~\\[0.4cm]
K(s)= \displaystyle c_0+c_1 s + c_2 s^2 + \frac{s^3}{\pi} \int_{s_0}^{\infty}
\frac{{\rm{Im}}K(z)}{z^3(z-s+i\epsilon)} dz
& ~~ 
\end{array}
\right.
\label{EE2}
\ee
\noindent where $s_0$ is the (direct) threshold mass squared, while  $s_c$ is the
(crossed) threshold mass squared. 

\subsection{The PP loop}
\indent \indent In the case of equal masses, $s_c=0$ and we have \cite{mixing}~:
\begin{equation}
\Pi(s)=\displaystyle 
\frac{g^2_{VP\overline{P}}}{48 \pi} \frac{s-s_0}{s} K(s) +P(s)
\label{EE3}
\end{equation}
and the solution is
\begin{equation}
\left \{
\begin{array}{lll}
~~& \Pi(s)=d_0+d_1s + Q(s) \\[0.4cm]
~~& Q(s)=\displaystyle \frac{g^2_{VP\overline{P}}}{24 \pi^2}
\left[ G(s) +s_0 \right]
 \\[0.4cm]
s \le 0~~~~: & G(s)= ~\displaystyle \frac{1}{2}\frac{(s_0-s)^{3/2}}{(- s)^{1/2}}
\ln \frac{(s_0-s)^{1/2} - (-s)^{1/2}}{(s_0-s)^{1/2} + (-s)^{1/2}} \\[0.4cm]
0 \le s \le s_0~~: &
G(s)=-\displaystyle \frac{(s_0-s)^{3/2}}{s^{1/2}}
\arctan{\sqrt{\frac{s}{(s_0-s)}}} \\[0.4cm]
s \ge s_0 ~~~:& G(s)=-\displaystyle \frac{1}{2} \frac{(s-s_0)^{3/2}}{s^{1/2}}
\left[ \ln \frac{s^{1/2}-(s-s_0)^{1/2}}{s^{1/2} + (s-s_0)^{1/2}}\right]\\[0.4cm]
~~ & -\displaystyle \frac{i \pi}{2} \displaystyle \frac{(s-s_0)^{3/2}}{s^{1/2}}\\[0.4cm]
\end{array}
\right.
\label{EE4}
\end{equation}
The behavior of $\Pi(s)$ near $s=0$ is simply ${\cal{O}}(s)$,
and $Q(s)$ behaves like ${\cal{O}}(s)$ near the origin. This result
coincides with the one of \cite{Klingl,mixing}. By performing more
subtractions, one could choose to fix externally the actual $s^n$ behavior
of the loop near the origin.

\subsection{The $PP^\prime$ Loop}
\indent \indent In this case, we have~:
\begin{equation}
\Pi(s)=\displaystyle 
\frac{g^2_{VP\overline{P}}}{48 \pi} \frac{(s-s_0)(s-s_c)}{s^2} K(s) +P(s)
\label{EE5}
\end{equation}
Let us define
\begin{equation}
\left \{
\begin{array}{lll}
s \le s_c~~~~: & \varphi(s)= \displaystyle \frac{1}{\pi} (s_0-s)^{1/2}(s_c- s)^{1/2}
\ln \frac{(s_0-s)^{1/2} - (s_c-s)^{1/2}}{(s_0-s)^{1/2} + (s_c-s)^{1/2}} \\[0.4cm]
s_c \le s \le s_0~~: &
\varphi(s)=~\displaystyle \frac{2}{\pi} (s_0-s)^{1/2}(s-s_c)^{1/2}
\arctan{\sqrt{\frac{s-s_c}{s_0-s}}} \\[0.4cm]
s \ge s_0 ~~~:& \varphi(s)=\displaystyle \frac{-1}{\pi} (s-s_0)^{1/2}(s-s_c)^{1/2}
\left[ \ln \frac{(s-s_c)^{1/2}-(s-s_0)^{1/2}}{(s-s_c)^{1/2} + (s-s_0)^{1/2}}\right]\\[0.4cm]
~~ & 
~~~~~~~~~~~ -\displaystyle i  \displaystyle (s-s_0)^{1/2}(s-s_c)^{1/2}\\[0.4cm]
\end{array}
\right.
\label{EE6}
\end{equation}
The solution for $K$ is obtained by subtracting a polynomial in such a way that
the behaviour of Eq. (\ref{EE5}) is not singular at origin~:
\be
K(s)=\varphi(s)-[c_0 + c_1 s + c_2 s^2]
\ee
with ($s_0=m_0^2$ and $s_c=m_c^2$)~: 
\begin{equation}
\left \{
\begin{array}{lll}
c_0= & \displaystyle \frac{m_0 m_c}{\pi} \ln{\frac{m_0-m_c}{m_0+m_c}} \\[0.4cm]
c_1=& \displaystyle -\frac{1}{2\pi} \ln{\frac{m_0-m_c}{m_0+m_c}} 
~\frac{(m_0-m_c)^2}{m_0m_c}\\[0.4cm]
c_2=& \displaystyle -\frac{1}{8\pi}\left[
 \frac{(m_0^2-m_c^2)^2}{m_0^3m_c^3} \ln{\frac{m_0-m_c}{m_0+m_c}}  
 +2 \frac{m_0^2+m_c^2}{m_0^2m_c^2} \right]\\
\end{array} 
\right.
\label{EE7}
\end{equation}
The exact behaviour for $\Pi(s)$ at origin is then detemined by the choice of $P(s)$.

\subsection{The $PP^\prime$ Loop In the Complex $s$--Plane}
\indent \indent The expressions in the two Subsections above give the value
of the loop functions for any real values of $s$. It is interesting to know how these functions
extend into the complex $s$--plane, {\it i.e.} for complex values of $s$. It is actually
in this manner that the expressions above have been derived. One can check that~:
\begin{equation}
\displaystyle \varphi(z) = -\frac{i}{\pi}(z-s_c)^{1/2} (s_0-z)^{1/2}
\ln{
\frac{(s_0-z)^{1/2}+i(z-s_c)^{1/2}}{(s_0-z)^{1/2}-i(z-s_c)^{1/2}}
}
\label{EE8}
\end{equation}
\noindent is -- up to a polynomial with real coefficients -- the (single) analytic function
of $z$ real for $s_c<z<s_0$, having as imaginary part for real $s > s_0$,  Im$K(s)$, given by
Eq. (\ref{EE2}). The most general solution to Eq. (\ref{EE2}), is then written:
\begin{equation}
K(z)=\varphi(z) + P_n(z)
\label{EE9}
\end{equation}
\noindent with a polynomial, $P_n(z)$, with real coefficients chosen in such a way that the
behaviour at $z=0$ is the required one. The other coefficients have to be fixed by
other external (renormalization) conditions. Eq. (\ref{EE8}) gives the loop function on the
so--called physical sheet of the Riemann surface. The expression for $\varphi(z)$ on the
unphysical sheet close to the physical region $s > s_0$ is obtained by a winding of $2 \pi$
radians around the threshold (branch--point) $s=s_0$.

\subsection{The Leptonic Loop}
\indent \indent In order to compute the photon vacuum polarization, one needs to
have at one's disposal the analytic expression of the $\ell^+ \ell^-$ loop. More
precisely, one needs the ratio of this loop (lepton contribution to the photon self--energy)
divided by $s$, the off--shell photon invariant mass. This can easily be derived from
the function $\Pi^0(s)$ given\footnote{In this reference, the computed loop is actually
the quark--antiquark one and therefore the color  factor  3 has to be removed. } 
in \cite{LeptonLoop} or computed from detailed information from \cite{LeptonLoop2}~:
   
\begin{equation}
\displaystyle \Pi(s)= \frac{\alpha}{4 \pi}
\left[
\frac{20}{9} +\frac{4}{3z} -\frac{4(1-z) (1+2z)}{3z} G(z)
\right]
\label{EE10}
\end{equation}
where ~:
\begin{equation}
\begin{array}{lll}
\displaystyle G(z)=\frac{2 u \ln{u}}{u^2-1}~~~, 
& \displaystyle u= \frac{\sqrt{1-1/z}-1}{\sqrt{1-1/z}+1}~~~,
& \displaystyle z= \frac{s}{4 m_\ell^2}
\end{array} 
\label{EE11}
\end{equation}
and $m_\ell$ is the lepton mass and $\alpha$ is the fine structure constant.

However, for explicit computation  in a minimization code, on needs
the explicit expression along the real axis. This is~:
\begin{equation}
\left \{
\begin{array}{lll}
~~& \Pi(z)= \displaystyle \frac{\alpha}{4 \pi} \left[\frac{20}{9} +\frac{4}{3z} + Q(z)\right] \\[0.4cm]
z \le 0~~~~: & Q(z)= ~\displaystyle
-\frac{2}{3} \frac{1+2z}{z^2} [-z(1-z)]^{1/2} 
\ln
\frac{(1-z)^{1/2}-(-z)^{1/2}}{(1-z)^{1/2} + (-z)^{1/2}}
\\[0.4cm]
0 \le z \le 1~~: &
Q(z)=\displaystyle 
-\frac{4}{3} \frac{1+2z}{z^2} [z(1-z)]^{1/2} \arctan{\sqrt{\frac{z}{1-z}}}
\\[0.4cm]
z \ge 1 ~~~:& 
Q(z)=\displaystyle 
\frac{2}{3} \frac{1+2z}{z^2} [z(z-1)]^{1/2} \left[
\ln
\frac{z^{1/2}-(z-1)^{1/2}}{z^{1/2} + (z-1)^{1/2}}
+i \pi
\right]
\end{array}
\right.
\label{EE12}
\end{equation}

This function will be  summed up with the (parametrized) hadronic vacuum polarization
provided to us by M. Davier \cite{DavierPriv} and H.Burkhardt \cite{HelmutPriv} for, respectively
the region above and below the 2--pion threshold. The function $\Pi(z)$ just defined is analytic
and vanishes at $z=0$. The term of order $\alpha^2$ can be derived from the 
function $\Pi^{(1)}$ given in \cite{LeptonLoop}  but is difficult to handle in fitting procedures.
Other expressions for the function we use can be found in \cite{IFSVP,RadCorr}.
Finally, even if possible in principle, we do not have the freedom of subtracting more the
function $ \Pi(z)$ as conditions at the $Z$ boson mass for the full photon vacuum 
polarization seem to fix it to be zero \cite{Bolek1,Bolek2,Bolek3}.

                     \bibliographystyle{h-physrev}
                     \bibliography{hlsdocRef}
\newpage

\begin{figure}[!ht]
\begin{minipage}{\textwidth}
\begin{center}
\resizebox{\textwidth}{!}
{\includegraphics*{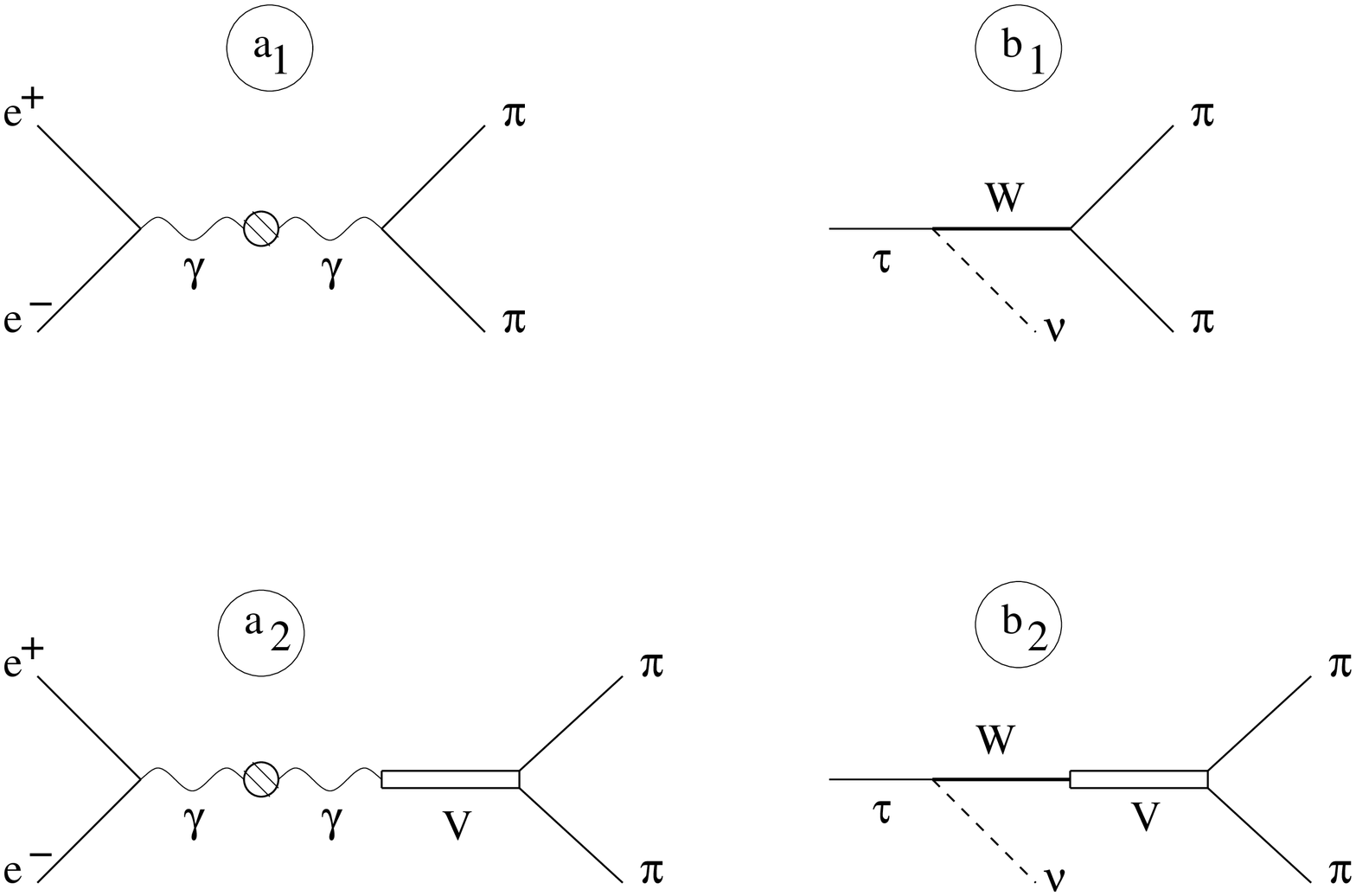}}
\end{center}
\end{minipage}
\begin{center}
\vspace{-0.3cm}
\caption{\label{PlotFF}
Schematic representation of the Feynman diagrams contributing to the pion form factor.
Left plots (referred to as a$_1$ and a$_2$) sketch the case of the pion form factor
in $e^+e^-$ annihilations, while right plots (referred to as b$_1$ and b$_2$)
figure out the $\tau$ decay. The upper plots show the non--resonant HLS specific diagrams,
the lower plots describe the resonance contributions. The shaded blobs represent
the photon vacuum polarization. the $\gamma V$ and $WV$ transitions are dressed
by $s$--dependent terms.
}
\end{center}
\end{figure}

\newpage

\begin{figure}[!ht]
\begin{minipage}{0.5\textwidth}
\begin{center}
\resizebox{\textwidth}{!}
{\includegraphics*{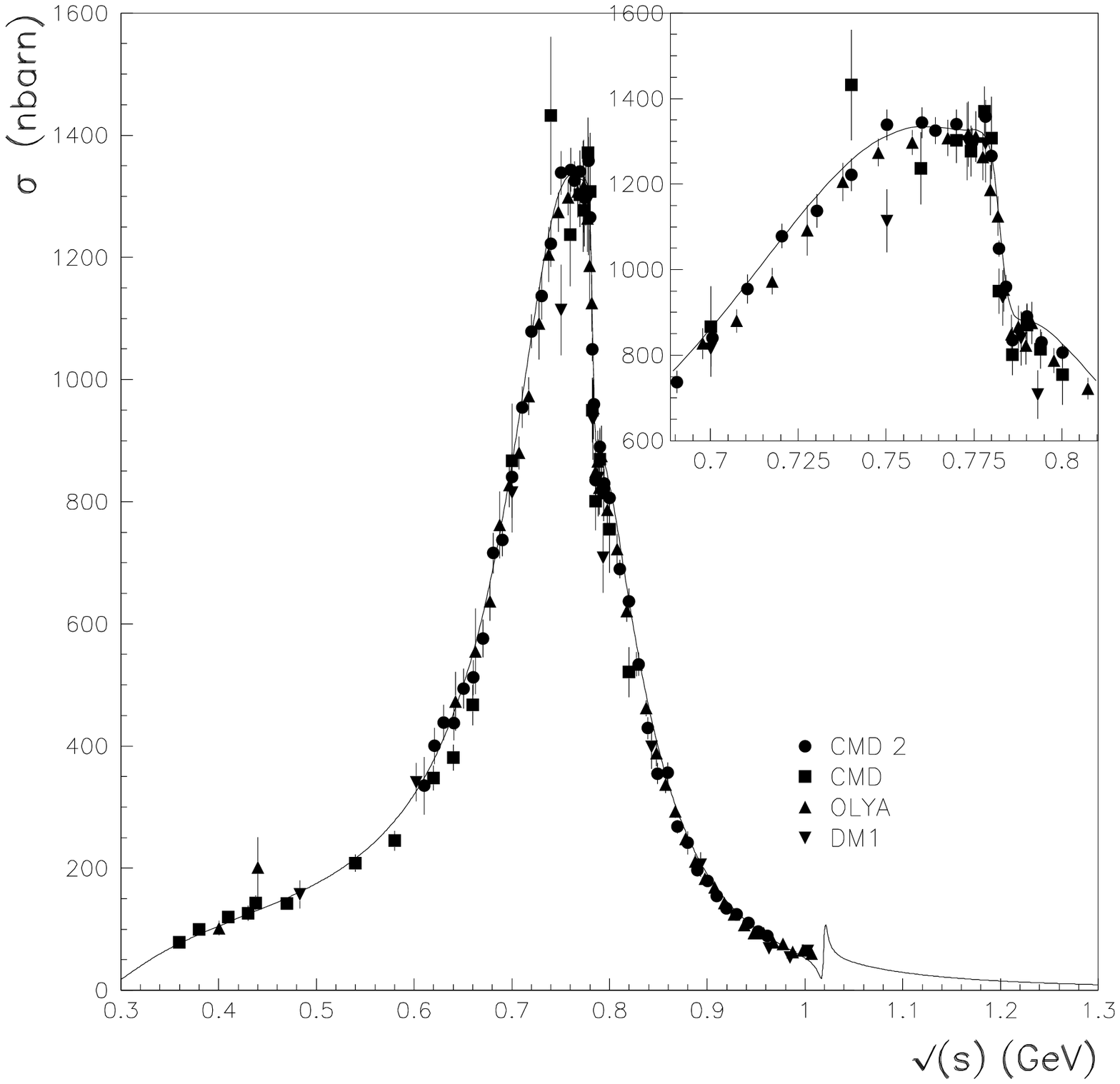}}
\end{center}
\end{minipage}
\begin{minipage}{0.5\textwidth}
\begin{center}
\resizebox{\textwidth}{!}
{\includegraphics*[width=5cm]{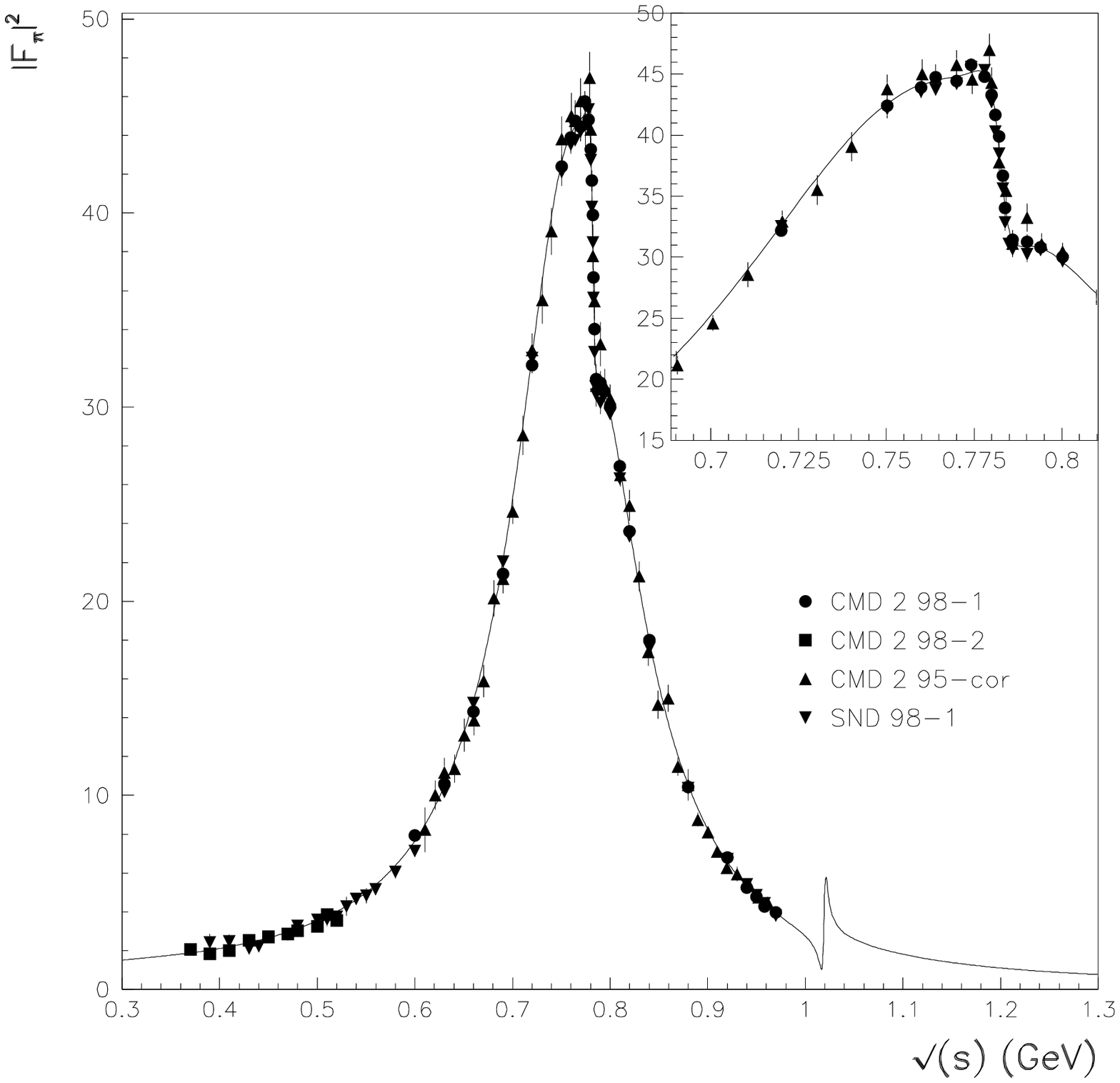}}
\end{center}
\end{minipage}
\begin{center}
\vspace{-0.8cm}
\caption{\label{NskFig}
Data and fits for the pion form factor in $e^+e^-$ timelike data.
Left figure gives the fit cross section with, superimposed,
the data from the Olya and CMD Collaborations \cite{Barkov}, the
data set \cite{DM1} from the DM1 Collaboration and the first (corrected) data from CMD2
\cite{CMD2-1995corr}. Right figure shows the form factor curve with superimposed 
all data sets  collected recently at Novosibirsk 
\cite{CMD2-1995corr,CMD2-1998-1,CMD2-1998-2,SND-1998}.
The $\phi$ region is commented upon in the body of the text.
}
\end{center}
\end{figure}

\begin{figure}[!ht]
\vspace{-1.5cm}
\begin{minipage}{0.5\textwidth}
\begin{center}
\resizebox{\textwidth}{!}
{\includegraphics*{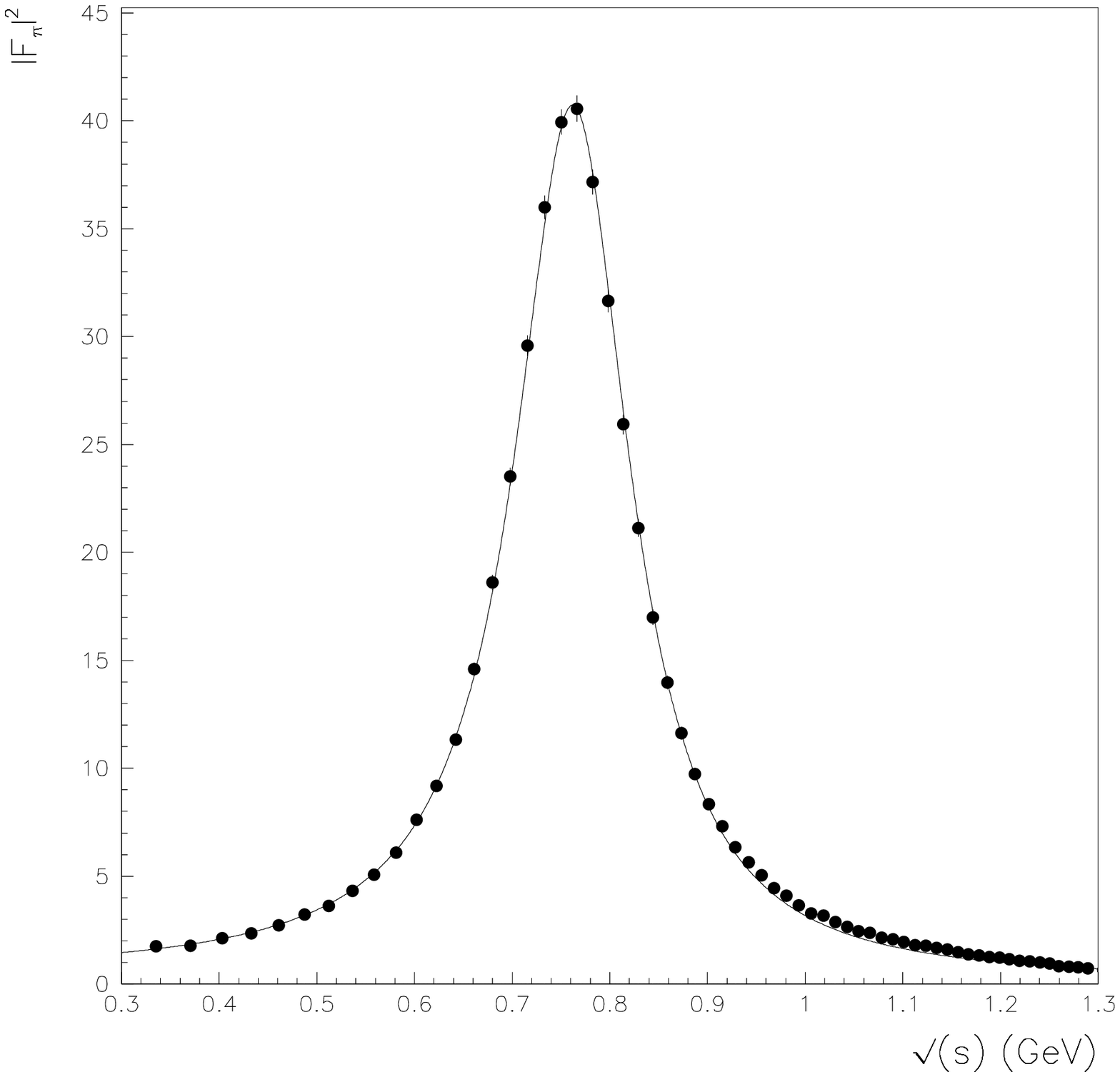}}
\end{center}
\end{minipage}
\begin{minipage}{0.5\textwidth}
\begin{center}
\resizebox{\textwidth}{!}
{\includegraphics*[width=5cm]{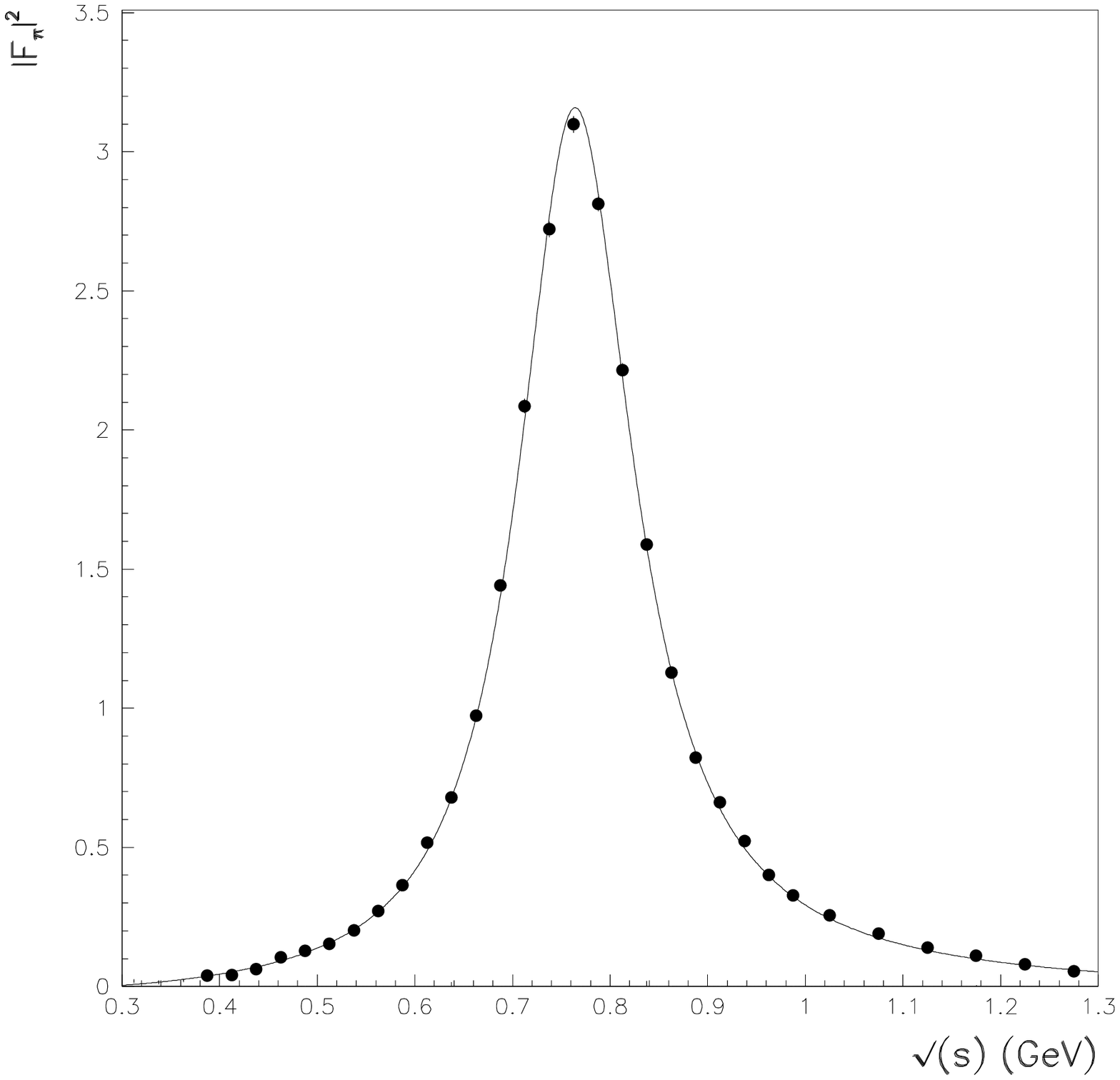}}
\end{center}
\end{minipage}
\begin{center}
\caption{\label{tauFig}
Data and fits for the pion form factor in $\tau$ decay. 
Left figure shows the case for ALEPH data \cite{Aleph},
right figure shows the case for CLEO data \cite{Cleo}.
}
\end{center}
\end{figure}

\newpage
\begin{figure}[!ht]
\begin{minipage}{0.5\textwidth}
\begin{center}
\resizebox{\textwidth}{!}
{\includegraphics*{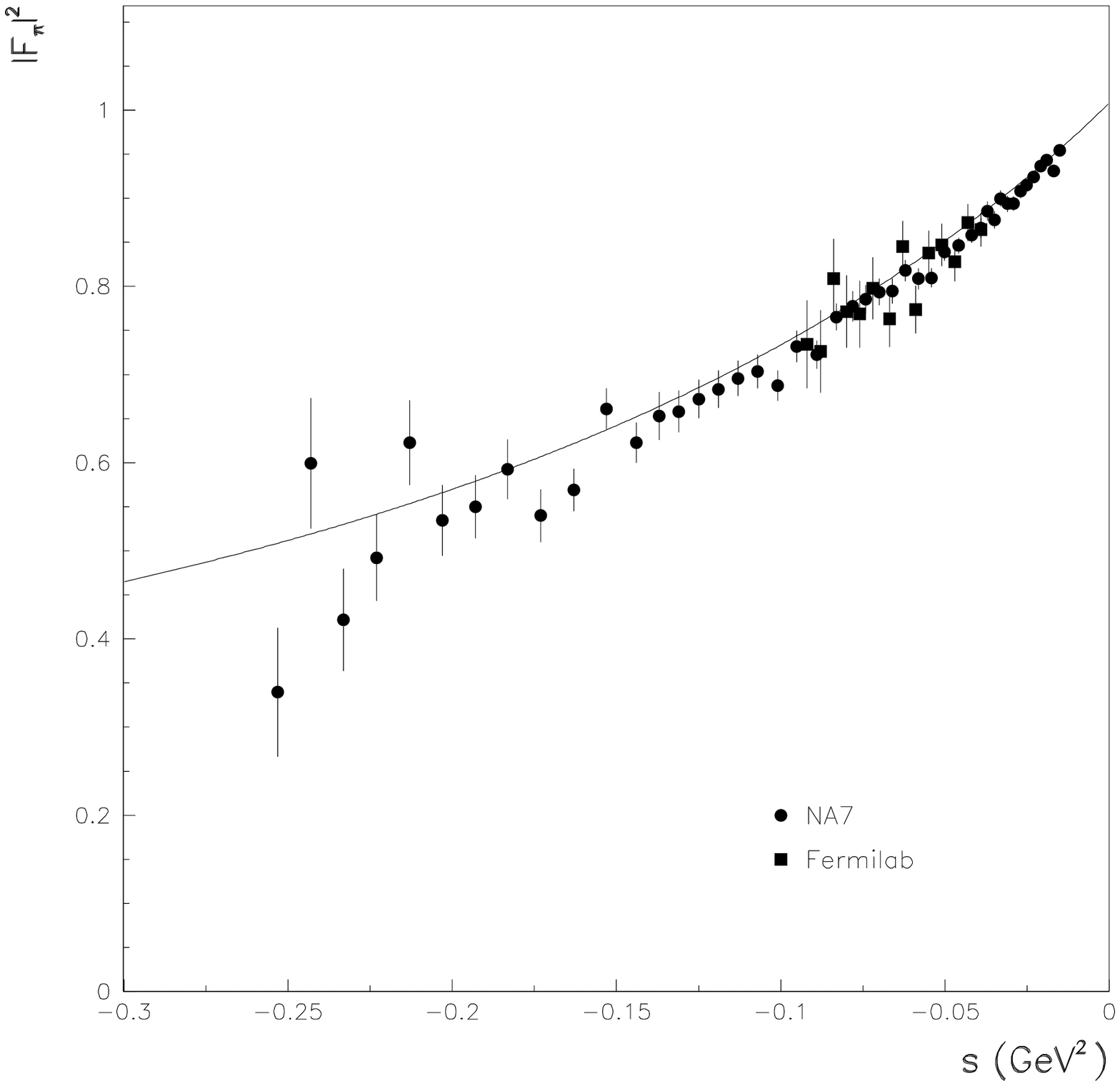}}
\end{center}
\end{minipage}
\begin{minipage}{0.5\textwidth}
\begin{center}
\resizebox{\textwidth}{!}
{\includegraphics*[width=5cm]{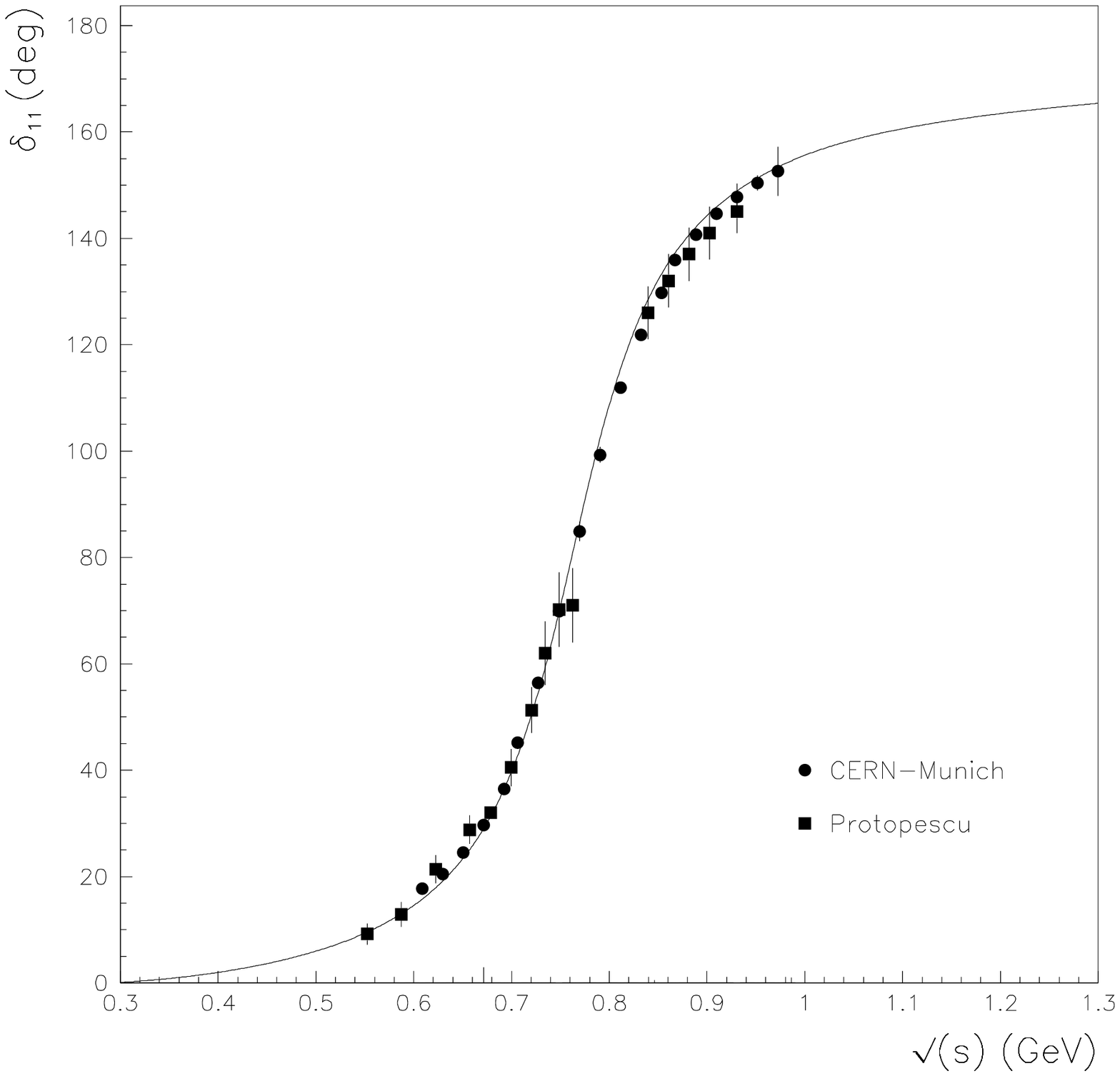}}
\end{center}
\end{minipage}
\begin{center}
\vspace{-0.8cm}
\caption{\label{NA7Fig}
Left Figure shows the fit in the spacelike region close to $s=0$
together with the data from NA7 \cite{NA7} and Fermilab \cite{fermilab2}~;
the fit scale factors (1.008 and 1.006 respectively) have been applied.
Rightside figure shows the $prediction$ for the $P_{11}$ phase shift
with the Cern-Munich data \cite{Ochs} and the data from \cite{Protopescu}
superimposed.}
\end{center}
\end{figure}

\begin{figure}[!ht]
\vspace{-1.8cm}
					\vspace{-0.9cm}
\begin{minipage}{0.8 \textwidth}
\begin{center}
\resizebox{\textwidth}{!}
{\includegraphics*{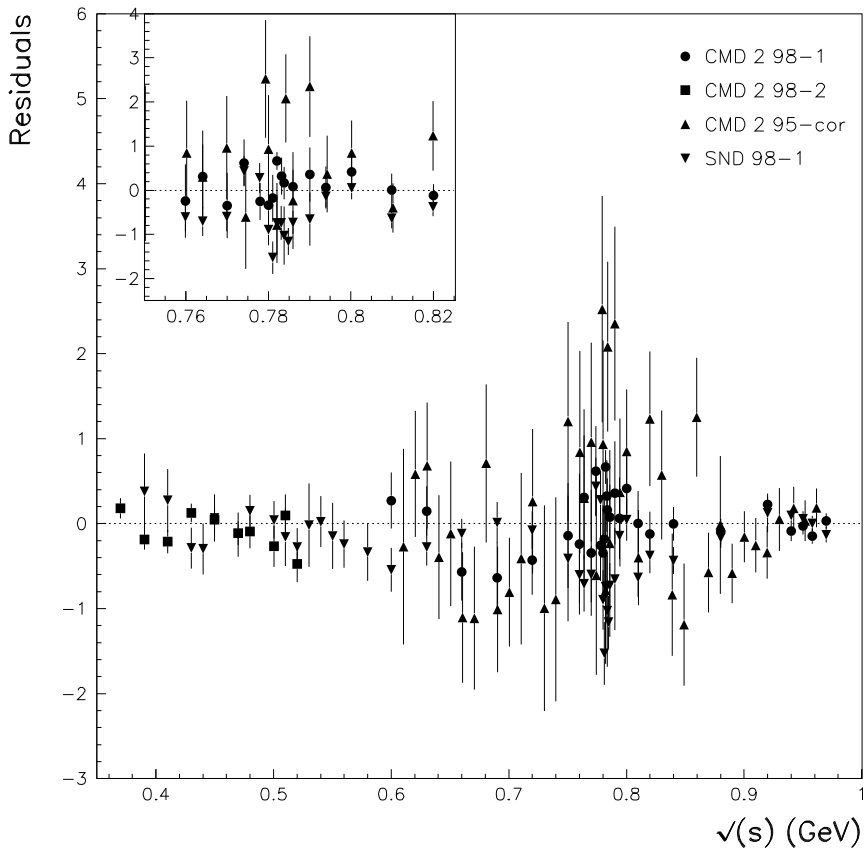}}
\end{center}
\end{minipage}
\vspace{-0.5cm}
\caption{\label{NskRes}
Residual distribution for all the $e^+e^-$ new timelike data over
the whole invariant mass interval. The inset magnifies the $\rho$ peak invariant mass region.}
\end{figure}

\begin{figure}[!ht]
\begin{minipage}{0.5\textwidth}
\begin{center}
\resizebox{\textwidth}{!}
{\includegraphics*{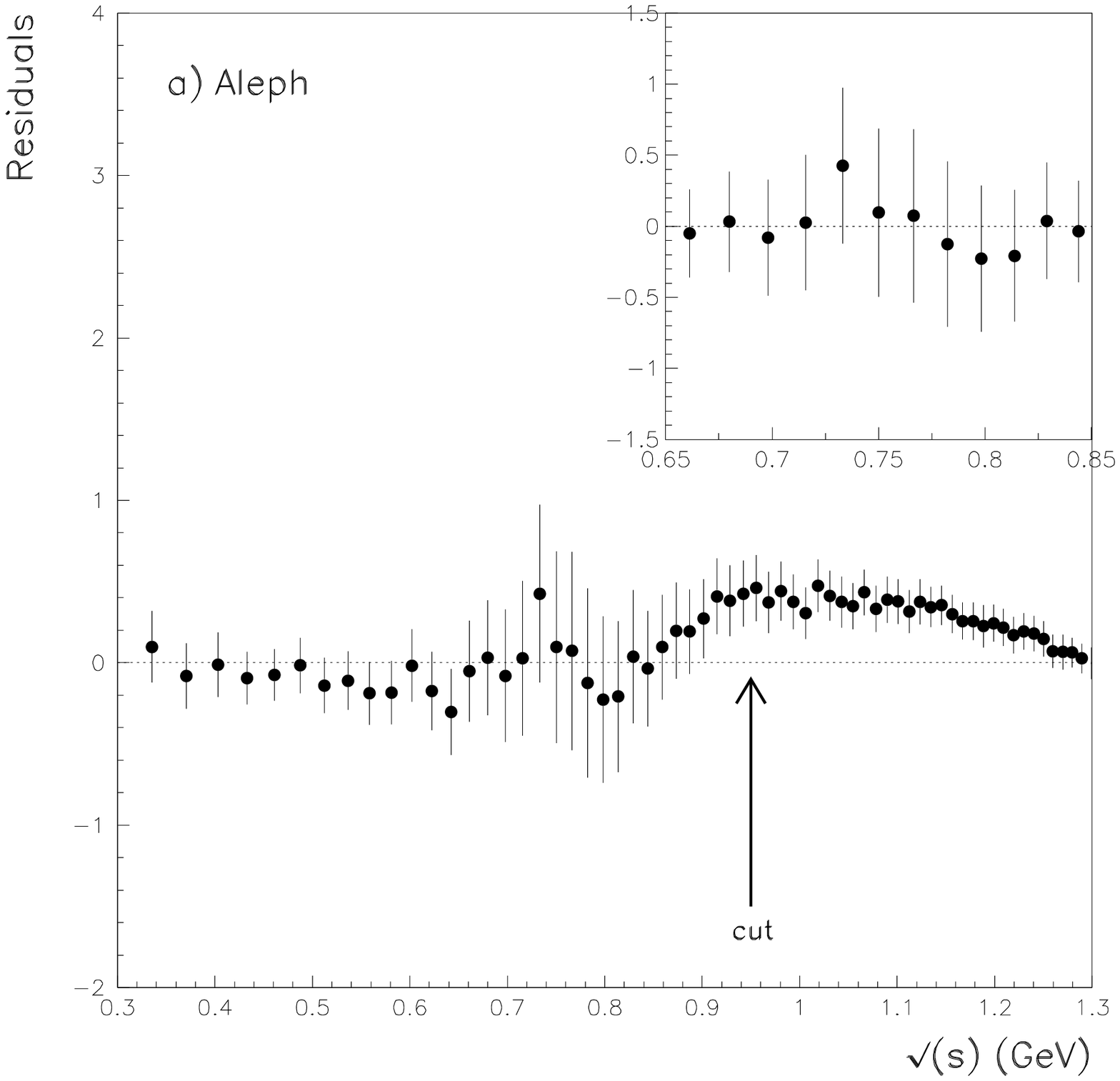}}
\end{center}
\end{minipage}
\begin{minipage}{0.5\textwidth}
\begin{center}
\resizebox{\textwidth}{!}
{\includegraphics*[width=5cm]{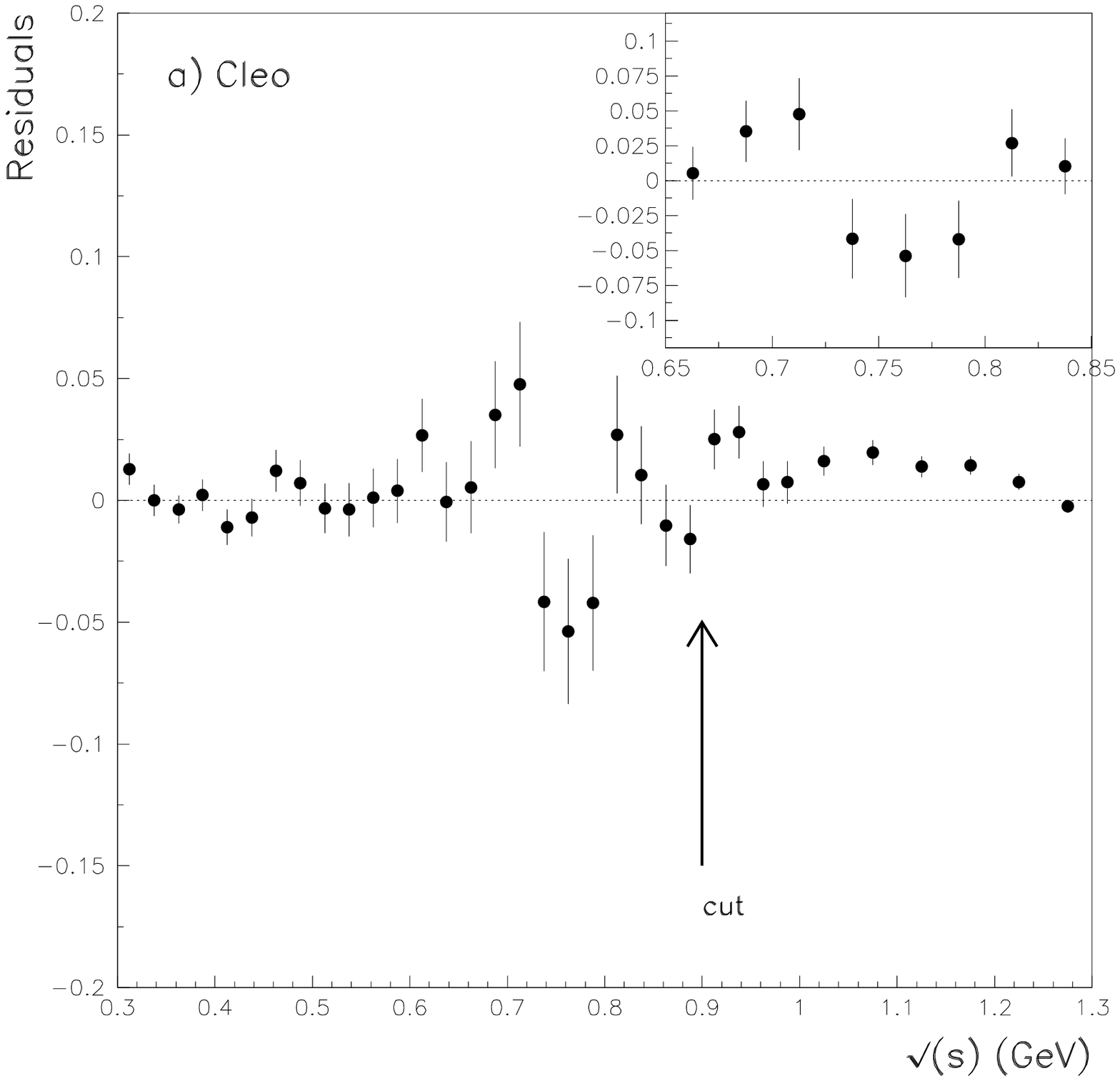}}
\end{center}
\end{minipage}
\begin{minipage}{0.5\textwidth}
\begin{center}
\resizebox{\textwidth}{!}
{\includegraphics*[width=5cm]{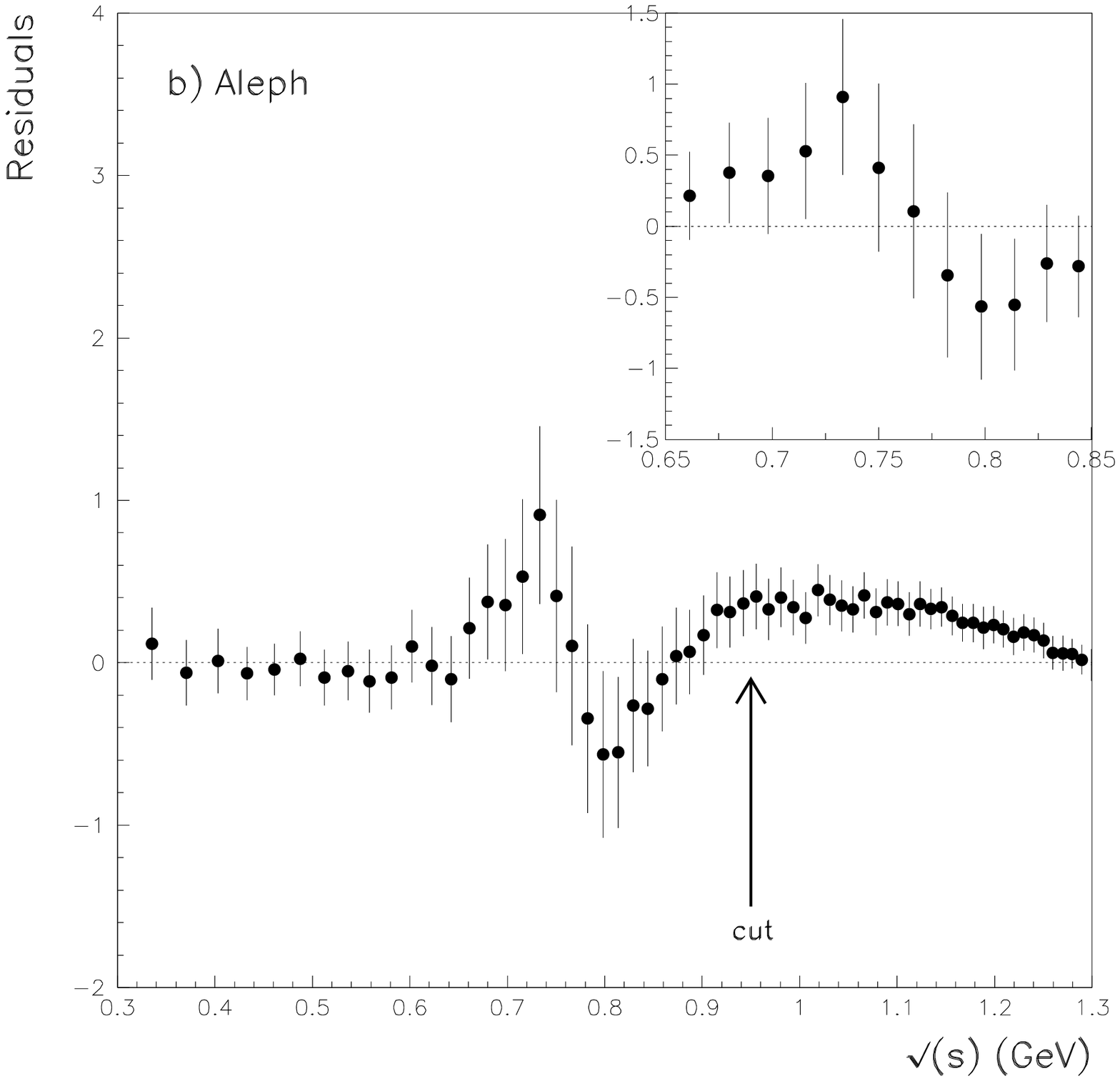}}
\end{center}
\end{minipage}
\begin{minipage}{0.5\textwidth}
\begin{center}
\resizebox{\textwidth}{!}
{\includegraphics*[width=5cm]{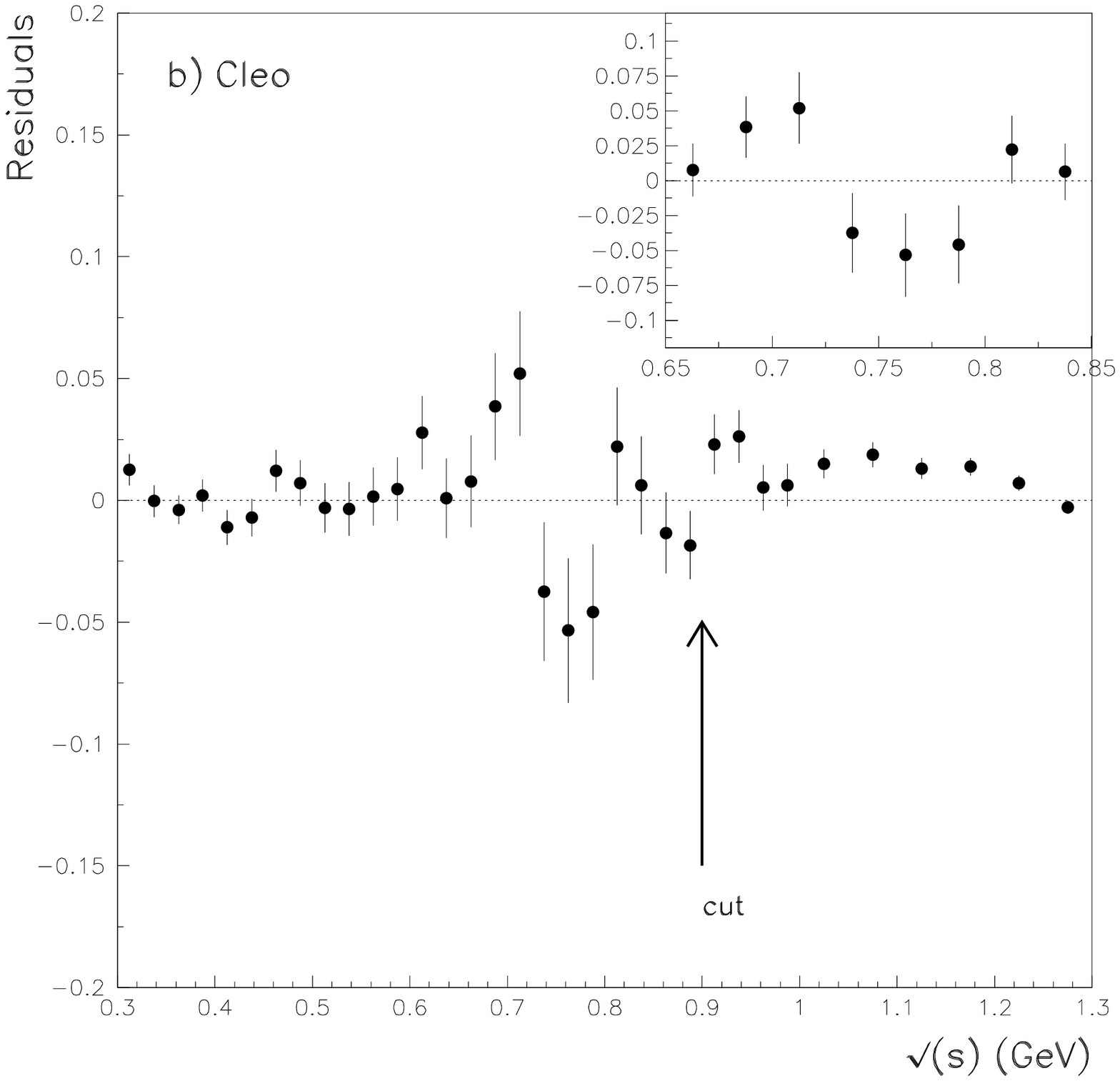}}
\end{center}
\end{minipage}
\begin{center}
\vspace{-0.5cm}
\caption{\label{TauRes}
Distribution of fit residuals for the $\tau$ data.
The upper two plots give the residuals for ALEPH and CLEO data
within the model presented. The lower plots shows the $\tau$ data residuals
when fitting without a $\rho^0 -\rho^\pm$ mass shift ({\it i.e.} $\delta m^2 \equiv 0$
is required)~; one should note the vanishing of the structure around the  $\rho$ peak
in ALEPH data produced by  the $\rho^0 -\rho^\pm$ mass shift. One should also note
that  the CLEO residual  is not modified. }
\end{center}
\end{figure}

\newpage
\begin{figure}[!ht]
\begin{minipage}{0.5 \textwidth}
\begin{center}
\resizebox{\textwidth}{!}
{\includegraphics*{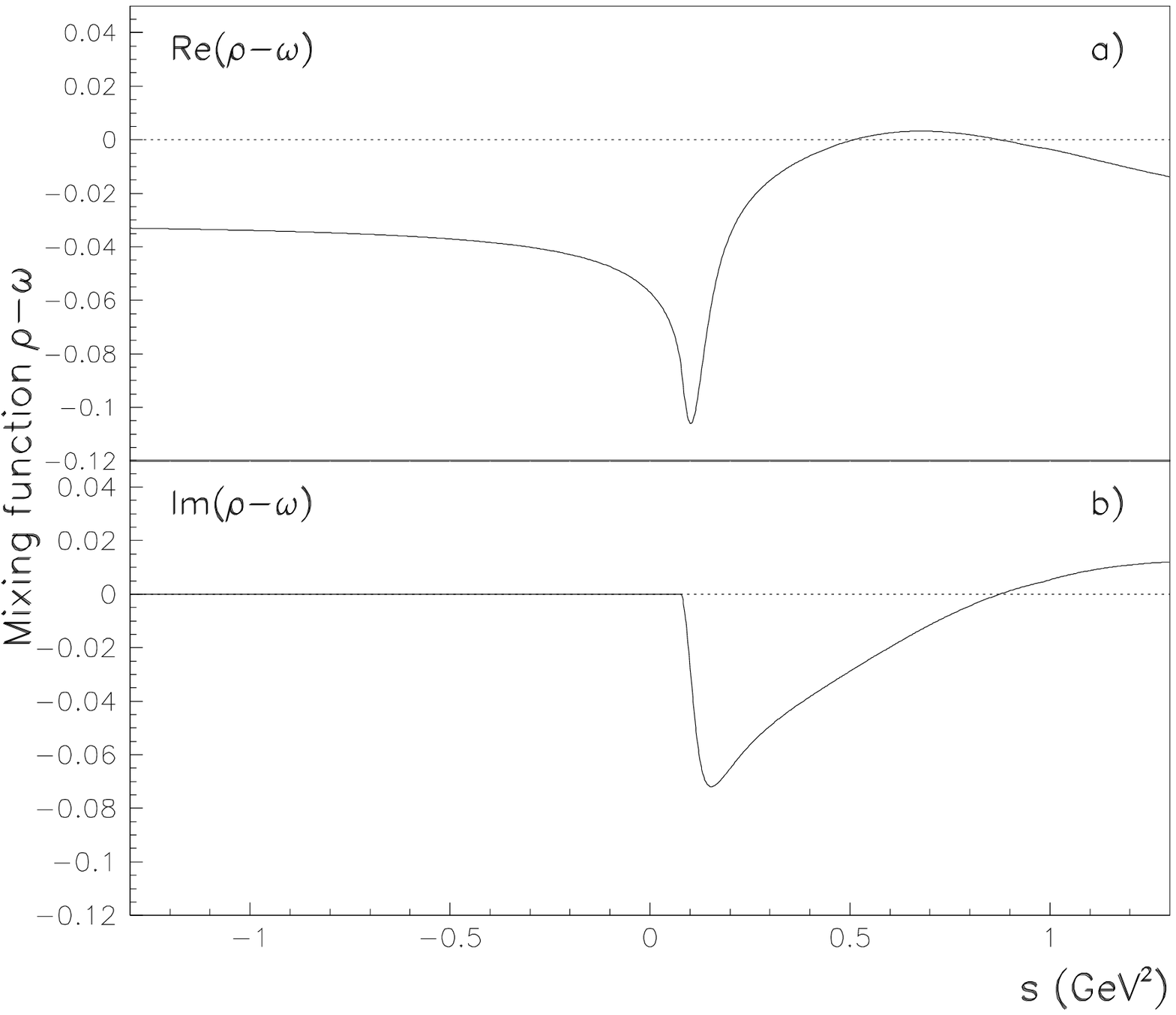}}
\end{center}
\end{minipage}
\begin{minipage}{0.5 \textwidth}
\begin{center}
\resizebox{\textwidth}{!}
{\includegraphics*{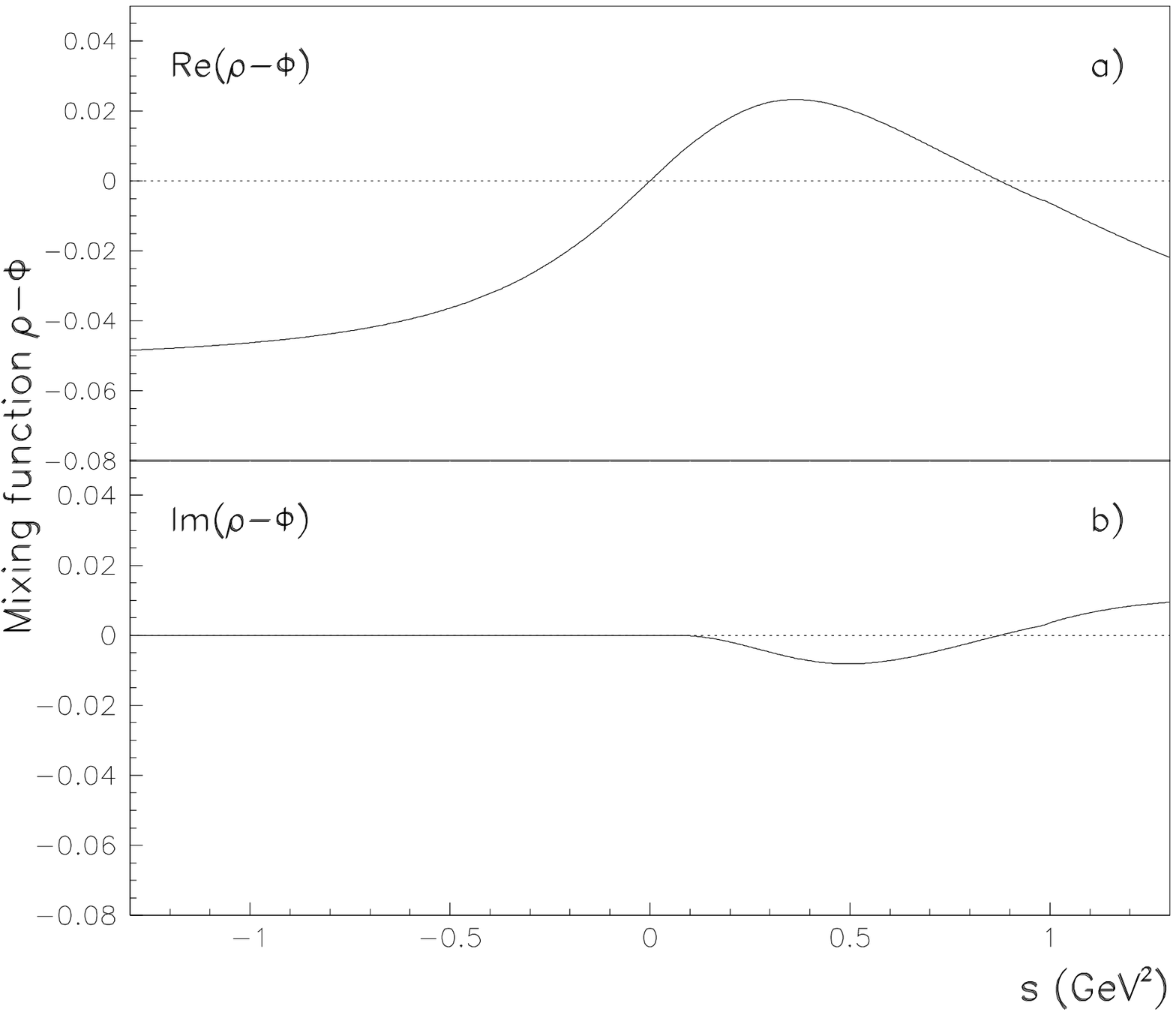}}
\end{center}
\end{minipage}
\begin{minipage}{0.8 \textwidth}
\vspace{-1.5cm}
\begin{center}
\resizebox{\textwidth}{!}
{\includegraphics*{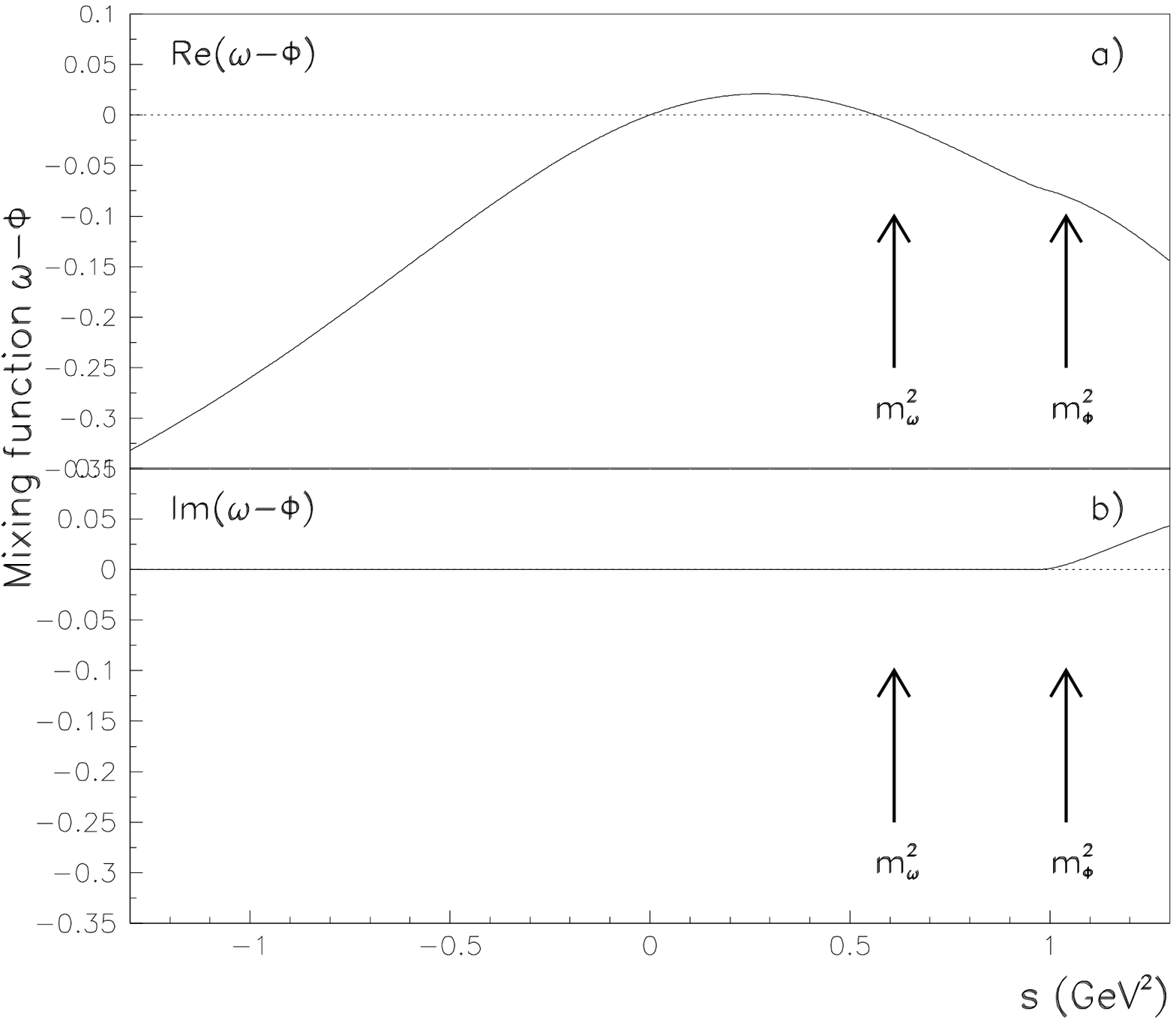}}
\end{center}
\end{minipage}
\vspace{-0.5cm}
\caption{\label{MixAng}
Matrix elements producing the neutral vector meson mixing.
The functions shown are those given in Eqs. (\ref{mixingAngles})
with their name recalled in each Figure.
The upper part of each plot gives the real part of the function, the
lower part, its imaginary part.}
\end{figure}

\newpage
\begin{figure}[!ht]
\begin{minipage}{0.8 \textwidth}
\begin{center}
\resizebox{\textwidth}{!}
{\includegraphics*{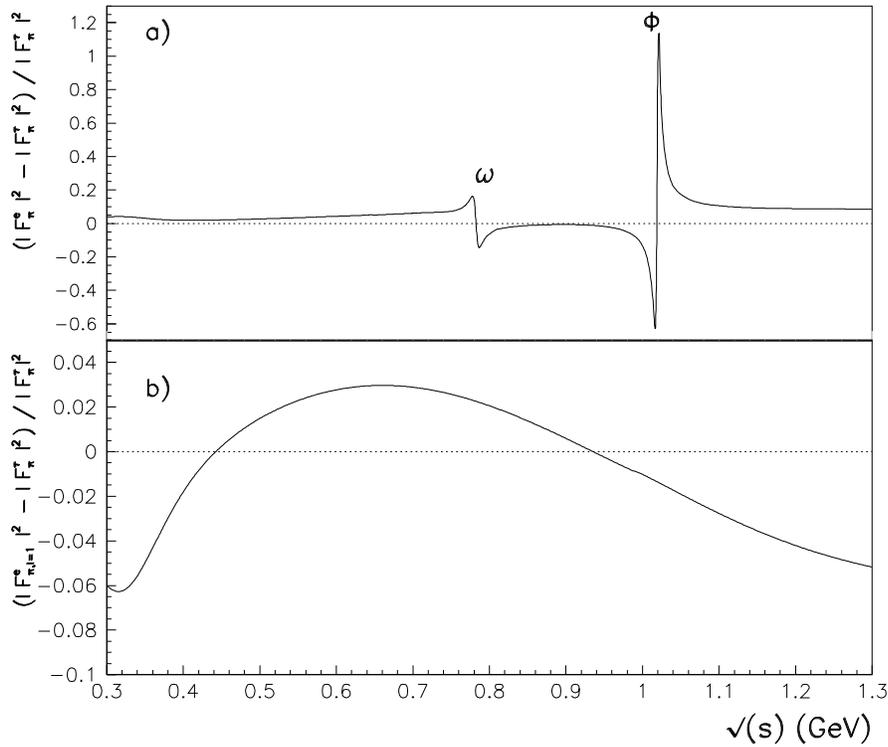}}
\end{center}
\end{minipage}
\vspace{-0.5cm}
\caption{\label{CVCFig}
Isospin symmetry breaking effects following from the 
$\rho^0-\omg-\phi$ mixing scheme. The upper plot
shows the difference between $|F_\pi^e(s)|^2$ and
$|F_\pi^\tau(s)|^2$ normalized to the latter.
The lower plot instead shows the $\rho$ part of 
$|F_\pi^e(s)|^2$. }
\end{figure}




\end{document}